\documentclass[fleqn,usenatbib]{mnras}

\usepackage[T1]{fontenc}

\usepackage[dvipsnames]{xcolor}

\DeclareRobustCommand{\VAN}[3]{#2}
\let\VANthebibliography\thebibliography
\def\thebibliography{\DeclareRobustCommand{\VAN}[3]{##3}\VANthebibliography}

\usepackage{graphicx}	
\usepackage{amsmath}	
\usepackage{amssymb}	
\usepackage{array}
\usepackage{adjustbox}

\usepackage{cprotect}

\usepackage{ulem}
\usepackage{txfonts}

\title[Integrated 3-point shear correlation function]{The integrated 3-point correlation function of cosmic shear}

\author[A. Halder et al.]{
Anik Halder,$^{1,2}$\thanks{E-mail: ahalder@usm.lmu.de}
Oliver Friedrich,$^{3,4}$ Stella Seitz,$^{1,2}$ Tamas N. Varga$^{1,2}$
\\
$^{1}$Universitäts-Sternwarte, Fakultät für Physik, Ludwig-Maximilians Universität München, Scheinerstr. 1, 81679 München, Germany\\
$^{2}$Max Planck Institute for Extraterrestrial Physics, Giessenbachstrasse 1, 85748 Garching, Germany\\
$^{3}$Kavli Institute for Cosmology, University of Cambridge, CB3 0HA Cambridge, United Kingdom\\
$^{4}$Churchill College, University of Cambridge, CB3 0DS Cambridge, United Kingdom
}

\date{Accepted XXX. Received YYY; in original form ZZZ}

\pubyear{2021}

\begin{document}
\label{firstpage}
\pagerange{\pageref{firstpage}--\pageref{lastpage}}
\maketitle

\begin{abstract}
We present the integrated 3-point shear correlation function $i\zeta_{\pm}$ --- a higher-order statistic of the cosmic shear field --- which can be directly estimated in wide-area weak lensing surveys without measuring the full 3-point shear correlation function, making this a practical and complementary tool to 2-point statistics for weak lensing cosmology. We define it as the 1-point aperture mass statistic $M_{\mathrm{ap}}$ measured at different locations on the shear field correlated with the corresponding \textit{local} 2-point shear correlation function $\xi_{\pm}$. Building upon existing work on the integrated bispectrum of the weak lensing convergence field, we present a theoretical framework for computing the integrated 3-point function in real space for any projected field within the flat-sky approximation and apply it to cosmic shear. Using analytical formulae for the non-linear matter power spectrum and bispectrum, we model $i\zeta_{\pm}$ and validate it on N-body simulations within the uncertainties expected from the sixth year cosmic shear data of the Dark Energy Survey. We also explore the Fisher information content of $i\zeta_{\pm}$ and perform a joint analysis with $\xi_{\pm}$ for two tomographic source redshift bins with realistic shape-noise to analyse its power in constraining cosmological parameters. We find that the joint analysis of $\xi_{\pm}$ and $i\zeta_{\pm}$ has the potential to considerably improve parameter constraints from $\xi_{\pm}$ alone, and can be particularly useful in improving the figure of merit of the dynamical dark energy equation of state parameters from cosmic shear data.
\end{abstract}




\section{Introduction}
Weak gravitational lensing involves the study of the cosmic shear field $\gamma$ --- coherent distortions imprinted in the shapes of background source galaxies by the gravitational lensing effect of foreground matter distribution in the Universe \citep{BartelmannSchneider2001,Schneider_2006, Kilbinger_2015}. Statistical analysis of the shear field facilitates the inference of various cosmological model parameters describing the foreground (late-time) matter field. The spatial distribution of these late-time matter density fluctuations consists of several, moderately underdense regions (e.g. voids) and relatively fewer, but highly overdense regions (e.g. galaxies, galaxy clusters) which have emerged through the interplay of gravitational and baryonic processes over billions of years. As a consequence, the late-time density fluctuations follow a positively skewed non-Gaussian distribution. However, most of the statistical analyses currently performed on cosmic shear data are focused on the evaluation of 2-point shear correlation functions $\xi_{\pm}$ \citep{Troxel_2018, Hamana_2020, Asgari_2021} which are insensitive to the information contained in the higher-order moments of the distribution. Therefore, the need for exploring methods beyond 2-point statistics in the vast amounts of observed shear data is of paramount importance. These higher-order statistics may not constrain cosmological parameters better than 2-point correlation functions. However, due to different dependence on the parameters, they hold the potential to break parameter degeneracies which appear in 2-point analyses.

The 3-point correlation function of cosmic shear ($\gamma$-3PCF), generalized third-order aperture mass statistics  \citep{Schneider_Lombardi_2003, Schneider2005, Kilbinger2005}, weak lensing convergence bispectrum \citep{Takada_2004, Kayo2013, Sato_2013} --- are examples of third-order statistics which can probe the full 3-point information of the observed weak lensing field.  Cosmological constraints using the $\gamma$-3PCF were first reported by \citealp{Semboloni_2010} in the COSMOS survey and by \citealp{Fu2014} in the CFHTLS survey\footnote{COSMOS - Cosmic Evolution Survey \url{https://cosmos.astro.caltech.edu}; CFHTLS - Canada France Hawaii Telescope Legacy Survey \url{https://www.cfht.hawaii.edu/Science/CFHTLS/} .}. However, in current weak lensing surveys (such as DES, KiDs, HSC\footnote{DES - Dark Energy Survey \url{https://www.darkenergysurvey.org}; KiDS - Kilo Degree Survey \url{http://kids.strw.leidenuniv.nl/index.php}; HSC - Hyper Suprime-Cam survey \url{https://hsc.mtk.nao.ac.jp/ssp/} .}) which span thousand square degrees and larger areas on the sky (much larger than COSMOS and CFHTLS), measuring and analysing the full $\gamma$-3PCF remains unexplored due to both theoretical and observational challenges.

Hence, in recent years, many alternate methods to probe parts of the higher-order information in the cosmic shear field have been proposed and some even measured in data, which although do not capture the full 3-point information, are easier to measure and model than $\gamma$-3PCF. Examples are shear peak statistics \citep{Kacprzak2016MNRAS}, shear peak counts and minima \citep{Zurcher_2021}, density split statistics \citep{Friedrich_2018, gruen_2018, Burger_2020}, lensing mass-map moments \citep{Chang2018, Gatti_2020}, and joint analyses of shear peaks with $\xi_{\pm}$ \citep{Martinet2021,Harnois-Deraps2020} to name a few. Most of them show potential in putting tighter constraints on cosmological parameters obtained from $\xi_{\pm}$ alone.

In this paper we propose another such statistic which can be measured directly from cosmic shear data, namely\footnote{The $i$ in $i\zeta_{\pm}$ stands for `integrated' and should not be confused with the complex imaginary unit $\sqrt{-1}$.}, \textit{the integrated 3-point shear correlation function} $i\zeta_{\pm}$. We define the statistic as the \textit{aperture mass} measured using a compensated filter at several locations, and correlate them with the \textit{position-dependent shear 2-point correlation function} measured within top-hat patches at the corresponding locations. Some key aspects that we explore in this paper are the following:
\begin{itemize}
    \item This statistic is the real space counterpart of the recently introduced \textit{integrated bispectrum} of the weak lensing convergence field $\kappa$ as studied by \citealp{munshi2020estimating, jung2021integrated}. In this paper, we build upon the existing work and formulate a theoretical model for our real space statistic on the shear field $\gamma$ and validate it on simulated cosmic shear maps.
    \item The most desirable feature of $i\zeta_{\pm}$ is that it can be easily measured from the observed shear field, a direct observable. This is possible because we define $i\zeta_{\pm}$ using an aperture mass --- a weighted measurement of the shear field at a given location using a compensated window which filters out a constant convergence mass sheet --- and the position-dependent 2-point shear correlation function which is intuitively the $\xi_{\pm}$ measured within top-hat patches (with area of a few square degrees). Our definition is different from \citealp{munshi2020estimating} who work with the convergence field in Fourier space and accordingly define the integrated convergence bispectrum $iB_{\kappa}$ using the local mean convergence measured within a top-hat patch instead of using a compensated filter. If one would want to measure $iB_{\kappa}$ then it would first be necessary to construct a convergence map from the observed shear field. This map-making process is not at all straight forward in the presence of complicated survey geometry and masks. 
    \item We investigate the information content of $i\zeta_{\pm}$ for a DES-sized tomographic survey in terms of Fisher constraints on cosmological parameters. This is the first work to perform such an analysis in the context of the integrated weak lensing bispectrum.
\end{itemize}
We organise the paper in the following manner. In chapter \ref{chap:theory_general_formalism} we formulate the integrated 3-point function statistic for any projected field within the flat-sky approximation and then apply it to the case for the cosmic shear field in chapter \ref{chap:theory_application}. In chapter \ref{chap:methods} we describe the simulations and numerical methods we use in order to measure and theoretically model the statistic. Finally, in chapter \ref{chap:results} we validate our theoretical model on the simulations and present Fisher constraints on cosmological parameters. 
Throughout this paper we assume flat cosmology i.e. $\Omega_{\mathrm{K}} = 0$. As we mainly work with projected 2D quantities, we differentiate them from 3D quantities by explicitly specifying the sub or super-script `3D' for the latter. 

\section{Theory I: General formalism}
\label{chap:theory_general_formalism}
In this chapter we formulate the general framework of equations required for describing the integrated 3-point function (in real-space) and the integrated bispectrum (in Fourier space) of any projected 2D field within the flat-sky approximation. For this chapter and the next, we provide a summary of this technical part of the paper at the end of chapter \ref{chap:theory_application}. Readers may feel free to skip these theoretical details and directly refer to the summary in section \ref{sec:theory_summary}.

\subsection{Projected fields}

Any cosmic field $f^{\mathrm{3D}}\big[\boldsymbol{\chi}, \eta \big]$ that we observe on our past light-cone at 3D comoving position $\boldsymbol{\chi}$ and corresponding conformal lookback time $\eta = \eta_0 - \chi$ (where $\eta_0$ is the conformal time today and $\chi$ the radial comoving distance), can be projected onto the 2D celestial sphere to obtain the weighted line-of-sight 2D quantity $f(\hat{\mathbf{n}})$ towards a radial unit direction $\hat{\mathbf{n}}$ \citep{BartelmannSchneider2001},
\begin{equation} 
\begin{split}
    f(\hat{\mathbf{n}}) & = \int \mathrm{d} \chi \; q_{\mathrm{f}}(\chi) f^{\mathrm{3D}}\big[\chi\hat{\mathbf{n}}, \eta_0 - \chi \big]
\end{split}
\end{equation}
where $q_{\mathrm{f}}(\chi)$ is a particular weighting kernel over which $f^{\mathrm{3D}}$ is projected. Examples are the projected galaxy number density or the weak lensing convergence field that we observe on the celestial sphere. Assuming that the angular extent of the field of view is small --- spanning an area of a few square degrees --- we can make the flat-sky approximation, where we denote the position on the sky as a 2D planar vector $\boldsymbol{\theta} = (\theta_x, \theta_y)$ and express $f$ as
\begin{equation} \label{eqn_projected_field_flat_sky}
\begin{split}
    f(\boldsymbol{\theta}) & = \int \mathrm{d} \chi \; q_{\mathrm{f}}(\chi) f^{\mathrm{3D}}\big[ (\chi \boldsymbol{\theta},\chi), \eta_0 - \chi \big] \ . 
\end{split}
\end{equation}

\subsection{The projected power spectrum and bispectrum}

The 2D power spectrum ($P_{\mathrm{gh}}$) and bispectrum ($B_{\mathrm{fgh}}$) of projected fields $f, g, h$ are defined as \citep{BartelmannSchneider2001}:
\begin{equation} \label{eq:projected_power_spectrum_definition}
    \big\langle g(\boldsymbol{l}_1) h(\boldsymbol{l}_2) \big\rangle \equiv (2\pi)^2 \delta_{D}(\boldsymbol{l}_1 + \boldsymbol{l}_2) \;  P_{\mathrm{gh}}(\boldsymbol{l}_1)
\end{equation} 
\begin{equation} \label{eq:projected_bispectrum_definition}
    \big\langle f(\boldsymbol{l}_1) g(\boldsymbol{l}_2) h(\boldsymbol{l}_3) \big\rangle \equiv (2\pi)^2 \delta_{D}(\boldsymbol{l}_1 + \boldsymbol{l}_2 + \boldsymbol{l}_3 ) \;  B_{\mathrm{fgh}}(\boldsymbol{l}_1, \boldsymbol{l}_2, \boldsymbol{l}_3)
\end{equation}
where $\langle ... \rangle$ denotes an ensemble average over different realizations of the Universe and $\delta_{D}$ denotes the Dirac delta function. $f(\boldsymbol{l})$ corresponds to the Fourier space representation\footnote{In this paper we do not use any distinguishing symbol (e.g. the commonly used tilde) for separately denoting the Fourier space representation of the field $f$. The Fourier representation is left understood when $f$ appears with argument $\boldsymbol{l}$ or $\boldsymbol{q}$ (2D Fourier wave-vectors).} of field $f(\boldsymbol{\theta})$ (see Appendix \ref{app:hankel}); and similarly for fields $g,h$. The bispectrum is defined for closed triangle configurations $\boldsymbol{l}_1 + \boldsymbol{l}_2 + \boldsymbol{l}_3 = 0$.

These 2D spectra can be expressed as line-of-sight projections of the power spectrum $P^{\mathrm{3D}}_{\mathrm{gh}}(\boldsymbol{k}, \eta)$, and bispectrum $B^{\mathrm{3D}}_{\mathrm{fgh}}(\boldsymbol{k}_1, \boldsymbol{k}_2, \boldsymbol{k}_3, \eta)$ of the 3D fields $f^{\mathrm{3D}}$, $g^{\mathrm{3D}}$, $h^{\mathrm{3D}}$ with $\boldsymbol{k}_i$ corresponding to 3D Fourier wave-vectors. This can be computed using the Limber approximation \citep{Limber1954, Kaiser1992, Buchalter_2000}:
\begin{equation}
\label{eq:projected_power_spectrum}
    P_{\mathrm{gh}}(\boldsymbol{l})= \int \mathrm{d}\chi \frac{q_{\mathrm{g}}(\chi)q_{\mathrm{h}}(\chi)}{\chi^2} P^{\mathrm{3D}}_{\mathrm{gh}}\left(\boldsymbol{k} = \frac{\boldsymbol{l}}{\chi}, \eta_0 - \chi \right)
\end{equation}
\begin{equation}
\label{eq:projected_bispectrum}
    B_{\mathrm{fgh}}(\boldsymbol{l}_1,\boldsymbol{l}_2,\boldsymbol{l}_3)= \int \mathrm{d} \chi \; \frac{q_{\mathrm{f}}(\chi)q_{\mathrm{g}}(\chi)q_{\mathrm{h}}(\chi)}{\chi^4} B^{\mathrm{3D}}_{\mathrm{fgh}}\left( \frac{\boldsymbol{l}_1}{\chi}, \frac{\boldsymbol{l}_2}{\chi}, \frac{\boldsymbol{l}_3}{\chi}, \eta_0 - \chi\right) 
\end{equation}
where $q_{\mathrm{f}}(\chi)$, $q_{\mathrm{g}}(\chi)$ and $q_{\mathrm{h}}(\chi)$ are the weighting kernels with which $f^{\mathrm{3D}}$, $g^{\mathrm{3D}}$, and $h^{\mathrm{3D}}$ are projected, respectively. Under the assumptions of an isotropic Universe, the power spectrum is independent of the direction of the wave-vector and the bispectrum does not depend on the orientation of the closed triangle of its wave-vectors. It should be noted again that these expressions are written assuming that the Universe is flat. However, it is straight forward to generalize these equations to a universe with non-zero spatial curvature (see \citealp{BartelmannSchneider2001, Schneider_2006}).

\subsection{The integrated 3-point function and integrated bispectrum of projected fields}

The integrated bispectrum $iB^{\mathrm{3D}}(k)$ of the 3D matter density contrast field was first studied by \citealp{Chiang_2014} who defined it as the correlation of the local mean density perturbation and the position-dependent power spectrum evaluated within 3D sub-volumes. They showed that this correlation can be expressed as integrals over different $\boldsymbol{k}$-modes of the full 3D matter density contrast bispectrum $B_{\delta}^{\mathrm{3D}}$. \citealp{Chiang_2015} studied the real space counterpart of $iB^{\mathrm{3D}}(k)$, namely the integrated 3-point function $i\zeta^{\mathrm{3D}}(r)$ which they showed to be the correlation of the local mean density perturbation and the position-dependent 2-point correlation function within 3D sub-volumes and presented the first detection of $i\zeta
^{\mathrm{3D}}(r)$ in the BOSS DR10 CMASS galaxy sample. The integrated bispectrum has also found other applications, for example in studying the Lyman alpha forest, quasars \citep{Doux2016, Chiang2017b, Chiang2017a} and also the 21 cm line in the epoch of reionization \citep{Giri2018}. Recently, \citealp{Munshi_2017, Munshi:2019csw, munshi2020estimating, Jung_2020, jung2021integrated} have extended the formalism to the integrated bispectrum $iB(l)$ of projected 2D fields. In particular, \citealp{munshi2020estimating}, studied this in the context of the weak lensing convergence field and developed various theoretical models for the same. In this section we build upon the mathematical formalism of the integrated bispectrum developed in these previous works and introduce its real space counterpart the integrated 3-point function $i\zeta(\theta)$ for any projected 2D field.

\subsubsection{Projected field within 2D window}

The central quantity to our discussion will be the projected field $f(\boldsymbol{\theta}; \boldsymbol{\theta}_C)$ at a given location $\boldsymbol{\theta}$ on the flat-sky, weighted by an azimuthally symmetric 2D window function $W$ (of a given size or characteristic scale) centred at $\boldsymbol{\theta}_C$
\begin{equation} \label{eq:2D_field_in_window_real}
    f(\boldsymbol{\theta}; \boldsymbol{\theta}_C) \equiv f(\boldsymbol{\theta})W(\boldsymbol{\theta}_C-\boldsymbol{\theta}) ,
\end{equation}
where $W(\boldsymbol{\theta}_C-\boldsymbol{\theta}) = W(\boldsymbol{\theta}-\boldsymbol{\theta}_C) = W(\vert\boldsymbol{\theta}_C-\boldsymbol{\theta} \vert)$.
For example, if the window function centred at $\boldsymbol{\theta}_C$ is a top-hat of size $\theta_{\mathrm{T}}$, then $f(\boldsymbol{\theta}; \boldsymbol{\theta}_C) = f(\boldsymbol{\theta})$ only when $\vert\boldsymbol{\theta}_C-\boldsymbol{\theta} \vert \leq \theta_{\mathrm{T}}$, otherwise $f(\boldsymbol{\theta}; \boldsymbol{\theta}_C) = 0$. Its local Fourier transform (see Appendix \ref{app:hankel}) is given by
\begin{equation} \label{eq:2D_field_in_window_Fourier}
\begin{split}
    f(\boldsymbol{l}; \boldsymbol{\theta}_C) \equiv \mathcal{F}_{\mathrm{2D}}[f(\boldsymbol{\theta}; \boldsymbol{\theta}_C)] & = \int \mathrm{d}^2 \boldsymbol{\theta} \; f(\boldsymbol{\theta})W(\boldsymbol{\theta}_C-\boldsymbol{\theta}) e^{-i\boldsymbol{l}\cdot\boldsymbol{\theta}} \\
    & = \int \frac{\mathrm{d}^2 \boldsymbol{l}_1}{(2\pi)^2} f(\boldsymbol{l}_1) W(\boldsymbol{l}_1-\boldsymbol{l}) e^{i(\boldsymbol{l}_1-\boldsymbol{l}) \cdot \boldsymbol{\theta}_C} 
\end{split}
\end{equation}
where$f$ can be any complex/real 2D field defined in a tomographic bin with projection kernel $q_{\mathrm{f}}$ e.g. projected galaxy density contrast, weak lensing convergence $\kappa$, weak lensing shear $\gamma$. $f(\boldsymbol{l})$ and $W(\boldsymbol{l})$ are the Fourier space representations of $f(\boldsymbol{\theta})$ and $W(\boldsymbol{\theta})$, respectively. If $f$ is a real field i.e. $f^*(-\boldsymbol{l}) = f(\boldsymbol{l})$ then we can easily see that $f^*(\boldsymbol{l}; \boldsymbol{\theta}_C) = f(-\boldsymbol{l}; \boldsymbol{\theta}_C)$.

\subsubsection{Position-dependent weighted mean of projected field}
We can now find the weighted mean of $f(\boldsymbol{\theta}; \boldsymbol{\theta}_C)$ defined within the 2D window\footnote{We use the subscript `1pt' for the window function in the equation for the weighted mean of a field inside the window $W_{\mathrm{1pt}}$ at a given location to distinguish it from the case when we compute the position dependent 2-point function within a different window $W$ at the same location (see equation \eqref{eq:position_dependent_2pt_function_2D_field_gh}.} $W_{\mathrm{1pt}}$ at $\boldsymbol{\theta}_C$ as
\begin{equation} \label{eq:weighted_mean_2D_field_in_window}
\begin{split}
    \Bar{f}( \boldsymbol{\theta}_C) \equiv \frac{1}{A_{\mathrm{1pt}}} \int \mathrm{d}^2 \boldsymbol{\theta} \; f(\boldsymbol{\theta}; \boldsymbol{\theta}_C) & = \frac{1}{A_{\mathrm{1pt}}} \int \mathrm{d}^2 \boldsymbol{\theta} \; f(\boldsymbol{\theta}) W_{\mathrm{1pt}}(\boldsymbol{\theta}_C-\boldsymbol{\theta}) \\
    & = \frac{1}{A_{\mathrm{1pt}}} \int \frac{\mathrm{d}^2 \boldsymbol{l}}{(2\pi)^2} \; f(\boldsymbol{l}) W_{\mathrm{1pt}}(\boldsymbol{l}) \; e^{i\boldsymbol{l}\cdot \boldsymbol{\theta}_C} \\
\end{split}
\end{equation}
where for the final equality we have used the convolution theorem and have defined the 1-point area normalisation term as
\begin{equation} \label{eq:area_normalisation_weighted_mean_2D_field_in_window}
    A_{\mathrm{1pt}} \equiv \int \mathrm{d}^2 \boldsymbol{\theta} \; W_{\mathrm{1pt}}(\boldsymbol{\theta}_C-\boldsymbol{\theta}) \ .
\end{equation}
Note that this normalisation term is a purely geometric factor independent of the location $\boldsymbol{\theta}_C$ of the window (evaluating it at any $\boldsymbol{\theta}_C$ gives the same result and for simplicity we evaluate it at $\boldsymbol{\theta}_C = \mathbf{0}$; see also footnote \ref{footnote:holes}). If we use a normalised window function i.e. $A_{\mathrm{1pt}} = 1$ or a compensated filter \citep{Schneider_2006}  --- which shall be important when we consider aperture masses (see section \ref{sec:shear_measures_2pt_Map}) --- then we do not need to consider this normalisation factor. From equations \eqref{eq:2D_field_in_window_Fourier} and \eqref{eq:weighted_mean_2D_field_in_window} we can see that
\begin{equation}
\begin{split}
    \Bar{f}( \boldsymbol{\theta}_C) & = \frac{1}{A_{\mathrm{1pt}}} \; f(\boldsymbol{l} = \mathbf{0}; \boldsymbol{\theta}_C)\ .
\end{split}
\end{equation}

\subsubsection{Position-dependent 2-point function of projected fields}
The 2-point correlation (as a function of the separation 2D vector $\boldsymbol{\alpha}$) of projected fields $g$ and $h$ is defined as
\begin{equation}
    \xi_{\mathrm{gh}}(\boldsymbol{\alpha}) \equiv \big\langle g(\boldsymbol{\theta}) h(\boldsymbol{\theta+\alpha})\big\rangle \ .
\end{equation}
This is the real space counterpart of the projected power spectrum  $P_{\mathrm{gh}}(\boldsymbol{l})$:
\begin{equation}
    \xi_{\mathrm{gh}}(\boldsymbol{\alpha}) =  \mathcal{F}_{\mathrm{2D}}^{-1}[P_{\mathrm{gh}}(\boldsymbol{l})] = \int \frac{\mathrm{d}^2 \boldsymbol{l}}{(2\pi)^2} \; P_{\mathrm{gh}}(\boldsymbol{l}) e^{i\boldsymbol{l}\cdot \boldsymbol{\alpha}} \ .
\end{equation}
Considering isotropic fields, this inverse 2D Fourier transformation becomes an inverse Hankel transform (see Appendix \ref{app:hankel}): $\xi_{\mathrm{gh}}(\alpha) =  \mathcal{F}_{\mathrm{2D}}^{-1}[P_{\mathrm{gh}}(l)]$ where the correlation function (power spectrum) is independent of the direction of the separation vector $\boldsymbol{\alpha}$ (Fourier mode $\boldsymbol{l}$).

For ergodic fields, we can write the expression for this 2-point correlation function evaluated within a finite region of area $A$ as
\begin{equation}
    \hat{\xi}_{\mathrm{gh}}(\boldsymbol{\alpha}) \equiv \frac{1}{A} \int \mathrm{d}^2 \boldsymbol{\theta} \; g(\boldsymbol{\theta}) h(\boldsymbol{\theta}+\boldsymbol{\alpha}) 
\end{equation}
where the integrand for a given separation $\boldsymbol{\alpha}$, is defined only for those points $\boldsymbol{\theta}$ for which both $\boldsymbol{\theta}$ and $\boldsymbol{\theta}+\boldsymbol{\alpha}$ lie within the boundary of the region under consideration. As $A \rightarrow \infty$, $\hat{\xi}_{\mathrm{gh}}(\boldsymbol{\alpha}) \rightarrow  \xi_{\mathrm{gh}}(\boldsymbol{\alpha})$. However, if the region spans only a small area (e.g. a small 2D aperture on the sky), then this limit does not hold and instead the expression $\hat{\xi}_{\mathrm{gh}}(\boldsymbol{\alpha})$ evaluates to a value which depends on the location of the aperture. Hence, we now formally define the expression for the \textit{position-dependent 2-point correlation function} $\hat{\xi}_{\mathrm{gh}}(\boldsymbol{\alpha};\boldsymbol{\theta}_C)$ of the projected fields $g$ and $h$ both defined within a 2D aperture $W$ centred at $\boldsymbol{\theta}_C$ as
\begin{equation} \label{eq:position_dependent_2pt_function_2D_field_gh}
\begin{split}
    \hat{\xi}_{\mathrm{gh}}(\boldsymbol{\alpha};\boldsymbol{\theta}_C) & \equiv \frac{1}{A_{\mathrm{2pt}}(\boldsymbol{\alpha})} \int \mathrm{d}^2 \boldsymbol{\theta} \; g(\boldsymbol{\theta};\boldsymbol{\theta}_C) h(\boldsymbol{\theta}+\boldsymbol{\alpha};\boldsymbol{\theta}_C)  \\
    & = \frac{1}{A_{\mathrm{2pt}}(\boldsymbol{\alpha})} \int \mathrm{d}^2 \boldsymbol{\theta} \; g(\boldsymbol{\theta}) W(\boldsymbol{\theta}_C-\boldsymbol{\theta})  \\ & \qquad \qquad \times  h(\boldsymbol{\theta}+\boldsymbol{\alpha})  W(\boldsymbol{\theta}_C-\boldsymbol{\theta}-\boldsymbol{\alpha}) \\
    & = \frac{1}{A_{\mathrm{2pt}}(\boldsymbol{\alpha})} \int \frac{\mathrm{d}^2 \boldsymbol{l}_1}{(2\pi)^2}  \int \frac{\mathrm{d}^2 \boldsymbol{l}_2}{(2\pi)^2} \int \frac{\mathrm{d}^2 \boldsymbol{q}}{(2\pi)^2} \; g(\boldsymbol{l}_1)
    h(\boldsymbol{l}_2) \\ & \quad \times  W(\boldsymbol{q})  W(\boldsymbol{l}_1+\boldsymbol{l}_2-\boldsymbol{q}) e^{i(\boldsymbol{l}_1+\boldsymbol{l}_2) \cdot \boldsymbol{\theta}_C} e^{i(\boldsymbol{q}-\boldsymbol{l}_1) \cdot \boldsymbol{\alpha}} \\  
\end{split}
\end{equation}
where $\boldsymbol{l}_i, \boldsymbol{q}$ are 2D Fourier wave-vectors. In the above equation $A_{\mathrm{2pt}}(\boldsymbol{\alpha})$ is the area normalisation for this projected position-dependent 2-point function and is given by
\begin{equation} \label{eq:area_normalisation_position_dependent_2pt_function_2D_field}
\begin{split}
   A_{\mathrm{2pt}}(\boldsymbol{\alpha}) & \equiv \int \mathrm{d}^2 \boldsymbol{\theta} \;  W(\boldsymbol{\theta}_C-\boldsymbol{\theta}) W(\boldsymbol{\theta}_C-\boldsymbol{\theta}-\boldsymbol{\alpha}) \\
   & = \int \frac{\mathrm{d}^2 \boldsymbol{q}}{(2\pi)^2} \; W(\boldsymbol{q})  W(-\boldsymbol{q}) e^{i \boldsymbol{q} \cdot \boldsymbol{\alpha}}
\end{split}
\end{equation}
which for simplicity we evaluate (using the first equality) at $\boldsymbol{\theta}_C = \mathbf{0}$ as this term is independent of the window's location\footnote{\label{footnote:holes}Of course, this is only true when we do not consider holes and masks in the data. To account for this, one may randomly throw away some points inside a window centred at $\boldsymbol{\theta}_C$ so as to have only those pairs of points $\{\boldsymbol{\theta},\boldsymbol{\theta}+\boldsymbol{\alpha}\}$ yielding the same effective area of another window at $\boldsymbol{\theta}_C'$ but which has masks and holes within its aperture.} $\boldsymbol{\theta}_C$. However, it is important to note that this area normalisation depends on the separation vector $\boldsymbol{\alpha}$ under consideration, unlike $A_{\mathrm{1pt}}$ defined in equation \eqref{eq:area_normalisation_weighted_mean_2D_field_in_window}. For azimuthally symmetric window functions that we are interested in, it follows from isotropy considerations that this normalisation term only depends on the magnitude $\alpha$ of the separation vector i.e. $A_{\mathrm{2pt}}(\boldsymbol{\alpha}) = A_{\mathrm{2pt}}(\alpha)$. Hence, one can evaluate this term for any polar angle $\phi_{\boldsymbol{\alpha}}$ (e.g. defined with respect to the x-axis of the flat-sky coordinate system). For simplicity, we shall consider $\phi_{\boldsymbol{\alpha}} = 0$.

On the other hand, for the position-dependent 2-point correlation function of field $g$ with the complex-conjugated field $h^*$ we have
\begin{equation} \label{eq:position_dependent_2pt_function_2D_field_gh*}
\begin{split}
    \hat{\xi}_{\mathrm{gh^*}}(\boldsymbol{\alpha};\boldsymbol{\theta}_C) & \equiv \frac{1}{A_{\mathrm{2pt}}(\boldsymbol{\alpha})} \int \mathrm{d}^2 \boldsymbol{\theta} \; g(\boldsymbol{\theta};\boldsymbol{\theta}_C) h^*(\boldsymbol{\theta}+\boldsymbol{\alpha};\boldsymbol{\theta}_C)  \\
    & = \frac{1}{A_{\mathrm{2pt}}(\boldsymbol{\alpha})} \int \frac{\mathrm{d}^2 \boldsymbol{l}_1}{(2\pi)^2}  \int \frac{\mathrm{d}^2 \boldsymbol{l}_2}{(2\pi)^2} \int \frac{\mathrm{d}^2 \boldsymbol{q}}{(2\pi)^2} \; g(\boldsymbol{l}_1)
    h^*(-\boldsymbol{l}_2) \\ & \quad \times  W(\boldsymbol{q})  W(\boldsymbol{l}_1+\boldsymbol{l}_2-\boldsymbol{q}) e^{i(\boldsymbol{l}_1+\boldsymbol{l}_2) \cdot \boldsymbol{\theta}_C} e^{i(\boldsymbol{q}-\boldsymbol{l}_1) \cdot \boldsymbol{\alpha}} \ .
\end{split}
\end{equation}
In case the field $h$ is real i.e. $h^*(-\boldsymbol{l}) = h(\boldsymbol{l})$, it follows from equation \eqref{eq:position_dependent_2pt_function_2D_field_gh} that $\hat{\xi}_{\mathrm{gh^*}}(\boldsymbol{\alpha};\boldsymbol{\theta}_C) = \hat{\xi}_{\mathrm{gh}}(\boldsymbol{\alpha};\boldsymbol{\theta}_C)$.

The position-dependent correlation function gives an unbiased estimate of the 2-point correlation function i.e. $\big\langle \hat{\xi}_{\mathrm{gh}}(\boldsymbol{\alpha};\boldsymbol{\theta}_C) \big\rangle = \xi_{\mathrm{gh}}(\boldsymbol{\alpha})$. Also, when we consider the fields and the window functions to be isotropic, then the above expressions only depend on the magnitude $\alpha$ of the separation vector i.e. $\hat{\xi}_{\mathrm{gh}}(\boldsymbol{\alpha};\boldsymbol{\theta}_C) = \hat{\xi}_{\mathrm{gh}}(\alpha;\boldsymbol{\theta}_C)$.

\subsubsection{Position-dependent power spectrum of projected fields}

The power spectrum is the forward Fourier transform of the 2-point correlation function. Hence, we define the Fourier space counterpart of $\hat{\xi}_{\mathrm{gh}}(\boldsymbol{\alpha};\boldsymbol{\theta}_C)$ as
\begin{equation} \label{eq:position_dependent_power_spectrum_2D_field_gh}
\begin{split}
    \hat{P}_{\mathrm{gh}}(\boldsymbol{l}; \boldsymbol{\theta}_C) & \equiv \mathcal{F}_{\mathrm{2D}}[A_{\mathrm{2pt}}(\boldsymbol{\alpha}) \hat{\xi}_{\mathrm{gh}}(\boldsymbol{\alpha};\boldsymbol{\theta}_C)] \\ 
    & = \int \mathrm{d}^2 \boldsymbol{\alpha} \; A_{\mathrm{2pt}}(\boldsymbol{\alpha}) \hat{\xi}_{\mathrm{gh}}(\boldsymbol{\alpha};\boldsymbol{\theta}_C) e^{-i\boldsymbol{l} \cdot \boldsymbol{\alpha}} \\
    & = \int \frac{\mathrm{d}^2 \boldsymbol{l}_1}{(2\pi)^2}  \int \frac{\mathrm{d}^2 \boldsymbol{l}_2}{(2\pi)^2} \; g(\boldsymbol{l}_1)  h(\boldsymbol{l}_2) \\ & \qquad \qquad \times W(\boldsymbol{l}_1+\boldsymbol{l}) W(\boldsymbol{l}_2-\boldsymbol{l}) e^{i(\boldsymbol{l}_1 + \boldsymbol{l}_2) \cdot \boldsymbol{\theta}_C} \\  
    & = g(-\boldsymbol{l}; \boldsymbol{\theta}_C) h(\boldsymbol{l}; \boldsymbol{\theta}_C) \ .
\end{split}
\end{equation}
This is slightly different from the \textit{position-dependent power spectrum} definition of \citealp{Chiang_2014} who define it for the 3D matter density contrast field in their equation 2.3 with a constant volume normalisation term. On the other hand, we factor out the scale-dependent area normalisation term $A_{\mathrm{2pt}}(\boldsymbol{\alpha})$ in our definition of $\hat{P}_{\mathrm{gh}}$.

Similarly, the Fourier space counterpart of  $\hat{\xi}_{\mathrm{gh^*}}(\boldsymbol{\alpha};\boldsymbol{\theta}_C)$ can be written as
\begin{equation} \label{eq:position_dependent_power_spectrum_2D_field_gh*}
\begin{split}
    \hat{P}_{\mathrm{gh^*}}(\boldsymbol{l}; \boldsymbol{\theta}_C) & \equiv \mathcal{F}_{\mathrm{2D}}[A_{\mathrm{2pt}}(\boldsymbol{\alpha}) \hat{\xi}_{\mathrm{gh^*}}(\boldsymbol{\alpha};\boldsymbol{\theta}_C)] \\ 
    & = \int \frac{\mathrm{d}^2 \boldsymbol{l}_1}{(2\pi)^2}  \int \frac{\mathrm{d}^2 \boldsymbol{l}_2}{(2\pi)^2} \; g(\boldsymbol{l}_1)  h^*(-\boldsymbol{l}_2) \\ & \qquad \qquad \times W(\boldsymbol{l}_1+\boldsymbol{l}) W(\boldsymbol{l}_2-\boldsymbol{l}) e^{i(\boldsymbol{l}_1 + \boldsymbol{l}_2) \cdot \boldsymbol{\theta}_C} \\  
    & = g(-\boldsymbol{l}; \boldsymbol{\theta}_C) h^*(-\boldsymbol{l}; \boldsymbol{\theta}_C) \ .
\end{split}
\end{equation}
When the field $h$ is real, $\hat{P}_{\mathrm{gh^*}}(\boldsymbol{l}; \boldsymbol{\theta}_C) = \hat{P}_{\mathrm{gh}}(\boldsymbol{l}; \boldsymbol{\theta}_C)$.

\subsubsection{Integrated 3-point function of projected fields}
\label{sec:i3pt_projected_fields}

We now define the integrated 3-point function of projected fields analogous to the 3D case \citep{Chiang_2015} --- the ensemble average (over different locations $\boldsymbol{\theta}_C$) of the product of the position-dependent weighted mean and the position-dependent 2-point function of projected fields:
\begin{equation} \label{eq:2D_iZ_flat_sky_raw}
\begin{split}
    i\zeta(\boldsymbol{\alpha}) & \equiv \Big\langle \Bar{f}( \boldsymbol{\theta}_C) \; \hat{\xi}_{\mathrm{gh}}(\boldsymbol{\alpha};\boldsymbol{\theta}_C) \Big\rangle \\
    & = \frac{1}{A_{\mathrm{1pt}} A_{\mathrm{2pt}}(\boldsymbol{\alpha})} \int \mathrm{d}^2 \boldsymbol{\theta}_1 \int \mathrm{d}^2 \boldsymbol{\theta}_2 \; \Big \langle f(\boldsymbol{\theta}_1) g(\boldsymbol{\theta}_2) h(\boldsymbol{\theta}_2+\boldsymbol{\alpha}) \Big \rangle  \\ & \qquad \qquad \times W_{\mathrm{1pt}}(\boldsymbol{\theta}_C-\boldsymbol{\theta}_1) W(\boldsymbol{\theta}_C-\boldsymbol{\theta}_2) W(\boldsymbol{\theta}_C-\boldsymbol{\theta}_2-\boldsymbol{\alpha}) \\
    & = \frac{1}{A_{\mathrm{1pt}} A_{\mathrm{2pt}}(\boldsymbol{\alpha})} \int \frac{\mathrm{d}^2 \boldsymbol{l}_1}{(2\pi)^2}  \int \frac{\mathrm{d}^2 \boldsymbol{l}_2}{(2\pi)^2} \int \frac{\mathrm{d}^2 \boldsymbol{l}_3}{(2\pi)^2} \int \frac{\mathrm{d}^2 \boldsymbol{q}}{(2\pi)^2} \\ & \qquad \qquad \qquad \times \Big\langle f(\boldsymbol{l}_1) g(\boldsymbol{l}_2) h(\boldsymbol{l}_3) \Big\rangle e^{i(\boldsymbol{l}_1+\boldsymbol{l}_2+\boldsymbol{l}_3) \cdot \boldsymbol{\theta}_C} \\ & \qquad \qquad \qquad \times W_{\mathrm{1pt}} (\boldsymbol{l}_1) W(\boldsymbol{q}) W(\boldsymbol{l}_2+\boldsymbol{l}_3-\boldsymbol{q}) e^{i(\boldsymbol{q}-\boldsymbol{l}_2) \cdot \boldsymbol{\alpha}} \ ,
\end{split}
\end{equation}
and for the case with complex-conjugated field $h^*$:
\begin{equation} \label{eq:2D_iZ*_flat_sky_raw}
\begin{split}
    i\zeta_*(\boldsymbol{\alpha}) & \equiv \Big\langle \Bar{f}( \boldsymbol{\theta}_C) \; \hat{\xi}_{\mathrm{gh^*}}(\boldsymbol{\alpha};\boldsymbol{\theta}_C) \Big\rangle \\
    & = \frac{1}{A_{\mathrm{1pt}} A_{\mathrm{2pt}}(\boldsymbol{\alpha})} \int \frac{\mathrm{d}^2 \boldsymbol{l}_1}{(2\pi)^2}  \int \frac{\mathrm{d}^2 \boldsymbol{l}_2}{(2\pi)^2} \int \frac{\mathrm{d}^2 \boldsymbol{l}_3}{(2\pi)^2} \int \frac{\mathrm{d}^2 \boldsymbol{q}}{(2\pi)^2} \\ & \qquad \qquad \qquad \times \Big\langle f(\boldsymbol{l}_1) g(\boldsymbol{l}_2) h^*(-\boldsymbol{l}_3) \Big\rangle e^{i(\boldsymbol{l}_1+\boldsymbol{l}_2+\boldsymbol{l}_3) \cdot \boldsymbol{\theta}_C} \\ & \qquad \qquad \qquad \times W_{\mathrm{1pt}} (\boldsymbol{l}_1) W(\boldsymbol{q}) W(\boldsymbol{l}_2+\boldsymbol{l}_3-\boldsymbol{q}) e^{i(\boldsymbol{q}-\boldsymbol{l}_2) \cdot \boldsymbol{\alpha}} \ .
\end{split}
\end{equation}
For a real field $h$, it follows that $i\zeta_*(\boldsymbol{\alpha}) = i\zeta(\boldsymbol{\alpha})$. 

\subsubsection{Integrated bispectrum of projected fields}

The Fourier space counterparts of the above equations can be written as
\begin{equation} \label{eq:2D_iB_flat_sky_raw}
\begin{split}
    iB(\boldsymbol{l}) & \equiv \mathcal{F}_{\mathrm{2D}}[A_{\mathrm{2pt}}(\boldsymbol{\alpha}) i\zeta(\boldsymbol{\alpha};\boldsymbol{\theta}_C)] \\
    & = \frac{1}{A_{\mathrm{1pt}}} \int \frac{\mathrm{d}^2 \boldsymbol{l}_1}{(2\pi)^2}  \int \frac{\mathrm{d}^2 \boldsymbol{l}_2}{(2\pi)^2} \int \frac{\mathrm{d}^2 \boldsymbol{l}_3}{(2\pi)^2} \Big\langle f(\boldsymbol{l}_1) g(\boldsymbol{l}_2) h(\boldsymbol{l}_3) \Big\rangle \\ & \qquad \qquad \times  e^{i(\boldsymbol{l}_1+\boldsymbol{l}_2+\boldsymbol{l}_3) \cdot \boldsymbol{\theta}_C} W_{\mathrm{1pt}} (\boldsymbol{l}_1) W(\boldsymbol{l}_2+\boldsymbol{l}) W(\boldsymbol{l}_3-\boldsymbol{l}) \\
    & = \Big\langle \Bar{f}( \boldsymbol{\theta}_C) \; \hat{P}_{\mathrm{gh}}(\boldsymbol{l};\boldsymbol{\theta}_C) \Big\rangle \ ,
\end{split}
\end{equation}
\begin{equation} \label{eq:2D_iB*_flat_sky_raw}
\begin{split}
    iB_*(\boldsymbol{l}) & \equiv \mathcal{F}_{\mathrm{2D}}[A_{\mathrm{2pt}}(\boldsymbol{\alpha}) i\zeta_*(\boldsymbol{\alpha};\boldsymbol{\theta}_C)] \\
    & = \frac{1}{A_{\mathrm{1pt}}} \int \frac{\mathrm{d}^2 \boldsymbol{l}_1}{(2\pi)^2}  \int \frac{\mathrm{d}^2 \boldsymbol{l}_2}{(2\pi)^2} \int \frac{\mathrm{d}^2 \boldsymbol{l}_3}{(2\pi)^2} \Big\langle f(\boldsymbol{l}_1) g(\boldsymbol{l}_2) h^*(-\boldsymbol{l}_3) \Big\rangle \\ & \qquad \qquad \times  e^{i(\boldsymbol{l}_1+\boldsymbol{l}_2+\boldsymbol{l}_3) \cdot \boldsymbol{\theta}_C} W_{\mathrm{1pt}} (\boldsymbol{l}_1) W(\boldsymbol{l}_2+\boldsymbol{l}) W(\boldsymbol{l}_3-\boldsymbol{l}) \\
    & = \Big\langle \Bar{f}( \boldsymbol{\theta}_C) \; \hat{P}_{\mathrm{gh^*}}(\boldsymbol{l};\boldsymbol{\theta}_C) \Big\rangle \ .
\end{split}
\end{equation}
where the last lines of both these equations show that the integrated bispectrum is the ensemble average of the position-dependent weighted mean and the position-dependent power spectrum of the projected fields.

From isotropy considerations (of the fields and of the symmetric window functions) we have $iB(\boldsymbol{l}) = iB(l)$ and $i\zeta(\boldsymbol{\alpha}) = i\zeta(\alpha)$. We can thereby relate the integrated 3-point function to the integrated bispectrum through an inverse Hankel transform:
\begin{equation}
    i\zeta(\alpha) = \frac{1}{A_{\mathrm{2pt}}(\alpha)} \mathcal{F}_{\mathrm{2D}}^{-1}[iB(l)] \ .
\end{equation}
The formalism for the integrated bispectrum and integrated 3-point function we have developed so far is very general and applicable to any projected field within the flat-sky approximation. For the curved-sky formulation of the projected integrated bispectrum the reader is referred to the work by \citealp{Jung_2020}.

In this paper, we shall look into only one application of our formalism for the integrated 3-point function --- on the cosmic shear field.

\section{Theory II: Application}
\label{chap:theory_application}
Having developed the general framework of equations for computing the integrated 3-point function for any projected field, we now apply it to the weak lensing shear field and formulate the equations for the \textit{integrated 3-point shear correlation function}.

\subsection{Weak lensing basics}
The light from background (source) galaxies is weakly deflected by the foreground (lens) intervening total matter distribution. This causes a coherent distortion pattern in the observed shapes of these background galaxies and is known as the cosmic shear field. This field can be interpreted as the shear caused by a weighted line-of-sight projection of the 3D matter density field --- known as the weak lensing convergence field. Statistical analysis of this shear field (directly observable) through the widely used 2-point shear correlation function allows one to infer about the projected power spectrum of the total matter distribution (theoretically predictable) and thereby constrain cosmological parameters.

Following equation \eqref{eqn_projected_field_flat_sky}, the weak lensing convergence field $\kappa(\boldsymbol{\theta})$ acting on source galaxies situated at the radial comoving distance $\chi_s$  can be written as a line-of-sight projection of the 3D matter density contrast field $\delta^{\mathrm{3D}}$:
\begin{equation} \label{eq:convergence_definition}
\begin{split}
    \kappa(\boldsymbol{\theta}) & = \int \mathrm{d} \chi \; q(\chi) \delta^{\mathrm{3D}} \big[(\chi\boldsymbol{\theta},\chi), \eta_0 - \chi \big]
\end{split}
\end{equation}
with projection kernel $q(\chi)$ (also known as lensing efficiency) written as\footnote{In this paper we only consider the case when all source galaxies are located in a Dirac-$\delta$ function like bin at $\chi_s$. However, it is straight forward to write $q(\chi)$ for a general distribution of source galaxies in a tomographic redshift bin (e.g. see \citealp{Schneider_2006}).}\citep{Kilbinger_2015}
\begin{equation} \label{eq:lensing_projection_kernel_single_zs}
    q(\chi) = \frac{3H_0^2 \Omega_{\mathrm{m}}}{2c^2} \frac{\chi}{a(\chi)} \frac{\chi_s - \chi}{\chi_s} \; ;  \qquad \text{with } \chi \leq \chi_s 
\end{equation}
where $\Omega_{\mathrm{m}}$ is the total matter density parameter of the Universe today, $H_0$ the Hubble parameter today, $a$ the scale factor and $c$ the speed of light. The convergence and the associated complex shear field are related to each other through second-order derivatives of the lensing potential $\psi(\boldsymbol{\theta})$ in the 2D sky-plane \citep{Schneider_2006}:
\begin{equation}
    \kappa(\boldsymbol{\theta}) = \frac{1}{2}\left(\partial_x^2 + \partial_y^2 \right) \psi(\boldsymbol{\theta}), \qquad \gamma(\boldsymbol{\theta}) = \frac{1}{2}\left(\partial_x^2 - \partial_y^2 + 2i \partial_x \partial_y\right) \psi(\boldsymbol{\theta})
\end{equation}
where $\psi(\boldsymbol{\theta})$ is the line-of-sight projection of the 3D Newtonian gravitational potential $\Phi \big[ (\chi\boldsymbol{\theta},\chi), \eta_0 - \chi \big]$ of the total matter distribution:
\begin{equation}
    \psi(\boldsymbol{\theta}) = \frac{2}{c^2} \int \mathrm{d} \chi \; \frac{\chi_s - \chi}{\chi_s \; \chi} \;  \Phi \big[ (\chi\boldsymbol{\theta},\chi), \eta_0 - \chi \big] \; ;  \qquad \text{with } \chi_s > \chi \ .     
\end{equation}
The shear $\gamma(\boldsymbol{\theta}) = \gamma_1(\boldsymbol{\theta}) + i \gamma_2(\boldsymbol{\theta})$ at a given location $\boldsymbol{\theta}$ is a complex quantity where the shear components $\gamma_1$ and $\gamma_2$ are specified in a chosen Cartesian frame (in 2D flat-sky). However, one is free to rotate the coordinates by any arbitrary angle $\beta$. With respect to this reference rotation angle $\beta$, one defines the \textit{tangential} and \textit{cross} components of the shear at position $\boldsymbol{\theta}$ as \citep{Schneider_2006}
\begin{equation}
    \gamma_{\mathrm{t}}(\boldsymbol{\theta},\beta) + i \gamma_{\times}(\boldsymbol{\theta},\beta) \equiv -e^{-2i\beta}\big[\gamma_1(\boldsymbol{\theta}) + i \gamma_2(\boldsymbol{\theta})\big] \ .
\end{equation}
Now, given a pair of points $\boldsymbol{\theta}_1$ and $\boldsymbol{\theta}_2$ on the field which are separated by the 2D vector $\boldsymbol{\alpha} \equiv \boldsymbol{\theta}_2 - \boldsymbol{\theta}_1$, one can write the tangential and cross components of the shear for this particular pair of points along the separation direction $\beta = \phi_{\boldsymbol{\alpha}}$ (polar angle of $\boldsymbol{\alpha}$) as
\begin{equation}
\begin{split}
    \gamma_{\mathrm{t}}(\boldsymbol{\theta}_j,\phi_{\boldsymbol{\alpha}}) + i \gamma_{\times}(\boldsymbol{\theta}_j,\phi_{\boldsymbol{\alpha}}) \equiv -e^{-2i\phi_{\boldsymbol{\alpha}}}\big[\gamma_1(\boldsymbol{\theta}_j) + i \gamma_2(\boldsymbol{\theta}_j)\big] \\
\end{split}
\end{equation}
where $j = 1,2$.

In the 2D Fourier plane, the shear $\gamma(\boldsymbol{l})$ is related to $\kappa(\boldsymbol{l})$ as \citep{Schneider_2006, Kilbinger_2015}
\begin{equation} \label{eq:shear_convergence_FS_relation}
    \gamma(\boldsymbol{l}) \; = \; \frac{(l_x + i \; l_y)^2}{l^2} \; \kappa(\boldsymbol{l}) \; = \; e^{2i\phi_{\boldsymbol{l}}} \kappa(\boldsymbol{l}) \; ;  \qquad \text{for } l \neq 0     
\end{equation}
where $l = \sqrt{l_x^2 + l_y^2}$ and $\phi_{\boldsymbol{l}} = \arctan \left( \frac{l_y}{l_x} \right)$ is the polar angle of $\boldsymbol{l}$.

The weak lensing convergence power spectrum $P_{\kappa,\mathrm{gh}}$ can be defined through  equation \eqref{eq:projected_power_spectrum_definition} --- $\big\langle \kappa_{\mathrm{g}}(\boldsymbol{l}_1) \kappa_{\mathrm{h}}(\boldsymbol{l}_2) \big\rangle \equiv (2\pi)^2 \delta_{D}(\boldsymbol{l}_1 + \boldsymbol{l}_2) P_{\kappa,\mathrm{gh}}(\boldsymbol{l}_1)$ for the convergence fields $\kappa_{\mathrm{g}}$ and $\kappa_{\mathrm{h}}$, each defined with projection kernels $q_{\mathrm{g}}(\chi)$ and $q_{\mathrm{h}}(\chi)$ for two different redshift bins (see equation \eqref{eq:convergence_definition}) with sources located at $\chi_{s,\mathrm{g}}$ and $\chi_{s,\mathrm{h}}$, respectively. It can be further expressed through equation \eqref{eq:projected_power_spectrum} as
\begin{equation}  \label{eq:convergence_power_spectrum}
    P_{\kappa,\mathrm{gh}}(\boldsymbol{l}) = \int \mathrm{d}\chi \frac{q_{\mathrm{g}}(\chi)q_{\mathrm{h}}(\chi)}{\chi^2} P^{\mathrm{3D}}_{\delta}\left(\boldsymbol{k} = \frac{\boldsymbol{l}}{\chi}, \eta_0 - \chi \right)
\end{equation}
where $P^{\mathrm{3D}}_{\delta}\left(\boldsymbol{k}, \eta \right)$ is the 3D matter density contrast power spectrum.

Similarly, the weak lensing convergence bispectrum defined through  equation \eqref{eq:projected_bispectrum_definition} --- $\big\langle \kappa_{\mathrm{f}}(\boldsymbol{l}_1) \kappa_{\mathrm{g}}(\boldsymbol{l}_2) \kappa_{\mathrm{h}}(\boldsymbol{l}_3)\big\rangle \equiv (2\pi)^2 \delta_{D}(\boldsymbol{l}_1 + \boldsymbol{l}_2 + \boldsymbol{l}_3) B_{\kappa,\mathrm{fgh}}(\boldsymbol{l}_1,\boldsymbol{l}_2,\boldsymbol{l}_3)$ of the convergence fields $\kappa_{\mathrm{f}}$, $\kappa_{\mathrm{g}}$ and $\kappa_{\mathrm{h}}$ with projection kernels $q_{\mathrm{f}}(\chi)$, $q_{\mathrm{g}}(\chi)$ and $q_{\mathrm{h}}(\chi)$ respectively can be expressed through equation \eqref{eq:projected_bispectrum} as 
\begin{equation} \label{eq:convergence_bispectrum}
    B_{\kappa,\mathrm{fgh}}(\boldsymbol{l}_1,\boldsymbol{l}_2,\boldsymbol{l}_3) = \int \mathrm{d}\chi \frac{q_{\mathrm{f}}(\chi)q_{\mathrm{g}}(\chi)q_{\mathrm{h}}(\chi)}{\chi^4} B^{\mathrm{3D}}_{\delta}\left(\frac{\boldsymbol{l}_1}{\chi},\frac{\boldsymbol{l}_2}{\chi},\frac{\boldsymbol{l}_3}{\chi}, \eta_0 - \chi \right)
\end{equation}
where $B^{\mathrm{3D}}_{\delta}(\boldsymbol{k}_1, \boldsymbol{k}_2, \boldsymbol{k}_3, \eta)$ is the 3D bispectrum of the matter density contrast field and $\boldsymbol{k}_i = \frac{\boldsymbol{l}_i}{\chi}$. From the statistical isotropy of the density contrast field, both $P^{\mathrm{3D}}_{\delta}$ and $B^{\mathrm{3D}}_{\delta}$ are independent of the direction of the $\boldsymbol{k}_i$ wave-vectors.

\subsection{Shear 2-point correlation function and aperture mass}
\label{sec:shear_measures_2pt_Map}
A widely used statistic to investigate the shear field $\gamma(\boldsymbol{\theta})$ is the 2-point shear correlation function. Using the notation $\gamma_{\mathrm{t},j} \equiv \gamma_{\mathrm{t}}(\boldsymbol{\theta}_j,\phi_{\boldsymbol{\alpha}})$ and $\gamma_{\times,j} \equiv \gamma_{\times}(\boldsymbol{\theta}_j,\phi_{\boldsymbol{\alpha}})$, the 2-point shear correlations (as a function of separation vector $\boldsymbol{\alpha}$) are defined as \citep{Schneider_Lombardi_2003, Jarvis_2004}:
\begin{equation} \label{eq:shear_2pt_correlation_definition}
\begin{split}
    \xi_+(\boldsymbol{\alpha}) & \equiv \big \langle \gamma_{\mathrm{t},1} \; \gamma_{\mathrm{t},2} \big \rangle + \big \langle \gamma_{\times,1} \; \gamma_{\times,2} \big \rangle  = \big\langle \gamma(\boldsymbol{\theta}_1) \gamma^*(\boldsymbol{\theta}_2)\big\rangle \ , \\
    \xi_-(\boldsymbol{\alpha}) & \equiv \big \langle \gamma_{\mathrm{t},1} \; \gamma_{\mathrm{t},2} \big \rangle - \big \langle \gamma_{\times,1} \; \gamma_{\times,2} \big \rangle = \big\langle \gamma(\boldsymbol{\theta}_1) \gamma(\boldsymbol{\theta}_2) e^{-4i\phi_{\boldsymbol{\alpha}}} \big \rangle  
\end{split}
\end{equation}
where the ensemble averages are over all pairs of points $\{\boldsymbol{\theta}_1, \boldsymbol{\theta}_2\}$ with $ \boldsymbol{\theta}_2 = \boldsymbol{\theta}_1 + \boldsymbol{\alpha}$.

Considering a pair of shear fields $\gamma_{\mathrm{g}}$, $\gamma_{\mathrm{h}}$ with projection kernels $q_{\mathrm{g}}(\chi)$ and $q_{\mathrm{h}}(\chi)$ respectively, the shear 2-point cross-correlations $\xi_{\pm,\mathrm{gh}}$ between the two fields can then be written as
\begin{equation} \label{eq:shear_2pt_correlation_cross}
\begin{split}
    \xi_{+,\mathrm{gh}}(\boldsymbol{\alpha}) & \equiv \big\langle \gamma_{\mathrm{g}}(\boldsymbol{\theta}) \gamma^*_{\mathrm{h}}(\boldsymbol{\theta+\alpha})\big\rangle \ , \\
    \xi_{-,\mathrm{gh}}(\boldsymbol{\alpha}) & \equiv \big\langle \gamma_\mathrm{g}(\boldsymbol{\theta}) \gamma_\mathrm{h}(\boldsymbol{\theta+\alpha}) e^{-4i\phi_{\boldsymbol{\alpha}}} \big\rangle \ .
\end{split}    
\end{equation}

In general, both the correlations are complex quantities but have vanishing imaginary parts only for the so-called E-mode shear fields (which we consider in this paper) \citep{Schneider_2002,Kilbinger_2015}. Moreover, from statistical isotropy of the fields it follows that $\xi_{\pm,\mathrm{gh}}(\boldsymbol{\alpha}) = \xi_{\pm,\mathrm{gh}}(\alpha)$. These shear correlations are related to the convergence power spectrum (equation \eqref{eq:convergence_power_spectrum}) through inverse Hankel transforms (see Appendix \ref{app:hankel}) \citep{Schneider_2006, Kilbinger_2015}:
\begin{equation} \label{eq:2pt_shear_correlation_power_spectrum}
\begin{split}
    \xi_{+,\mathrm{gh}}(\alpha) = \mathcal{F}^{-1}_{\mathrm{2D}}[P_{\kappa,\mathrm{gh}}(l)] & = \int \frac{\mathrm{d} l\; l}{2\pi}  \;P_{\kappa,\mathrm{gh}}(l) \;J_0(l \alpha) \ , \\
    \xi_{-,\mathrm{gh}}(\alpha) = \mathcal{F}^{-1}_{\mathrm{2D}}[P_{\kappa,\mathrm{gh}}(l) e^{-4i\phi_{\boldsymbol{l}}}]  &=  \int \frac{\mathrm{d} l\; l}{2\pi}  \;P_{\kappa,\mathrm{gh}}(l) \;J_4(l \alpha)
\end{split}
\end{equation}
where $J_0(x)$, $J_4(x)$ are the zeroth and fourth-order Bessel functions of the first kind, respectively.

We can now write the position-dependent 2-point correlation functions $\hat{\xi}_{\pm,\mathrm{gh}}(\boldsymbol{\alpha};\boldsymbol{\theta}_C)$ of the shear field within a 2D window $W$ centred at position $\boldsymbol{\theta}_C$. Using equations \eqref{eq:position_dependent_2pt_function_2D_field_gh*}, \eqref{eq:shear_convergence_FS_relation} and the first line of equation \eqref{eq:shear_2pt_correlation_cross}, we can write the $\hat{\xi}_{+,\mathrm{gh}}(\boldsymbol{\alpha};\boldsymbol{\theta}_C)$ correlation as
\begin{equation} \label{eq:position_dependent_2pt_function_2D_field_xip}
\begin{split}
    \hat{\xi}_{+,\mathrm{gh}}(\boldsymbol{\alpha};\boldsymbol{\theta}_C) & \equiv \frac{1}{A_{\mathrm{2pt}}(\boldsymbol{\alpha})} \int \mathrm{d}^2 \boldsymbol{\theta} \; \gamma_{\mathrm{g}}(\boldsymbol{\theta};\boldsymbol{\theta}_C) \gamma_{\mathrm{h}}^*(\boldsymbol{\theta}+\boldsymbol{\alpha};\boldsymbol{\theta}_C)  \\
    & = \frac{1}{A_{\mathrm{2pt}}(\boldsymbol{\alpha})} \int \frac{\mathrm{d}^2 \boldsymbol{l}_1}{(2\pi)^2}  \int \frac{\mathrm{d}^2 \boldsymbol{l}_2}{(2\pi)^2} \int \frac{\mathrm{d}^2 \boldsymbol{q}}{(2\pi)^2} \; \kappa_{\mathrm{g}}(\boldsymbol{l}_1)
    \kappa_{\mathrm{h}}(\boldsymbol{l}_2) \times \\ & e^{2i(\phi_1 - \phi_2)}  W(\boldsymbol{q})  W(\boldsymbol{l}_1+\boldsymbol{l}_2-\boldsymbol{q}) e^{i(\boldsymbol{l}_1+\boldsymbol{l}_2) \cdot \boldsymbol{\theta}_C} e^{i(\boldsymbol{q}-\boldsymbol{l}_1) \cdot \boldsymbol{\alpha}} 
\end{split}
\end{equation}
where $\phi_1$ and $\phi_2$ are the polar angles of the Fourier modes $\boldsymbol{l}_1$ and $\boldsymbol{l}_2$ respectively.

Taking into account the phase factor $e^{-4i\phi_{\boldsymbol{\alpha}}}$ present in the second line of equation \eqref{eq:shear_2pt_correlation_cross}, we can write the  $\hat{\xi}_{-,\mathrm{gh}}(\boldsymbol{\alpha};\boldsymbol{\theta}_C)$ using equations \eqref{eq:position_dependent_2pt_function_2D_field_gh} and \eqref{eq:shear_convergence_FS_relation} as
\begin{equation} \label{eq:position_dependent_2pt_function_2D_field_xim}
\begin{split}
    \hat{\xi}_{-,\mathrm{gh}}(\boldsymbol{\alpha};\boldsymbol{\theta}_C) & \equiv \frac{1}{A_{\mathrm{2pt}}(\boldsymbol{\alpha})} \int \mathrm{d}^2 \boldsymbol{\theta} \; \gamma_{\mathrm{g}}(\boldsymbol{\theta};\boldsymbol{\theta}_C) \gamma_{\mathrm{h}}(\boldsymbol{\theta}+\boldsymbol{\alpha};\boldsymbol{\theta}_C) e^{-4i\phi_{\boldsymbol{\alpha}}} \\
    & = \frac{1}{A_{\mathrm{2pt}}(\boldsymbol{\alpha})} \int \frac{\mathrm{d}^2 \boldsymbol{l}_1}{(2\pi)^2}  \int \frac{\mathrm{d}^2 \boldsymbol{l}_2}{(2\pi)^2} \int \frac{\mathrm{d}^2 \boldsymbol{q}}{(2\pi)^2} \; \kappa_{\mathrm{g}}(\boldsymbol{l}_1)
    \kappa_{\mathrm{h}}(\boldsymbol{l}_2) \\ 
    & \qquad \times  e^{2i(\phi_1 + \phi_2)}  W(\boldsymbol{q})  W(\boldsymbol{l}_1+\boldsymbol{l}_2-\boldsymbol{q}) e^{i(\boldsymbol{l}_1+\boldsymbol{l}_2) \cdot \boldsymbol{\theta}_C} \\ 
    & \qquad \times  e^{i(\boldsymbol{q}-\boldsymbol{l}_1) \cdot \boldsymbol{\alpha}} e^{-4i\phi_{\boldsymbol{\alpha}}}\ .
\end{split}
\end{equation}
For isotropic window functions $W$, both the estimators are independent of the direction of $\boldsymbol{\alpha}$ i.e. $\hat{\xi}_{\pm,\mathrm{gh}}(\boldsymbol{\alpha};\boldsymbol{\theta}_C) = \hat{\xi}_{\pm,\mathrm{gh}}(\alpha;\boldsymbol{\theta}_C)$. Moreover, taking the ensemble average of the above equations we can see that $\langle \hat{\xi}_{\pm,\mathrm{gh}}(\boldsymbol{\alpha};\boldsymbol{\theta}_C)  \rangle = \xi_{\pm,\mathrm{gh}}(\boldsymbol{\alpha})$.

Along these lines we can also define the position-dependent shear power spectra expressions as the Fourier space counterparts of the above equations. Using equations \eqref{eq:position_dependent_power_spectrum_2D_field_gh*} and \eqref{eq:position_dependent_power_spectrum_2D_field_gh} respectively (with an extra phase factor  $e^{4i\phi_{\boldsymbol{\alpha}}}$ in the latter), we get
\begin{equation}
\begin{split}
    \hat{P}_{+,\mathrm{gh}}(\boldsymbol{l}; \boldsymbol{\theta}_C) & \equiv \mathcal{F}_{\mathrm{2D}}[A_{\mathrm{2pt}}(\boldsymbol{\alpha}) \hat{\xi}_{+,\mathrm{gh}}(\boldsymbol{\alpha};\boldsymbol{\theta}_C)] \\
    & = \gamma_{\mathrm{g}}(-\boldsymbol{l}; \boldsymbol{\theta}_C) \gamma_{\mathrm{h}}^*(-\boldsymbol{l}; \boldsymbol{\theta}_C) \\
    & = \int \frac{\mathrm{d}^2 \boldsymbol{l}_1}{(2\pi)^2}  \int \frac{\mathrm{d}^2 \boldsymbol{l}_2}{(2\pi)^2} \; \kappa_{\mathrm{g}}(\boldsymbol{l}_1)  \kappa_{\mathrm{h}}(\boldsymbol{l}_2) e^{2i(\phi_1-\phi_2)} \\ & \qquad \qquad \times W(\boldsymbol{l}_1+\boldsymbol{l}) W(\boldsymbol{l}_2-\boldsymbol{l}) e^{i(\boldsymbol{l}_1 + \boldsymbol{l}_2) \cdot \boldsymbol{\theta}_C}
\end{split}
\end{equation}
and
\begin{equation}
\begin{split}
    \hat{P}_{-,\mathrm{gh}}(\boldsymbol{l}; \boldsymbol{\theta}_C) & \equiv \mathcal{F}_{\mathrm{2D}}[A_{\mathrm{2pt}}(\boldsymbol{\alpha}) \hat{\xi}_{-,\mathrm{gh}}(\boldsymbol{\alpha};\boldsymbol{\theta}_C)e^{4i\phi_{\boldsymbol{\alpha}}}] \\
    & = \gamma_{\mathrm{g}}(-\boldsymbol{l}; \boldsymbol{\theta}_C) \gamma_{\mathrm{h}}(\boldsymbol{l}; \boldsymbol{\theta}_C) \\
    & = \int \frac{\mathrm{d}^2 \boldsymbol{l}_1}{(2\pi)^2}  \int \frac{\mathrm{d}^2 \boldsymbol{l}_2}{(2\pi)^2} \; \kappa_{\mathrm{g}}(\boldsymbol{l}_1)  \kappa_{\mathrm{h}}(\boldsymbol{l}_2) e^{2i(\phi_1+\phi_2)} \\ & \qquad \qquad \times W(\boldsymbol{l}_1+\boldsymbol{l}) W(\boldsymbol{l}_2-\boldsymbol{l}) e^{i(\boldsymbol{l}_1 + \boldsymbol{l}_2) \cdot \boldsymbol{\theta}_C} \ .
\end{split}
\end{equation}

In this paper, we shall use a top-hat (disc) window function $W$ of radius $\theta_{\mathrm{T}}$ inside which we shall evaluate the 2-point shear correlations:
\begin{equation}
    W(\boldsymbol{\theta}) = W(\theta)= \left\{
        \begin{array}{ll}
            1 & \quad \theta \leq \theta_T, \\
            0 & \quad \theta > \theta_T
        \end{array}
    \right.
\end{equation}
and the Fourier transform  of this window function reads
\begin{equation} \label{eq:tophat_window_function_unnormalised}
\begin{split}
    W(\boldsymbol{l}) = W(l) & = \int \mathrm{d}^2 \boldsymbol{\theta} \; W(\theta) e^{-i\boldsymbol{l}\cdot \boldsymbol{\theta}} = 2\pi \theta_{\mathrm{T}}^2 \; \frac{J_1(l\theta_{\mathrm{T}})}{l\theta_{\mathrm{T}}} 
\end{split}
\end{equation}
where $J_1$ is the first-order ordinary Bessel function of the first kind. One should note that this form of the top-hat window function is not normalised since $\int \mathrm{d}^2 \boldsymbol{\theta} \; W(\boldsymbol{\theta}) = \pi \theta_{\mathrm{T}}^2$.

Another statistic used for investigating the convergence/shear field is the aperture mass $M_{\mathrm{ap}}( \boldsymbol{\theta}_C)$ which measures
the weighted $\kappa$ --- a projected surface mass --- inside an aperture $U$ located at a given point $\boldsymbol{\theta}_C$ \citep{Kaiser1995, Schneider1996, Schneider_2006}:
\begin{equation} \label{eq:aperture_mass_convergence}
\begin{split}
    M_{\mathrm{ap}}( \boldsymbol{\theta}_C)
    & = \int \mathrm{d}^2 \boldsymbol{\theta} \;  \kappa(\boldsymbol{\theta}) \; U(\boldsymbol{\theta}_C-\boldsymbol{\theta}) \\
    & = \int \frac{\mathrm{d}^2 \boldsymbol{l}}{(2\pi)^2} \; \kappa(\boldsymbol{l}) U(\boldsymbol{l}) \; e^{i\boldsymbol{l}\cdot \boldsymbol{\theta}_C} \\
\end{split}
\end{equation}
where the azimuthally symmetric aperture $U(\boldsymbol{\theta}) = U(\theta)$ has a characteristic size scale $\theta_{\mathrm{ap}}$ and in the second line we have expanded the equation with Fourier space expressions (see equation \eqref{eq:weighted_mean_2D_field_in_window}). Furthermore, if $U$ is a compensated window function i.e. its integral over its support vanishes $\int \mathrm{d}^2 \boldsymbol{\theta} \; U(\boldsymbol{\theta}_C-\boldsymbol{\theta}) = 0$ then a very interesting property of the aperture mass is that it can be directly evaluated from the shear field as a weighted tangential shear within an azimuthally symmetric aperture $Q$ (of size $\theta_{\mathrm{ap}}$) located at $\boldsymbol{\theta}_C$ \citep{Kaiser1995, Schneider1996, Schneider_2006}:
\begin{equation} \label{eq:aperture_mass_shear}
\begin{split}
    M_{\mathrm{ap}}( \boldsymbol{\theta}_C) & = \int \mathrm{d}^2 \boldsymbol{\theta} \;  \gamma_{\mathrm{t}}(\boldsymbol{\theta}, \phi_{\boldsymbol{\theta}_C-\boldsymbol{\theta}}) \; Q(\boldsymbol{\theta}_C-\boldsymbol{\theta})
\end{split}
\end{equation}
where the tangential shear $\gamma_{\mathrm{t}}(\boldsymbol{\theta}, \phi_{\boldsymbol{\theta}_C-\boldsymbol{\theta}})$ at any given location $\boldsymbol{\theta}$ is defined with respect to $\phi_{\boldsymbol{\theta}_C-\boldsymbol{\theta}}$ which is the polar angle of the separation vector between $\boldsymbol{\theta}$ and the centre of the aperture $\boldsymbol{\theta}_C$. The azimuthally symmetric aperture $Q$ has the form \citep{Schneider_2006}
\begin{equation}
    Q(\boldsymbol{\theta}) = Q(\theta) = -U(\theta) + \frac{2}{\theta^2}\int_0^{\theta} \mathrm{d} \theta' \; \theta' U(\theta')  \ .
\end{equation}
The aperture mass statistic can be interpreted as a position-dependent weighted mean of the shear/convergence field (see equation \eqref{eq:weighted_mean_2D_field_in_window}) with $W_{\mathrm{1pt}} = U$. However, as we define it using a compensated filter, an area normalisation term for this statistic is irrelevant (see equation \eqref{eq:area_normalisation_weighted_mean_2D_field_in_window}).

For the filter functions $Q$ and $U$, several choices have been investigated. In this paper we use the forms proposed by \citealp{Crittenden2002} (see also \citealp{Kilbinger2005}, \citealp{Schneider2005}):
\begin{equation}
\begin{split}
    U(\theta) & = \frac{1}{2\pi\theta_{\mathrm{ap}}^2}\left( 1-\frac{\theta^2}{2\theta_{\mathrm{ap}}^2}\right) \exp{\left(-\frac{\theta^2}{2\theta_{\mathrm{ap}}^2}\right)}\\
    Q(\theta) & = \frac{
    \theta^2}{4\pi\theta_{\mathrm{ap}}^4} \; \exp{\left(-\frac{\theta^2}{2 \theta_{\mathrm{ap}}^2}\right)} \ .
\end{split}
\end{equation}

We shall also work closely with the Fourier space representation of $U$ for our theoretical modelling:
\begin{equation}
    U(\boldsymbol{l}) = U(l) = \int \mathrm{d}^2 \boldsymbol{\theta} \; U(\theta) e^{-i\boldsymbol{l}\cdot \boldsymbol{\theta}} = \frac{l^2\theta_{\mathrm{ap}}^2}{2} \; \exp{\left(-\frac{l^2\theta_{\mathrm{ap}}^2}{2}\right)} \ .
\end{equation}

\subsection{Integrated 3-point shear correlation function}

We now have all the necessary ingredients to define the integrated 3-point function (see section \ref{sec:i3pt_projected_fields}) of the cosmic shear field as follows:
\begin{equation} \label{eq:iZ_statistic}
    i\zeta_{\pm,\mathrm{fgh}}(\boldsymbol{\alpha}) \equiv \Big\langle M_{\mathrm{ap,f}}(\boldsymbol{\theta}_C) \; \hat{\xi}_{\pm,\mathrm{gh}}(\boldsymbol{\alpha};\boldsymbol{\theta}_C) \Big\rangle 
\end{equation}
where $M_{\mathrm{ap,f}}(\boldsymbol{\theta}_C)$ is the aperture mass at location $\boldsymbol{\theta}_C$ (see equations \eqref{eq:aperture_mass_convergence}, \eqref{eq:aperture_mass_shear}) evaluated from the shear field $\gamma_{\mathrm{f}}$ with projection kernel $q_{\mathrm{f}}(\chi)$ and $\hat{\xi}_{\pm,\mathrm{gh}}(\boldsymbol{\alpha};\boldsymbol{\theta}_C)$ are the position-dependent shear 2-point correlation functions (see equations \eqref{eq:position_dependent_2pt_function_2D_field_xip},\eqref{eq:position_dependent_2pt_function_2D_field_xim}) computed inside a top-hat patch centred at $\boldsymbol{\theta}_C$ from fields $\gamma_{\mathrm{g}}$, $\gamma_{\mathrm{h}}$ with projection kernels $q_{\mathrm{g}}(\chi)$ and $q_{\mathrm{h}}(\chi)$ respectively. Note again that each of these projection kernels indicate source redshifts corresponding to different comoving distances $\chi_{s,\mathrm{f}}, \chi_{s,\mathrm{g}}$, and $\chi_{s,\mathrm{h}}$ respectively.

Using equations \eqref{eq:2D_iZ*_flat_sky_raw}, \eqref{eq:aperture_mass_convergence} and \eqref{eq:position_dependent_2pt_function_2D_field_xip}, we can write the expression for the $i\zeta_+$ correlation function as
\begin{equation}
\begin{split}
    i\zeta_{+,\mathrm{fgh}}(\boldsymbol{\alpha}) & \equiv \Big\langle M_{\mathrm{ap,f}}(\boldsymbol{\theta}_C) \; \hat{\xi}_{+,\mathrm{gh}}(\boldsymbol{\alpha};\boldsymbol{\theta}_C)\Big\rangle  \\
    & = \frac{1}{A_{\mathrm{2pt}}(\boldsymbol{\alpha})} \int \mathrm{d}^2 \boldsymbol{\theta}_1 \int \mathrm{d}^2 \boldsymbol{\theta}_2 \; \Big \langle \kappa_{\mathrm{f}}(\boldsymbol{\theta}_1) \gamma_{\mathrm{g}}(\boldsymbol{\theta}_2) \gamma_{\mathrm{h}}^*(\boldsymbol{\theta}_2+\boldsymbol{\alpha}) \Big \rangle  \\ & \qquad \qquad \times U(\boldsymbol{\theta}_C-\boldsymbol{\theta}_1) W(\boldsymbol{\theta}_C-\boldsymbol{\theta}_2) W(\boldsymbol{\theta}_C-\boldsymbol{\theta}_2-\boldsymbol{\alpha}) \\
    & = \frac{1}{ A_{\mathrm{2pt}}(\boldsymbol{\alpha})} \int \frac{\mathrm{d}^2 \boldsymbol{l}_1}{(2\pi)^2}  \int \frac{\mathrm{d}^2 \boldsymbol{l}_2}{(2\pi)^2}  \int \frac{\mathrm{d}^2 \boldsymbol{q}}{(2\pi)^2} \\ & \qquad \times B_{\kappa,\mathrm{fgh}}(\boldsymbol{l}_1,\boldsymbol{l}_2,-\boldsymbol{l}_1-\boldsymbol{l}_2) e^{2i(\phi_2 - \phi_{-1-2})} \\ & \qquad \times U(\boldsymbol{l}_1) W(\boldsymbol{q}) W(-\boldsymbol{l}_1-\boldsymbol{q}) e^{i(\boldsymbol{q}-\boldsymbol{l}_2) \cdot \boldsymbol{\alpha}} \\
\end{split}
\end{equation}
where $\phi_{-1-2}$ is the polar angle of the $-\boldsymbol{l}_1-\boldsymbol{l}_2$ 2D Fourier-mode and in the last equality we have used the definition of the convergence bispectrum $B_{\kappa}$ which can be further expressed in terms of a line-of-sight projection of the 3D matter density bispectrum using equation \eqref{eq:convergence_bispectrum} to obtain:
\begin{equation} \label{eq:integrated_3_point_function_plus_bispectrum_relation}
\begin{split}
    i\zeta_{+,\mathrm{fgh}}(\boldsymbol{\alpha}) & = \frac{1}{ A_{\mathrm{2pt}}(\boldsymbol{\alpha})} \int \mathrm{d}\chi \frac{q_{\mathrm{f}}(\chi)q_{\mathrm{g}}(\chi)q_{\mathrm{h}}(\chi)}{\chi^4} \int \frac{\mathrm{d}^2 \boldsymbol{l}_1}{(2\pi)^2}  \int \frac{\mathrm{d}^2 \boldsymbol{l}_2}{(2\pi)^2} \\ & \times \int \frac{\mathrm{d}^2 \boldsymbol{q}}{(2\pi)^2} \; B^{\mathrm{3D}}_{\delta}\left(\frac{\boldsymbol{l}_1}{\chi},\frac{\boldsymbol{l}_2}{\chi},\frac{-\boldsymbol{l}_1-\boldsymbol{l}_2}{\chi}, \eta_0 - \chi \right)  e^{2i(\phi_2 - \phi_{-1-2})} \\ & \qquad \times U(\boldsymbol{l}_1) W(\boldsymbol{q}) W(-\boldsymbol{l}_1-\boldsymbol{q}) e^{i(\boldsymbol{q}-\boldsymbol{l}_2) \cdot \boldsymbol{\alpha}} \ .
\end{split}
\end{equation}
Similarly, using equation \eqref{eq:position_dependent_2pt_function_2D_field_xim} the $i\zeta_-$ correlation reads
\begin{equation}
\label{eq:integrated_3_point_function_minus_bispectrum_relation}
\begin{split}
    i\zeta_{-,\mathrm{fgh}}(\boldsymbol{\alpha}) & \equiv \Big\langle M_{\mathrm{ap,f}}(\boldsymbol{\theta}_C) \; \hat{\xi}_{-,\mathrm{gh}}(\boldsymbol{\alpha};\boldsymbol{\theta}_C)\Big\rangle  \\
    & = \frac{1}{A_{\mathrm{2pt}}(\boldsymbol{\alpha})} \int \mathrm{d}^2 \boldsymbol{\theta}_1 \int \mathrm{d}^2 \boldsymbol{\theta}_2 \; \Big \langle \kappa_{\mathrm{f}}(\boldsymbol{\theta}_1) \gamma_{\mathrm{g}}(\boldsymbol{\theta}_2) \gamma_{\mathrm{h}}(\boldsymbol{\theta}_2+\boldsymbol{\alpha}) \Big \rangle  \\ & \qquad \times e^{-4i\phi_{\boldsymbol{\alpha}}}  U(\boldsymbol{\theta}_C-\boldsymbol{\theta}_1) W(\boldsymbol{\theta}_C-\boldsymbol{\theta}_2) W(\boldsymbol{\theta}_C-\boldsymbol{\theta}_2-\boldsymbol{\alpha}) \\
    & = \frac{1}{ A_{\mathrm{2pt}}(\boldsymbol{\alpha})} \int \mathrm{d}\chi \frac{q_{\mathrm{f}}(\chi)q_{\mathrm{g}}(\chi)q_{\mathrm{h}}(\chi)}{\chi^4} \int \frac{\mathrm{d}^2 \boldsymbol{l}_1}{(2\pi)^2}  \int \frac{\mathrm{d}^2 \boldsymbol{l}_2}{(2\pi)^2} \\ & \times \int \frac{\mathrm{d}^2 \boldsymbol{q}}{(2\pi)^2} \; B^{\mathrm{3D}}_{\delta}\left(\frac{\boldsymbol{l}_1}{\chi},\frac{\boldsymbol{l}_2}{\chi},\frac{-\boldsymbol{l}_1-\boldsymbol{l}_2}{\chi}, \eta_0 - \chi \right)  e^{2i(\phi_2 + \phi_{-1-2})} \\ & \qquad \times U(\boldsymbol{l}_1) W(\boldsymbol{q}) W(-\boldsymbol{l}_1-\boldsymbol{q}) e^{i(\boldsymbol{q}-\boldsymbol{l}_2) \cdot \boldsymbol{\alpha}} e^{-4i\phi_{\boldsymbol{\alpha}}} \ .
\end{split}
\end{equation}

As stated before, for isotropic window functions, these correlations are independent of the direction of $\boldsymbol{\alpha}$ i.e. $ i\zeta_{\pm,\mathrm{fgh}}(\boldsymbol{\alpha}) = i\zeta_{\pm,\mathrm{fgh}}(\alpha)$. One thing to note is the similarity between the expressions of $i\zeta_{\pm,\mathrm{fgh}}$ and the generalized third-order aperture mass statistics with different compensated filter radii as proposed by of \citealp{Schneider2005} (see their Section 6). Our expressions can be interpreted as a special-case of these generalized aperture mass-statistics where we use 2 top-hat filters of same radii and 1 compensated filter with a different size instead of using 3 compensated filters.

Computationally, it is more convenient to arrive at these expressions for the integrated 3-point shear correlation functions from the inverse Fourier transforms of the integrated shear bispectra which we define as
\begin{equation}
\begin{split}
    iB_{+,\mathrm{fgh}}(\boldsymbol{l}) & \equiv \mathcal{F}_{\mathrm{2D}}[A_{\mathrm{2pt}}(\boldsymbol{\alpha}) i\zeta_{+,\mathrm{fgh}}(\boldsymbol{\alpha})] \ , \\
    iB_{-,\mathrm{fgh}}(\boldsymbol{l}) & \equiv \mathcal{F}_{\mathrm{2D}}[A_{\mathrm{2pt}}(\boldsymbol{\alpha}) i\zeta_{-,\mathrm{fgh}}(\boldsymbol{\alpha}) \; e^{4i\phi_{\boldsymbol{\alpha}}}] \ .
\end{split}
\end{equation}
Upon simplification, the expressions for these integrated bispectra read:
\begin{equation} \label{eq:integrated_bispectra}
\begin{split}
    iB_{\pm,\mathrm{fgh}}(\boldsymbol{l}) & = \int \mathrm{d}\chi \frac{q_{\mathrm{f}}(\chi)q_{\mathrm{g}}(\chi)q_{\mathrm{h}}(\chi)}{\chi^4} \int \frac{\mathrm{d}^2 \boldsymbol{l}_1}{(2\pi)^2}  \int \frac{\mathrm{d}^2 \boldsymbol{l}_2}{(2\pi)^2} \\ & \times \; B^{\mathrm{3D}}_{\delta}\left(\frac{\boldsymbol{l}_1}{\chi},\frac{\boldsymbol{l}_2}{\chi},\frac{-\boldsymbol{l}_1-\boldsymbol{l}_2}{\chi}, \eta_0 - \chi \right)  e^{2i(\phi_2 \mp \phi_{-1-2})} \\ & \times \; U(\boldsymbol{l}_1) W(\boldsymbol{l}_2+\boldsymbol{l}) W(-\boldsymbol{l}_1-\boldsymbol{l}_2-\boldsymbol{l}) \\
    & = \Big\langle M_{\mathrm{ap,f}}(\boldsymbol{\theta}_C) \; \hat{P}_{\pm,\mathrm{gh}}(\boldsymbol{l};\boldsymbol{\theta}_C) \Big\rangle
\end{split}
\end{equation}
where the last equality confirms our expectation (see equations \eqref{eq:2D_iB_flat_sky_raw} and \eqref{eq:2D_iB*_flat_sky_raw}) that the integrated bispectrum of the shear field is the correlation of the aperture mass and the position-dependent shear power spectrum.

Due to the isotropy argument, we have $iB_{\pm,\mathrm{fgh}}(\boldsymbol{l}) = iB_{\pm,\mathrm{fgh}}(l)$ i.e. the above equation is true for any polar angle $\phi_{\boldsymbol{l}}$ and the integrated 3-point functions are then inverse Hankel transforms of these integrated bispectra:
\begin{equation} \label{eq:integrated_3pt_shear_correlations_bispectrum}
\begin{split}
    i\zeta_{+,\mathrm{fgh}}(\alpha) & = \frac{1}{A_{\mathrm{2pt}}(\alpha)} \mathcal{F}^{-1}_{\mathrm{2D}}[iB_{+,\mathrm{fgh}}(l)] \\
    & = \frac{1}{A_{\mathrm{2pt}}(\alpha)} \int \frac{\mathrm{d} l\; l}{2\pi}  \;iB_{+,\mathrm{fgh}}(l) \;J_0(l \alpha) \ , \\
    i\zeta_{-,\mathrm{fgh}}(\alpha) & = \frac{1}{A_{\mathrm{2pt}}(\alpha)} \mathcal{F}^{-1}_{\mathrm{2D}}[iB_{-,\mathrm{fgh}}(l) e^{-4i\phi_{\boldsymbol{l}}}] \\
    & = \frac{1}{A_{\mathrm{2pt}}(\alpha)} \int \frac{\mathrm{d} l\; l}{2\pi}  \;iB_{-,\mathrm{fgh}}(l) \;J_4(l \alpha) \ .
\end{split}
\end{equation}
The $J_0(l \alpha)$ filter puts more weight on low-$l$ values of the integrated bispectrum than the $J_4(l \alpha)$ filter at a given angular separation $\alpha$. Hence, $i\zeta_{+}(\alpha)$ is more sensitive to large scale fluctuations (lower-$l$) than $i\zeta_{-}(\alpha)$ at the same angular separation $\alpha$.

\subsection{Summary}
\label{sec:theory_summary}
So far, we have developed the following:
\begin{enumerate}
    \item The integrated 3-point shear correlation function $i\zeta_{\pm}$ can be estimated from the cosmic shear field by measuring the aperture mass statistic (with a compensated filter) at different locations and then correlating it with the position-dependent 2-point shear correlation function (evaluated inside top-hat apertures) located at the corresponding locations (see equation \eqref{eq:iZ_statistic}).
    \item Given a prescription of the 3D matter density bispectrum $B_{\delta}^{\mathrm{3D}}(k_1,k_2,k_3,\eta)$ for a set of cosmological parameters, we can theoretically predict the $i\zeta_{\pm}$ through an inverse Hankel transform of the integrated shear bispectrum $iB_{\pm}$ --- an integral of the convergence bispectrum (see equations \eqref{eq:convergence_bispectrum}, \eqref{eq:integrated_bispectra} and \eqref{eq:integrated_3pt_shear_correlations_bispectrum}). This is analogous to the way in which one obtains the shear 2-point correlation function $\xi_{\pm}$ from the convergence power spectrum which is in turn related to the 3D matter density power spectrum $P_{\delta}^{\mathrm{3D}}(k,\eta)$ through a line-of-sight projection (see equations \eqref{eq:convergence_power_spectrum} and \eqref{eq:2pt_shear_correlation_power_spectrum}).
    \item In chapter \ref{chap:theory_general_formalism} we provide a general framework of equations for the integrated 3-point function (equations \eqref{eq:2D_iZ_flat_sky_raw}, \eqref{eq:2D_iZ*_flat_sky_raw}) and the integrated bispectrum (equations \eqref{eq:2D_iB_flat_sky_raw}, \eqref{eq:2D_iB*_flat_sky_raw}) which can be extended to the analysis of any projected field within the flat-sky approximation.
\end{enumerate}
We shall now proceed to measure the $\xi_{\pm}$ and $i\zeta_{\pm}$ statistics on simulated cosmic shear data and also perform theoretical calculations for the same using the equations mentioned above. We will test the accuracy of our models on the simulations and then investigate their constraining power on cosmological parameters.

\section{Simulations, Measurements and numerical methods for theoretical modelling}
\label{chap:methods}

In this chapter we describe the simulations (sections \ref{sec:T17_sims} and \ref{sec:flask_sims}) we use in order to measure our data vector and the data-covariance matrix (section \ref{sec:data_vector}). We will then discuss the methods we use in order to theoretically model the data vector in section \ref{sec:model_vector}.

\subsection{T17 N-body simulations}
\label{sec:T17_sims}
We use the publicly available simulated data sets from \citealp{Takahashi2017}\footnote{The data products of the simulation are available at \url{http://cosmo.phys.hirosaki-u.ac.jp/takahasi/allsky_raytracing/} .} cosmological simulations (hereafter T17 simulations). The simulations were generated primarily for the gravitational lensing studies for the HSC Survey. In this paper, we use the full-sky light cone weak lensing shear and convergence maps of the simulation suite.

These data sets were obtained from a cold dark matter (CDM) only cosmological N-body simulation in periodic cubic boxes. The simulation setting consisted of 14 boxes of increasing side lengths $L, 2L, 3L, ..., 14L$ (with $L=450 \; \mathrm{Mpc/h}$), nested around a common vertex (see Figure 1 of \citealp{Takahashi2017}). Each box contained $2048^3$ particles (smaller boxes hence have better spatial and mass resolution) and their initial conditions were set with second-order Lagrangian perturbation theory \citep{Crocce2006} with an initial power spectrum computed for a flat $\Lambda$CDM cosmology with the following parameters\footnote{The density parameter for species $\mathrm{X}$ is defined at $\eta = \eta_0$ i.e. $\Omega_{\mathrm{X}} \equiv \Omega_{\mathrm{X},0}$.}: $\Omega_{\mathrm{cdm}} = 0.233,\; \Omega_{\mathrm{b}} = 0.046,\; \Omega_{\mathrm{m}} = \Omega_{\mathrm{cdm}} + \Omega_{\mathrm{b}} = 0.279,\; \Omega_{\mathrm{de}} = \Omega_{\Lambda} = 0.721,\; h = 0.7,\; \sigma_8 = 0.82\; \mathrm{and}\; n_s = 0.97$. We adopt this set of parameters as our fiducial cosmology. The particles in each box were then made to evolve from the initial conditions using the N-body gravity solver code \verb|GADGET2|  \citep{Springel2001, Springel2005}. The evolved particle distribution of the different nested boxes were combined in layers of shells, each $150 \; \mathrm{Mpc/h}$ thick, to obtain full-sky light cone matter density contrast inside the shells. The simulation boxes were also ray traced using the multiple-lens plane ray-tracing algorithm \verb|GRAYTRIX| \citep{Hamana2015, Shirasaki2015} to obtain full-sky weak lensing convergence/shear maps (in \verb|Healpix| format \citealp{Gorski2005, Zonca2019}) for several Dirac-$\delta$ like source redshift bins. Multiple simulations were run to produce 108 realizations for each of their data products. The authors report that the average matter power spectra from their several realizations of the simulations agreed with the theoretical \verb|revised Halofit| power spectrum (\citealp{Smith2003}, later revised by \citealp{Takahashi2012}) to within 5 (10) per cent for $k < 5 (6) \; \mathrm{h/Mpc}$  at $z < 1$. They also provide correction formulae for their 3D and angular power spectra in order to account for the discrepancies stemming from the finite shell thickness, angular resolution and finite simulation box size effect in their simulations. We refer the reader to our Appendix \ref{app:takahashi} for a summary of those corrections.

In this paper, for validating the 2-point and integrated 3-point shear correlation functions (see section \ref{sec:data_vector}) we use the 108 full-sky weak lensing convergence and shear maps from the simulation suite. These maps come in the \verb Healpix  format \citep{Gorski2005, Zonca2019} for various angular resolutions. We only use the maps with \verb NSIDE = 4096 (angular pixel scale of $0.82'$) at source redshifts $z_1 = 0.5739$ and $z_2 = 1.0334$. For reference, these two redshifts correspond closely to the mean redshifts of the second and fourth photometric source redshift bins which have been used in the cosmic shear 2-point analyses of the Dark Energy Survey (DES) \citep{Troxel_2018}.

\subsection{FLASK Lognormal simulations}
\label{sec:flask_sims}
A crucial part of any cosmological analysis involves the calculation of the covariance matrix of a data vector --- which for us shall consist of 2-point and integrated 3-point correlations of the shear field (see section \ref{sec:data_vector}). The estimation of the inverse of this data-covariance, namely the precision matrix, is particularly important for forecasting cosmological parameter constraints (e.g. see section  \ref{sec:results_Fisher}). Although an analytically modelled data-covariance matrix can be inverted easily as it is inherently noise-free, it needs to be modelled sufficiently accurately. An easier approach is to estimate the covariance for a desired data vector from an ensemble of realistic N-body simulations. However, this comes at a cost that the sample covariance suffers from noise when estimated from a finite number of mock simulations. The inversion of such a noisy matrix comes with its own challenges. In order to beat down this noise in the precision matrix one therefore needs a large ensemble of independent simulations --- with the number of simulations required to be much larger than the size of the data vector (see \citealp{Taylor2013}). Unfortunately, for our purpose, we have only 108 independent T17 simulations to estimate the data-covariance of our quite high-dimensional data vector which will result in a noisy covariance matrix estimate (see Appendix \ref{app:precision_matrix_expansion}). Hence we need another way to estimate the covariance. Many possible methods to circumvent the problem have been suggested in literature such as re-sampling techniques for estimating the covariance matrix using a few mocks \citep{Escoffier2016}, shrinkage estimators \citep{Joachimi_2016} or to use lognormal simulations to name a few.

We choose the option of simulating a large ensemble of full-sky lognormal random fields for estimating the data-covariance matrix. Lognormal random fields have been extensively studied in the cosmological context \citep{Coles1991} and have been shown to be a very good approximation for the 1-point probability density function (PDF) of the weak lensing convergence/shear \citep{hilbert_2011, xavier2016} or the distribution of the late time matter density contrast fields \citep{Friedrich_2018, gruen_2018}. This assumption has been confirmed from the DES Science Verification data for the convergence field \citep{clerkin2017} and most recently been used to compute covariances for the 2-point shear correlations for the third year data analysis of the DES \citep{friedrich2020}. We further discuss and test the validity of modelling the data-covariance matrix with lognormal simulations in Appendix \ref{app:precision_matrix_expansion}. We show that our lognormal data-covariance and its inverse is indeed a good model as the Fisher parameter constraints shown in section \ref{sec:results_Fisher} are hardly affected when we correct the lognormal model with residual terms measured from the T17 simulations.

We use the publicly available \verb|FLASK| tool\footnote{currently hosted at \url{http://www.astro.iag.usp.br/~flask/} .} (Full-sky Lognormal Astro-fields
Simulation Kit) \citep{xavier2016} which can be used to create realisations of correlated lognormal fields on the celestial sphere at different redshifts. Concisely, \verb|FLASK| draws from a lognormal variable $\kappa$ with the PDF \citep{xavier2016}
\begin{equation}
\label{eq:pdf_lognormal_univariate}
p(\kappa) = \left\{
        \begin{array}{ll}
            \frac{\exp \left( -\frac{1}{2\sigma^2} [\ln (\kappa + \lambda) - \mu]^2\right)}{\sqrt{2\pi}\sigma(\kappa+\lambda)}  & \qquad \kappa > -\lambda, \\
            0 & \qquad \text{otherwise.}
        \end{array}
    \right.
\end{equation}
where $\mu$ and $\sigma^2$ are the mean and variance of the associated normal variable and $\lambda$ is the lognormal shift parameter marking the lower limit for possible values that $\kappa$ can realise. Using \verb|FLASK| we create lognormal mocks of the T17 convergence/shear fields which approximately follow the 1-point PDFs of the T17 maps at redshifts $z_1$ and $z_2$ respectively. As input to \verb|FLASK|, one needs to provide the convergence power spectra $P_{\kappa,\mathrm{gh}}(l)$ and the lognormal-shift parameters $\lambda_\mathrm{i}$ for the two redshifts (with $g,h,i$ = 1,2). We obtain the power spectra by projecting the the 3D matter density contrast power spectrum $P_{\delta}^{\mathrm{3D}}(k,\eta)$ along the line-of-sight as described in equation \eqref{eq:convergence_power_spectrum}. We use the open-source Boltzmann solver code \verb|CLASS|\footnote{currently hosted at \url{http://class-code.net} . We use version v2.9.4 of the code.} \citep{lesgourgues2011, blas2011} for computing the non-linear $P_{\delta}^{\mathrm{3D}}(k,\eta)$ in the fiducial T17 cosmology for which we use the \verb|revised halofit| prescription for the non-linear matter power spectrum \citep{Takahashi2012, Bird2012, Smith2003} which is included in \verb|CLASS|. For obtaining the lognormal shift parameters we follow the strategy of \citealp{hilbert_2011} and fit the above form of the lognormal PDF to the 1-point PDF of the T17 maps at both redshifts and get the following values\footnote{Precisely, we only fit the PDF to the first of the 108 T17 maps at both redshifts to obtain the quoted $\lambda_\mathrm{i}$ values. We have also tested the fits on other maps at each redshift and the values for the logshift parameters differ in only the third decimal place whose effect on the summary statistics evaluated from the corresponding FLASK maps is insignificant. The values for the other fit parameters are: $\mu_1 = -4.578, \mu_2 = -3.565$ and $\sigma_1^2 = 0.351, \sigma_2^2 = 0.205$.}: $\lambda_1 = 0.012$ and $\lambda_2 = 0.031$. Using these settings we generate 1000 correlated \verb|FLASK| pairs (each pair consists of 2 maps at source redshifts $z_1$ and $z_2$ respectively) of full-sky shear maps in \verb|Healpix| format with \verb|NSIDE| = 4096.

The T17 simulations are pure gravitational lensing shear/convergence maps without any noise. In real data, the shear is obtained from the measured ellipticities of background galaxies which --- besides the gravitational shear effect --- are subject to different sources of noise such as non-circular intrinsic ellipticities of the galaxies, measurement noise, noise from point-spread-function correction etc. In our covariance matrix we want to include the effect of this \textit{shape-noise}. This is important when we want to forecast realistic constraints on cosmological parameters. In principle, this can be modelled by adding a complex noise term $N(\boldsymbol{\theta}) = N_1(\boldsymbol{\theta}) + i N_2(\boldsymbol{\theta})$ to the shear field $\gamma\boldsymbol{\theta}) = \gamma_1(\boldsymbol{\theta}) + i \gamma_2(\boldsymbol{\theta})$ \citep{Pires_2020} where $\boldsymbol{\theta}$ represents a pixel on the \verb|Healpix| shear map. The noise components $N_1, N_2$ can both be modelled as uncorrelated Gaussian variables with zero-mean and variance
\begin{equation}
    \sigma_N^2 = \frac{\sigma_{\epsilon}^2}{n_g \cdot A_{pix}}
\end{equation}
where $A_{pix}$ is the area of the pixel at the given \verb|NSIDE|, $\sigma_{\epsilon}$ is the dispersion of intrinsic galaxy ellipticities which we set to be 0.3 as found for weak lensing surveys \citep{Leauthaud2007,Schrabback2018}, $n_g$ is the number of observed galaxies per square arcminute for which we assume a value of 5 at each redshift bin. Note that for the two Dirac-$\delta$ source redshift bins we consider, this adds up to give 10 galaxies per square arcminutes which is in accordance with the expected number density of galaxies for the full DES Year 6 cosmic shear data. To every pixel in a \verb|FLASK| generated shear map we add an independent draw of each Gaussian noise term. We then convert these noisy shear maps into noisy convergence maps on the curved sky using a Kaiser-Squires (KS) \citep{KaiserSquires1993} mass map reconstruction method as described in section 2.1 of \citealp{Gatti_2020} (see also \citealp{Chang2018}). This process of first adding noise to the shear field and then converting it to a convergence map is more accurate than the usually prevalent way of adding independent Gaussian noise to the pixels of the noiseless convergence field. This is because convergence at a given pixel is a convolution of the shear in several pixels around the desired location. This makes the noise in the convergence at a given pixel be correlated with the noise in neighbouring pixels. Although the KS method ensures this, the approach where uncorrelated Gaussian noise is added to the pixels of a noiseless convergence map directly does not account for it and is therefore not entirely accurate.

\subsection{Measurements: data vector and data-covariance matrix}
\label{sec:data_vector}
We carry out measurements of the position-dependent 2-point shear correlations $\hat{\xi}_{\pm,\mathrm{gh}}(\alpha;\boldsymbol{\theta}_C)$ on the T17 and \verb|FLASK| shear maps at source redshifts $z_1$ and $z_2$ (i.e. $g,h = 1,2$) within top-hat windows $W$ with radius $\theta_{\mathrm{T}} = 75'$. Approximately, this results in a circular patch of area 5 square degrees (which is small enough for the flat-sky approximation to hold). We use the publicly available code \verb|TreeCorr|\footnote{currently hosted at: \url{https://rmjarvis.github.io/TreeCorr/_build/html/index.html\#} .} \citep{Jarvis_2004} to measure these correlations in 20 log-spaced bins with angular separations $5' < \alpha < 140'$. To be precise, we execute \verb|TreeCorr| on those pixels of the map which lie within a disc of radius $\theta_{\mathrm{T}}$ centred at a given location $\boldsymbol{\theta}_C$ in order to obtain $\hat{\xi}_{\pm,\mathrm{gh}}(\alpha;\boldsymbol{\theta}_C)$.

For computing the aperture mass $M_{\mathrm{ap,f}}(\boldsymbol{\theta}_C)$ (with $f = 1,2$) we use a compensated window $U$ with an aperture scale\footnote{We found that for $\theta_{\mathrm{ap}} = 70'$ the amplitude of the $iB_{+}$ signal was larger than other aperture scales when measured in combination with the top-hat patch of $\theta_{\mathrm{T}} = 75'$. Optimization of the filter sizes remains an interesting avenue to explore.} $\theta_{\mathrm{ap}} = 70'$. From a convergence map at a given source redshift $z_{\mathrm{f}}$, we measure the aperture mass at location $\boldsymbol{\theta}_C$ through a convolution of the $U$ filter with pixels in the neighbourhood of $\boldsymbol{\theta}_C$ (see equation \eqref{eq:aperture_mass_convergence}). Note that it is completely equivalent to compute the aperture mass from the corresponding shear map at $z_{\mathrm{f}}$ by convolving shear pixels with the $Q$ filter (with the same aperture scale size as that of $U$, see equation \eqref{eq:aperture_mass_shear}) and completely skip the KS convergence map making procedure (see \citealp{Harnois-Deraps2020}). Hence, the way in which we compute the aperture mass using the convergence field is redundant. As we are working in a simulated setting and do not consider holes and masks in our data, the map making procedure is straight forward. However, this is not the case in real data and it is then practical to evaluate the aperture mass from the shear map directly.

In the 108 T17 noiseless simulation maps, we evaluate the above statistics at locations distributed over the full-sky. We do not do this for every pixel in the \verb|Healpix| map but rather choose well separated pixels (about $2\theta_{\mathrm{T}}$ apart --- the diameter of $W$) for which the top-hat patches at those chosen pixels only slightly overlap with the patches centred at neighbouring chosen pixels. The overlap is not a problem and allows us to maximise the area over which we evaluate the statistics. For computing the 2-point shear correlations $\xi_{\pm,\mathrm{gh}}(\alpha)$ in a given map we take the average of the position-dependent shear correlations evaluated at all chosen patches on the map (see the discussion after equation \eqref{eq:position_dependent_2pt_function_2D_field_xim}). The integrated 3-point shear correlations $i\zeta_{\pm,\mathrm{fgh}}(\alpha)$ are evaluated by first taking the product of the aperture mass and the position-dependent shear correlation at a chosen location and then performing an average of this product evaluated at all other locations (see equation \eqref{eq:iZ_statistic}) for a specific realization.

We perform the same measurements on the \verb|FLASK| maps (with shape-noise). Unlike the T17 maps, we do not distribute patches over the whole sky but rather cut out two big circular footprints of 5000 square degrees (approximately the size of the DES footprint) in each hemisphere of a \verb|FLASK| map and restrict the distribution of patches to within the extent of each footprint. In each \verb|FLASK| map, the two footprints are widely separated which allows us to treat each region as an independent survey realization. This helps to maximize the use of our \verb|FLASK| simulations and allows us to have a total of 2000 DES-like realizations (from 1000 \verb|FLASK| maps) which we consider sufficient for the estimation of the covariance matrix of our data-vector for a DES-sized survey as the number of realizations is much larger than the maximum size of our data vector which we discuss next.

For the two source redshifts $z_1$ and $z_2$, our data vector $D_i$ evaluated from the $i-$th simulation realization (T17 or \verb|FLASK|) consists of the  2-point shear cross-correlations and the integrated 3-point shear cross-correlations (each correlation function evaluated at 20 angular separations $\alpha$) as depicted below:
\begin{equation} \label{eq:data_vector}
    D_i \equiv \big( \xi_{\pm,11},\; \xi_{\pm,22},\; \xi_{\pm,12},\; i\zeta_{\pm,111},\; i\zeta_{\pm,222},\; i\zeta_{\pm,122},\; i\zeta_{\pm,211} \big)^{\mathrm{T}}    
\end{equation}
where T stands for transpose. This gives a data-vector of size $N_{d} = 7 \times 2 \times 20 = 280$ elements. The mean data vector is obtained by taking an average of the individual data vectors obtained from each of the $N_{r}$ realizations:
\begin{equation}
    \overline{D} = \frac{1}{N_{r}} \sum_{i=1}^{N_{r}}  D_i \ .
\end{equation}
On the other hand, we evaluate our covariance matrix of the data-vector as
\begin{equation} \label{eq:data_covoriance_independent_realisations}
    \mathbf{\hat{C}} = \frac{1}{N_{r}-1} \sum_{j=1}^{N_{r}} \left( D_j - \overline{D} \right) \left(D_j - \overline{D}\right)^\mathrm{T}  
\end{equation}
thus resulting in an $N_{d} \times N_{d} = 280\times280$ matrix. For validating our theoretical model for the data vector we compare it with the mean data vector from the 108 T17 noiseless maps. For obtaining our DES-like data-covariance matrix (with impact of shape-noise) we evaluate it from the $N_r = 2000$ footprints cut out from the \verb|FLASK| simulations. 

\subsection{Methods for theoretical modelling}
\label{sec:model_vector}
In this section we detail the numerical recipes that go into the theoretical computation of the constituents of the model vector $M$ which we evaluate for the fiducial T17 cosmology (see section \ref{sec:T17_sims}). For modelling the 2-point shear correlations $\xi_{\pm,\mathrm{gh}}(\alpha)$ (see equation \eqref{eq:2pt_shear_correlation_power_spectrum}) we need to compute the convergence power spectrum $P_{\kappa,\mathrm{gh}}(l)$. As already stated before, we use the public Boltzmann solver code \verb|CLASS|\footnote{To be precise, we use the c++ wrapper of the code (version v2.9.4) which can be obtained from the official repository, currently hosted at: \url{https://github.com/lesgourg/class_public} .} to compute the nonlinear \verb|revised halofit| 3D matter power spectrum $P_{\delta}^{\mathrm{3D}}(k,\eta)$ which we integrate along the line-of-sight to obtain $P_{\kappa,\mathrm{gh}}(l)$. We use the 1-dimensional adaptive quadrature integration routine from the GNU Scientific Library \verb|gsl|\footnote{currently hosted at: \url{http://www.gnu.org/software/gsl/} .} \citep{gough2009gnu} to perform the integration. To partly correct for the flat-sky and the Limber approximation that goes into the derivation for the expressions of the shear correlations, we multiply the convergence power spectrum by an $l$-dependent correction factor proposed by \citealp{Kitching_2017}:
\begin{equation}
    C_{\kappa,\mathrm{gh}}(l) \equiv \frac{(l+2)(l+1)l(l-1)}{\left(l+\frac{1}{2}\right)^4}P_{\kappa,\mathrm{gh}}\left(l+\frac{1}{2}\right)  \ .
\end{equation}
Moreover, instead of performing the inverse Hankel transform $l$-integrals (i.e. the $\mathcal{F}_{\mathrm{2D}}^{-1}[...]$ operations in equation \eqref{eq:2pt_shear_correlation_power_spectrum}) for converting the Fourier space power spectra to shear correlations, we use expressions with summation over $l$ as given in \citealp{friedrich2020} (see also \citealp{Stebbins1996}):
\begin{equation} \label{eq:Stebbins_conversion}
    \xi_{\pm,\mathrm{gh}}(\alpha) = \sum_{l > 2} \frac{2l+1}{4\pi} \; \frac{2 \left( G^+_{l,2}(\cos \alpha) \pm G^-_{l,2}(\cos \alpha) \right)}{l^2(l+1)^2}  C_{\kappa,\mathrm{gh}}(l) 
\end{equation}
where the functions $G^{\pm}_{l,2}(x)$ can be expressed in terms of 2nd order associated Legendre polynomials $\mathcal{P}_{l,2}(x)$ \citep{friedrich2020, Stebbins1996}:
\begin{equation}
\begin{split}
   G^+_{l,2}(x) \pm G^-_{l,2}(x)&  = \mathcal{P}_{l,2}(x) \left(\frac{4-l\pm2x(l-1)}{1-x^2} - \frac{l(l-1)}{2} \right) \\  & \qquad +  \mathcal{P}_{l-1,2}(x) \frac{(l+2)(x\mp2)}{1-x^2} \ . 
\end{split}
\end{equation}
These equations are exact for a curved-sky treatment and more accurate than the inverse Hankel transforms; the latter resulting in increasing errors for larger angular separations \citep{Kitching_2017}. The expressions can be easily evaluated using the \verb|gsl| library.

For computing the integrated 3-point functions $i\zeta_{\pm,\mathrm{fgh}}(\alpha)$, we first need to evaluate the integrated shear bispectra $iB_{\pm,\mathrm{fgh}}(l)$ (see equation \eqref{eq:integrated_bispectra}). We use the fitting formula for the 3D dark matter bispectrum $B^{\mathrm{3D}}_{\delta}(k_1, k_2, k_3, \eta)$ by \citealp{Gil_Marin_2012} (hereafter GM, see more in Appendix \ref{app:3D_bispectrum_modelling}) with the \verb|revised halofit| non-linear power spectrum implementation in \verb|CLASS| which we then integrate over the $\boldsymbol{l}_i$-modes and along the line-of-sight to obtain $iB_{\pm,\mathrm{fgh}}(l)$. For numerically computing the 5-dimensional integration in equation \eqref{eq:integrated_bispectra} we use the publicly available adaptive multi-dimensional integration package \verb|cubature|\footnote{currently hosted at: \url{https://github.com/stevengj/cubature} .} and evaluate each integrated bispectrum for 157 $l$-modes log-spaced in the range $1 \leq l \leq 20000$. In converting to real-space, we again replace the required inverse Hankel integrals of $iB_{\pm,\mathrm{fgh}}(l)$ in equation \eqref{eq:integrated_3pt_shear_correlations_bispectrum} by the summation over $l$ expressions in equation \eqref{eq:Stebbins_conversion} to obtain $i\zeta_{\pm,\mathrm{fgh}}(\alpha)$. In order to do so, we first linearly interpolate the $iB_{\pm,\mathrm{fgh}}(l)$ between the 157 log-spaced $l$-modes to get the $iB_{\pm,\mathrm{fgh}}(l)$ for every integer-$l$ multipole within the range specified above. We then use the interpolated value at every multipole to perform the summation.

In order to validate the theoretical model for the 2-point and integrated 3-point shear correlations on the T17 simulations, we also need to account for the effects in the simulations due to limited angular resolution of the maps, finite simulation box size and finite thickness of the lens shells as reported by \citealp{Takahashi2017}. We include these corrections in our theory power spectra as summarized in Appendix \ref{app:takahashi}.

\section{Results and Discussion}
\label{chap:results}
\begin{figure}
    \centering
	\includegraphics[width=0.8\columnwidth]{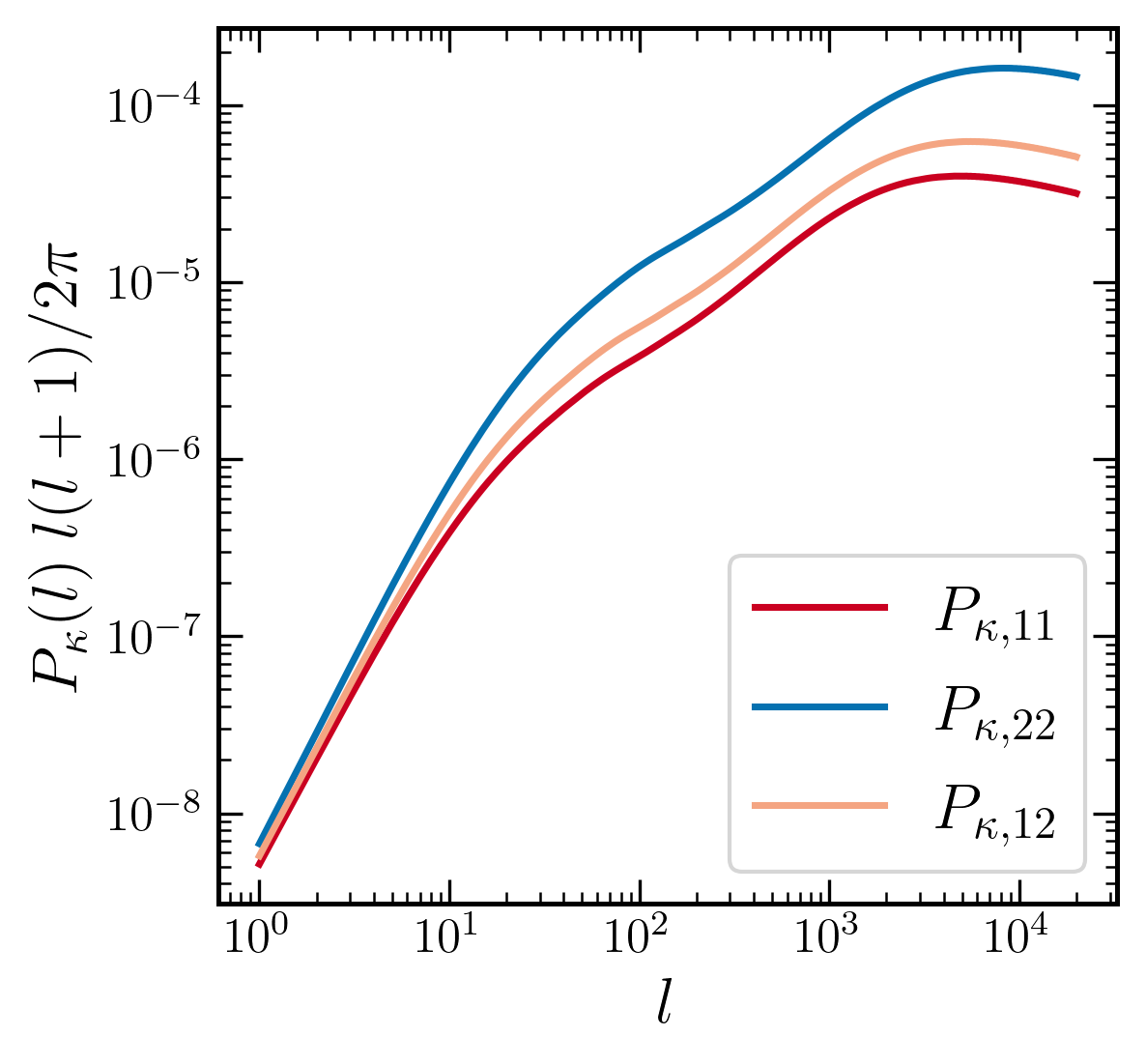}
    \caption{The scaled convergence auto and cross power spectra $P_{\kappa}(l)$ for two tomographic source redshift bins $z_1 = 0.5739$ and $z_2 = 1.0334$.}
    \label{fig:P_kappa}
\end{figure}
\begin{figure}
	\centering
	\includegraphics[width=0.8\columnwidth]{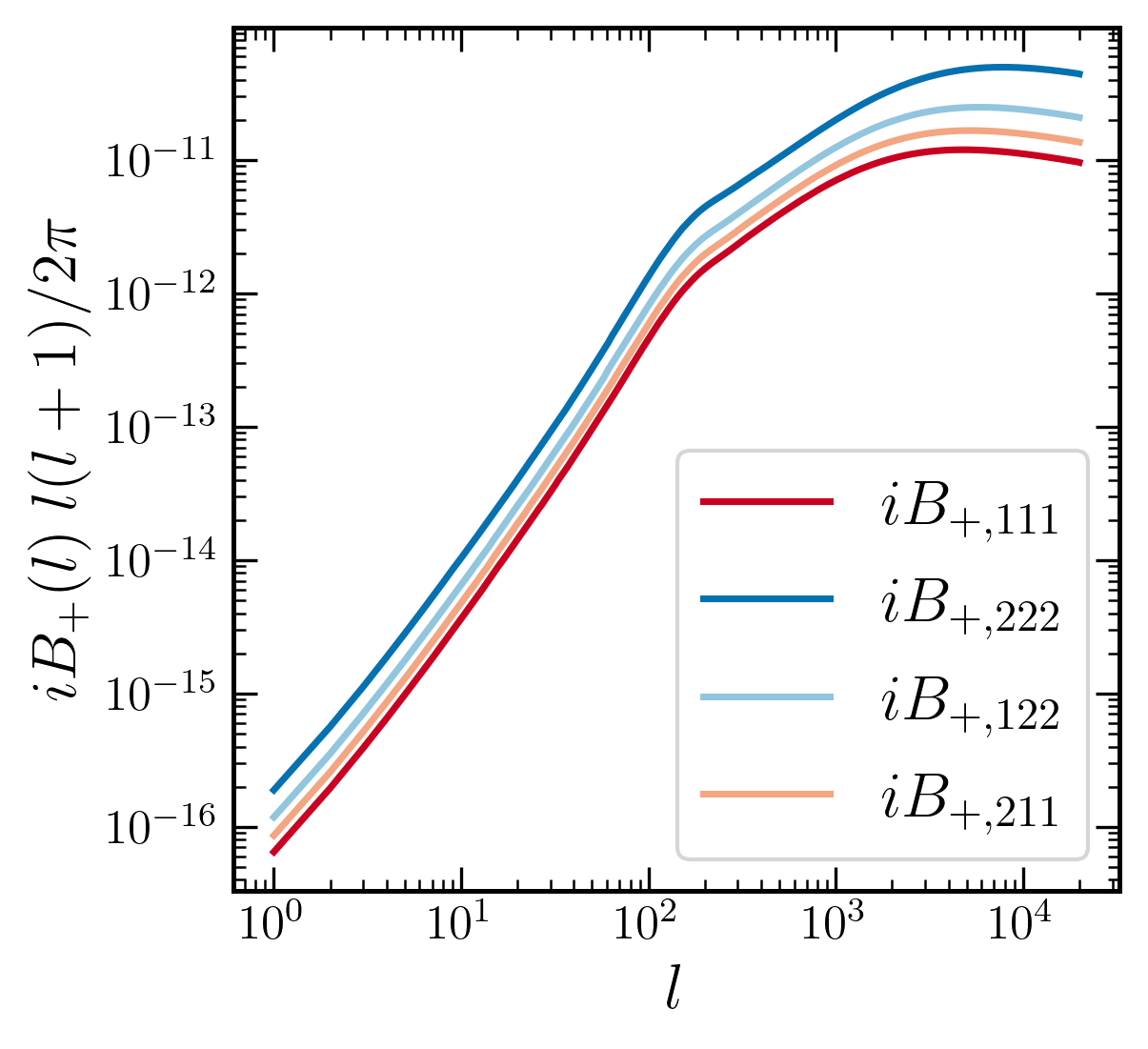}
    \caption{The scaled integrated bispectra $iB_{+}(l)$ for two tomographic source redshift bins $z_1 = 0.5739$ and $z_2 = 1.0334$. These spectra have been computed using a compensated filter of size $\theta_{\mathrm{ap}} = 70'$ and top-hat window of radius $\theta_{\mathrm{T}} = 75'$.}
    \label{fig:iB_plus}
\end{figure}

We now present the results of our measurements and theory calculations. In section \ref{sec:results_validation}, we test the accuracy of our model in describing the T17 data vector within the uncertainties expected from the sixth year cosmic shear data of the DES using the \verb|FLASK| covariance matrix. And in section \ref{sec:results_Fisher}, we explore the Fisher constraining power on cosmological parameters which can be obtained on performing a joint analysis of $\xi_{\pm}$ and $i\zeta_{\pm}$.

The results of the theory computation of the convergence power spectra $P_{\kappa,\mathrm{gh}}(l)$ for source redshifts $z_1 = 0.5739$ and $z_2 = 1.0334$ (where $g,h = 1,2$) are shown in Figure \ref{fig:P_kappa}. It is clear from the Figure that the convergence power spectrum for sources at higher redshift i.e. $P_{\kappa,22}$ is larger than the lower redshift power spectrum $P_{\kappa,11}$ indicating the presence of more amount of deflecting material between the observer and the source at larger redshifts; in other words, a larger lensing efficiency for sources situated at a higher redshift (see equation \eqref{eq:lensing_projection_kernel_single_zs}). Also, the spectra are smooth as features like the  baryonic acoustic oscillations which are prominent in the 3D matter power spectrum are smeared out due to the mixing of 3D $k$-modes into 2D $l$-modes through the line-of-sight projection (see equation \eqref{eq:convergence_power_spectrum}). In Figure \ref{fig:iB_plus} we show the integrated bispectra $iB_{+,\mathrm{fgh}}(l)$ for the two source redshifts $z_1$ and $z_2$ (where $f,g,h = 1,2$). As mentioned before, the integrated bispectra are evaluated using a compensated filter of size $\theta_{\mathrm{ap}} = 70'$ and two top-hat windows of radii $\theta_{\mathrm{T}} = 75'$. Other cross-combinations besides the four cross-spectra shown in the Figure, e.g. $iB_{+,\mathrm{112}}(l)$ and $iB_{+,\mathrm{212}}(l)$ are the same as $iB_{+,\mathrm{211}}(l)$ and $iB_{+,\mathrm{122}}(l)$, respectively (e.g. this can be easily verified from equation \eqref{eq:integrated_bispectra}). Hence, they add no extra information and we only consider these four. The $iB_{-,\mathrm{fgh}}$ spectra look similar to $iB_{+,\mathrm{fgh}}$ and are not shown separately.

It should be noted here that the high-$l$ end of the integrated shear bispectra pick up significant contributions from squeezed configurations of the convergence bispectrum $B_{\kappa}$ since the high-$l$ values correspond to computing the position-dependent correlation function in real space on angular scales much smaller than the size of the patch ($l \gg 2\pi/2\theta_T \approx 145$). As shown before for the 3D integrated bispectrum by \citealp{Chiang_2014} and for the 2D convergence bispectrum by \citet{Barreira_2019, munshi2020estimating}, this in turn corresponds to picking up the squeezed bispectrum configurations. However, it should be noted that the low-$l$ end of $iB$ picks up contribution from triangle configurations other than squeezed as the angular scales that the low-$l$ correspond to are close to the diameter of the patch where the squeezed limit does not hold (see Figure \ref{fig:iB_plus_tests} and discussion in Appendix \ref{app:tests_on_iB} for more details).

\subsection{Validation on T17 simulations}
\label{sec:results_validation}
\begin{figure}
	\includegraphics[width=\columnwidth]{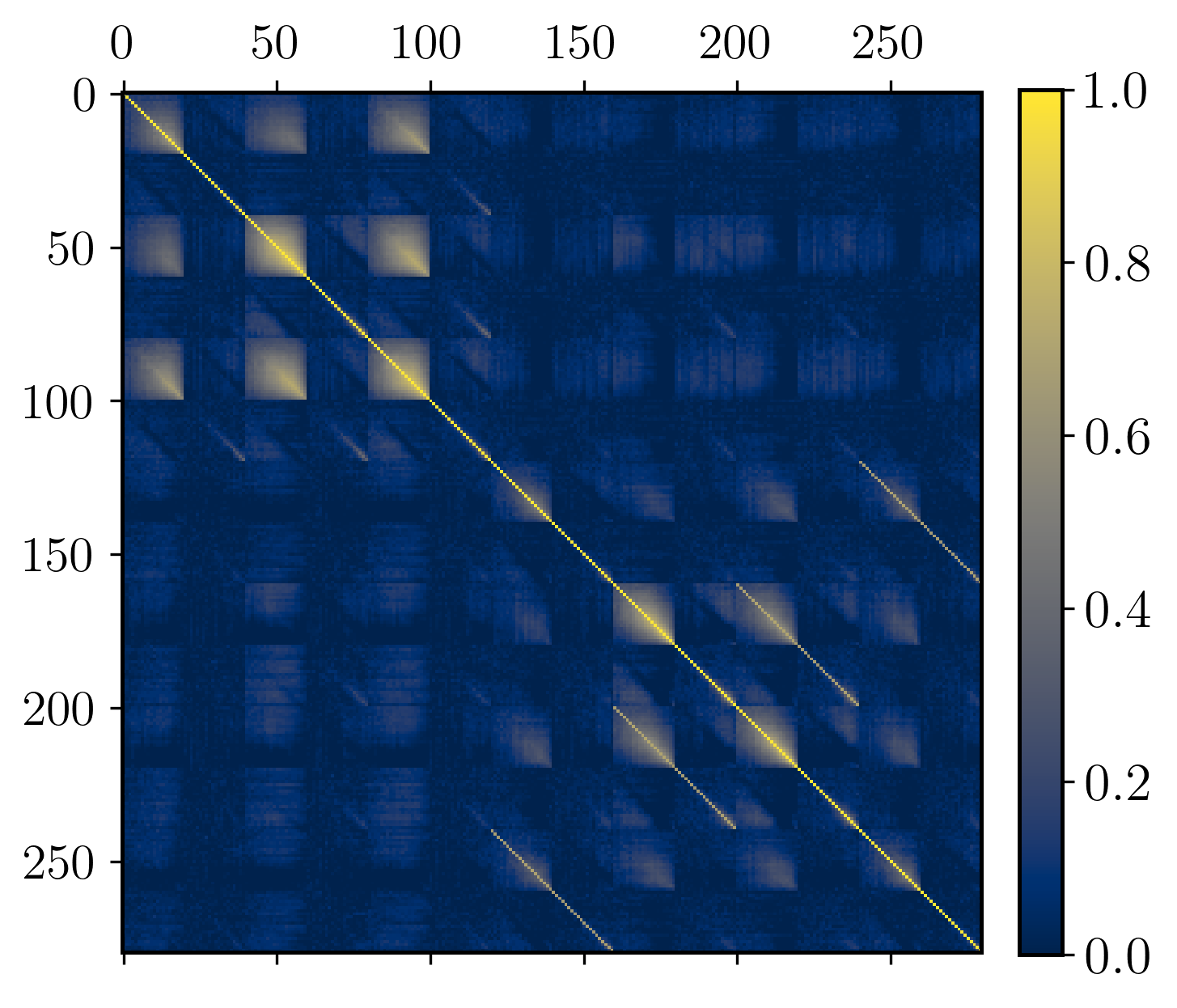}
    \caption{The $280\times280$ data-correlation matrix (normalised version of $\mathbf{\hat{C}}$, see equation \eqref{eq:data_covoriance_independent_realisations}) estimated from 2000 DES Year 6 sized footprints in FLASK lognormal sky-maps which include realistic shape-noise for two tomographic source redshift bins $z_1 = 0.5739$ and $z_2 = 1.0334$. Each $20\times20$ box around the diagonal indicates the correlation matrix for the 20 separation bins $\alpha$ of each of the 14 components of the data vector $D$ = $\left(\xi_{+,11}(\alpha), \xi_{-,11}(\alpha), \xi_{+,22}(\alpha), \xi_{-,22}(\alpha), ... , i\zeta_{+,211}(\alpha), i\zeta_{-,211}(\alpha) \right)^{\mathrm{T}}$ (see equation \eqref{eq:data_vector}). The off-diagonal boxes indicate the cross-correlations between the angular bins of different correlation functions.}
    \label{fig:covariance_flask}
\end{figure}

\begin{figure*} 
	\includegraphics[width=\textwidth]{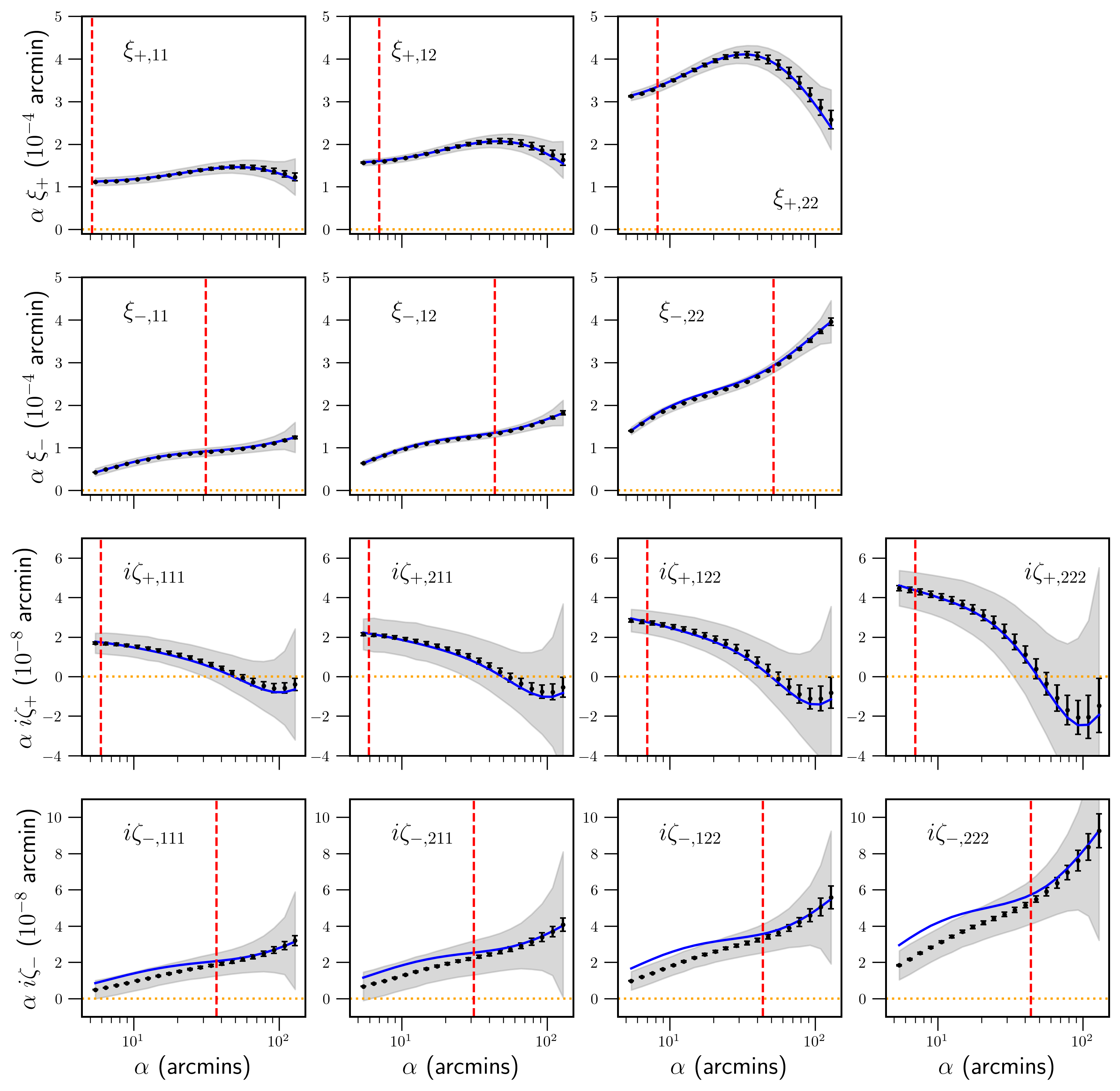}
    \caption{The 2-point shear correlation functions $\xi_{\pm}(\alpha)$ and the integrated 3-point shear correlation functions $i\zeta_{\pm}(\alpha)$ for two tomographic source redshift bins $z_1 = 0.5739$ and $z_2 = 1.0334$. The black dots with the error bars show the mean and the 1-sigma standard deviation of the measurements from the 108 T17 simulation maps, respectively. The grey shaded regions show the 1-sigma standard deviation for these statistics in DES Year 6 sized footprints obtained from the data-covariance matrix estimated using FLASK lognormal simulations with realistic shape-noise. The blue curves show the theoretical model predictions for the statistics. The theory curves include the corrections needed to account for finite angular resolution, simulation box size and shell thickness effects in the T17 simulations (see Appendix \ref{app:takahashi}). The integrated 3-point functions have been computed using a compensated filter of size $\theta_{\mathrm{ap}} = 70'$ and top-hat window of radius $\theta_{\mathrm{T}} = 75'$. The red-dashed lines denote the angular scale-cuts imposed on the data/model vectors using a $\chi^2$ criterion (see text). The angular bins smaller than the scale-cuts are not included in further analyses.}
    \label{fig:xi_iZ}
\end{figure*}

In Figure \ref{fig:xi_iZ} we show each component of the data vector $\overline{D}$ (black dots) evaluated from the mean of 108 T17 simulated maps for the two source redshifts. The error bars on the data points indicate the standard deviation over the 108 maps (note that these are noiseless simulations). The grey shaded region is the 1-sigma standard deviation computed from the data-covariance matrix $\mathbf{\hat{C}}$ estimated from 2000 DES Year 6 sized footprints in \verb|FLASK| lognormal sky-maps which include realistic shape-noise (see Figure \ref{fig:covariance_flask}). The model vector for each statistic is also shown in the plots (in blue) where we also include the corrections proposed by \citealp{Takahashi2017} to account for the various resolution effects of the T17 simulation (see Appendix \ref{app:takahashi}). The $\xi_{\pm}$ models are in good agreement with the T17 measurements within both the scatter of the simulations and the DES error bars. This is another confirmation of the result already reported by \citealp{Takahashi2017} that the convergence power spectrum (that we obtain using the \verb|revised halofit| 3D matter power spectrum) matches with the T17 simulations after taking into account the resolution corrections (see Appendix \ref{app:takahashi}). Our model predictions for the $i\zeta_{+}$ statistic also agrees well on all angular scales with the T17 simulations not only within the grey DES error bars but also within the scatter of the T17 simulations (black error bars). However, this is not the case for $i\zeta_{-}$ models as they are seen to be in agreement with the T17 simulations only on larger angular scales but over predict the simulations on smaller scales. This stems from an inaccuracy of the GM bispectrum fitting formula. At the small angular scales, the $i\zeta_{-}$ with its fourth-order Bessel function $J_{4}$ (see equation \eqref{eq:integrated_3pt_shear_correlations_bispectrum} and discussion after the equation) is most sensitive to the very high-$l$ values of the integrated bispectrum. At these very high-$l$ values, the integrated bispectrum signal is mostly due to the contributions from the highly squeezed configurations of the convergence bispectrum (see discussion in Appendix \ref{app:tests_on_iB}). The GM formula on the other hand, is known to overestimate these highly squeezed bispectrum configurations \citep{Sato_2013,Namikawa2018, Takahashi_2020} and hence causes the overestimation of the $i\zeta_{-}$ signal on the small angular scales. On the other hand, $i\zeta_{+}$ has a zeroth-order Bessel function $J_0$ weighting which is more sensitive to lower $l$ values (for a given angular scale) of the integrated bispectrum compared to $i\zeta_{-}$. At low to moderate-$l$, the integrated bispectrum receives contribution from not so highly squeezed and other bispectrum triangle configurations where the GM fitting function works reasonably well. In Appendix \ref{app:tests_on_iB}, we show the results of using a more accurate bispectrum fitting function \verb|bihalofit| \citep{Takahashi_2020} which correctly estimates the squeezed configurations and allows for an improved modelling of the $i\zeta_{-}$ correlations down to smaller angular scales (see Figure \ref{fig:iZ_GM_bihalofit}). However, \verb|bihalofit| is currently only applicable to $w$CDM cosmologies (i.e. $w_0$ = constant and $w_a$ = 0) whereas one of the major goals of our analysis is to investigate the constraining power of $i\zeta_{\pm}$ for cosmologies with dynamical dark energy $w_a \neq 0$ (see section \ref{sec:results_Fisher} and the discussion in Appendix \ref{app:tests_on_iB}). On the other hand, the GM fitting function is applicable to cosmologies with dynamical dark energy (as previously shown by \citealp{Sato_2013}) and hence we choose it as our fiducial bispectrum model instead of \verb|bihalofit|.

To compare how well the model vector $M$ describes the data vector $\overline{D}$ of a given statistic quantitatively, we compute the $\chi^2$ value as
\begin{equation} \label{eq:chi_squared}
    \chi^2 = (\overline{D} - M)^{\mathrm{T}} \mathbf{C}^{-1} (\overline{D} - M) 
\end{equation}
where $\mathbf{C}^{-1}$ is an unbiased estimate of the inverse data-covariance matrix $\mathbf{\hat{C}}^{-1}$ measured from $N_{r}$ realizations for a data vector containing $N_{d}$ elements \citep{Hartlap2007}:
\begin{equation} \label{eq:Hartlap_correction}
    \mathbf{C}^{-1} = \frac{N_{r} - N_{d} - 2}{N_{r} - 1} \; \mathbf{\hat{C}}^{-1} \ .
\end{equation}
Note that the estimation of this unbiased inverse data-covariance matrix requires $N_r > N_d + 2$. Moreover, a relatively small number of $N_r$ compared to $N_d$ results in a highly noisy inverse covariance estimate (see Appendix \ref{app:precision_matrix_expansion}). Hence, one usually needs $N_r \gg N_d$.

Using the $\chi^2$ value, computed using the \verb|FLASK| covariance matrix (see Figure \ref{fig:covariance_flask}), we make angular scale-cuts for every individual statistic (see the red-dashed vertical lines in Figure \ref{fig:xi_iZ}). For making a scale-cut we impose two conditions. Firstly, the $\chi^2$ value of a given statistic using all angular bins larger than the scale-cut must be lower than a threshold value of 0.15. Secondly, the fractional change in the $\chi^2$ value when ignoring the smallest bin right after the scale-cut, should be less than 15 per cent. For further analyses, this enables us to include only those parts of the model vectors which agree very well with the simulations with respect to the DES-like uncertainties.

In Table \ref{tab:S_N_ratio} we report the signal-to-noise ratio ($S/N$) of the various statistics after imposing the angular scale-cuts. The $S/N$ is computed as \citep{Chang2018_4_surveys}:
\begin{equation}
    S/N = \sqrt{ \; \overline{D}^{\mathrm{T}} \mathbf{C}^{-1} \overline{D} }
\end{equation}
and it indicates the statistical significance of the data vector. We also report the corresponding $\chi^2$ values for the data vectors. Although we require the $\chi^2$ for each individual statistic e.g. $\xi_{+,11}$ etc. to be below 0.15, there is no such restriction for the joint data vectors. The low $\chi^2$ value of 1.08 for the entire data vector (after the scale-cuts) confirms that the model is in good agreement with the simulations within the DES uncertainties. We also check in Appendix \ref{app:systematic_check} (see Figure \ref{fig:fisher_joint_best_fit_params}) whether any remaining systematic offset between the T17 data vector and our model vector after imposing the scale-cuts can cause any large parameter biases in our Fisher forecasts (see next section). We verify that the systematic offset for each parameter from the corresponding fiducial parameter value is smaller than one-third of the 1-sigma constraints expected from the Fisher analysis of the entire data vector.
\begin{table}
\caption{Signal-to-noise ratio $S/N$ for the T17 simulation data vectors computed with the FLASK covariance matrix. The $\chi^2$ values for the theory model with respect to the data vector are also reported along with the length of the data vector. All reported quantities are evaluated after imposing angular scale-cuts (see Figure \ref{fig:xi_iZ}). The data vector for each statistic includes all auto and cross-correlations for both tomographic bins $z_1 = 0.5739$ and $z_2 = 1.0334$ e.g. $\xi_{+} = \left( \xi_{+,11}, \xi_{+,12}, \xi_{+,22}\right)^{\mathrm{T}}$, $i\zeta_{+} = \left( i\zeta_{+,111}, i\zeta_{+,222}, i\zeta_{+,122}, i\zeta_{+,211}\right)^{\mathrm{T}}$ etc.}
\label{tab:S_N_ratio}
\centering
\begin{tabular}{|c|c|c|c|}
\hline
Data vector & Length of data vector & $S/N$ & $\chi^2$ \\
\hline
\hline
$\xi_{+}$ & 55 & 43.65 & 0.26 \\
$\xi_{-}$ & 22 & 36.77 & 0.30 \\
$\xi_{\pm}$ & 77 & 47.26 & 0.57 \\
\hline
\hline
$i\zeta_{+}$ & 74 & 8.06 & 0.16 \\
$i\zeta_{-}$ & 31 & 7.91 & 0.26 \\
$i\zeta_{\pm}$ & 105 & 9.41 & 0.51 \\
\hline
\hline
$\xi_{\pm}$ and $i\zeta_{\pm}$ & 182 & 48.40 & 1.08 \\
\hline
\end{tabular}
\end{table}

Although the $S/N$ of the $i\zeta_{\pm}$ is not as high as $\xi_{\pm}$ for a DES-like survey, the non-zero signals measured from the simulations without having had to compute the full 3-point correlation function shows the ease of measurement and also the potential of the integrated 3-point shear correlation function to probe higher-order information of the highly non-Gaussian late-time matter density field.  

\subsection{Fisher forecast on cosmological parameter constraints}
\label{sec:results_Fisher}
\begin{figure*}
    \centering
    \includegraphics[width=0.8\linewidth]{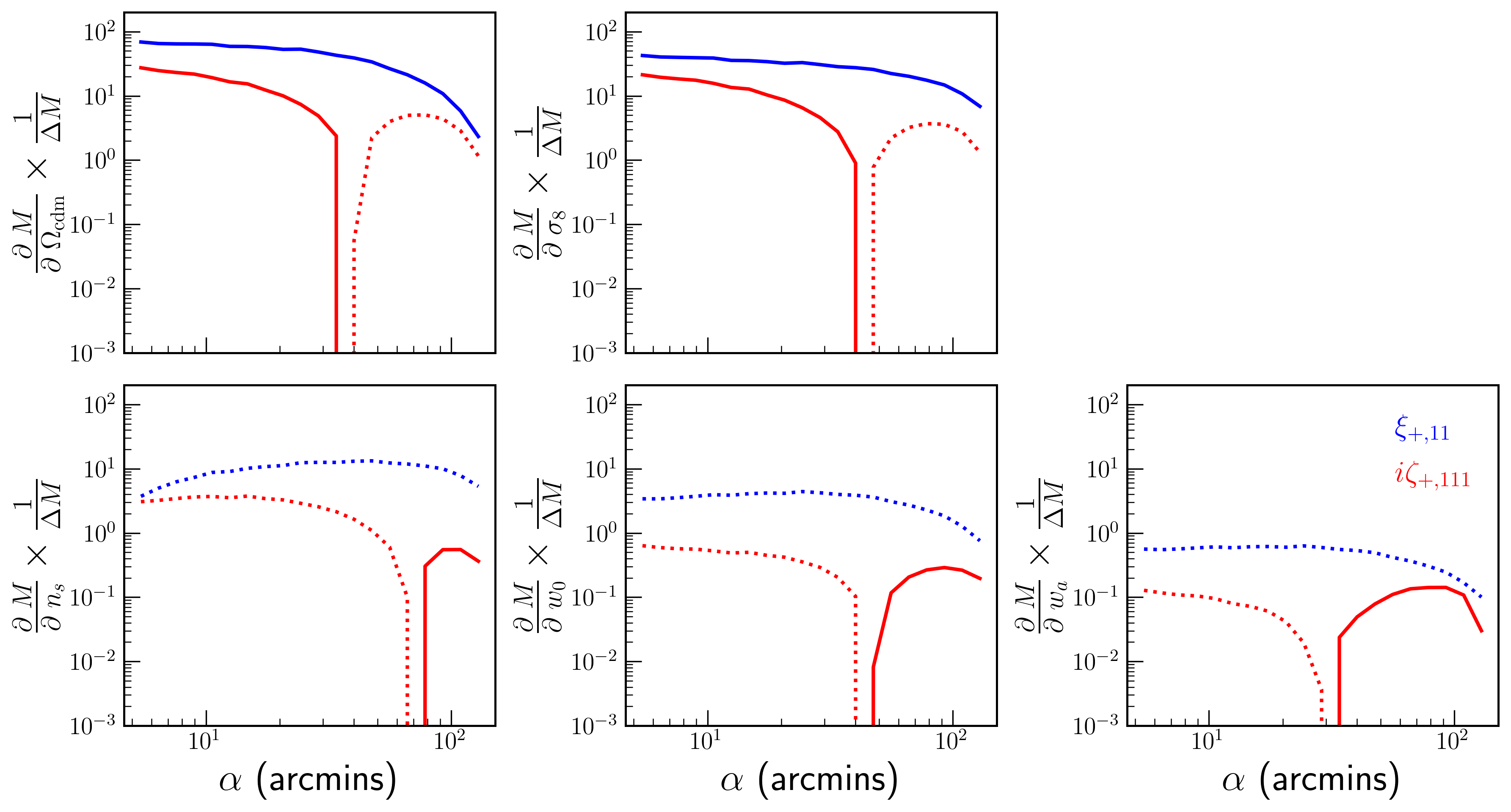}
    \caption{The derivatives $\frac{\partial M}{\partial \pi_{i}}$ of 2 components of the model vector $M$ --- $\xi_{+,11}$ (blue) and $i\zeta_{+,111}$ (red) shear correlation functions for source redshift bin $z_1 = 0.5739$ --- with respect to the 5 cosmological parameters $\boldsymbol{\pi} = \{ \Omega_{\mathrm{cdm}}, \sigma_8, n_s, w_0, w_a \}$ and normalised by the corresponding 1-sigma standard deviation $\Delta M (\alpha)$ (from the FLASK covariance matrix) at a given $\alpha$. The derivatives are negated (indicated with dotted lines) where $\frac{\partial M}{\partial \pi_{i}} < 0$.}
    \label{fig:derivatives_z1}
\end{figure*}
\begin{figure*}
    \centering
    \includegraphics[width=0.8\linewidth]{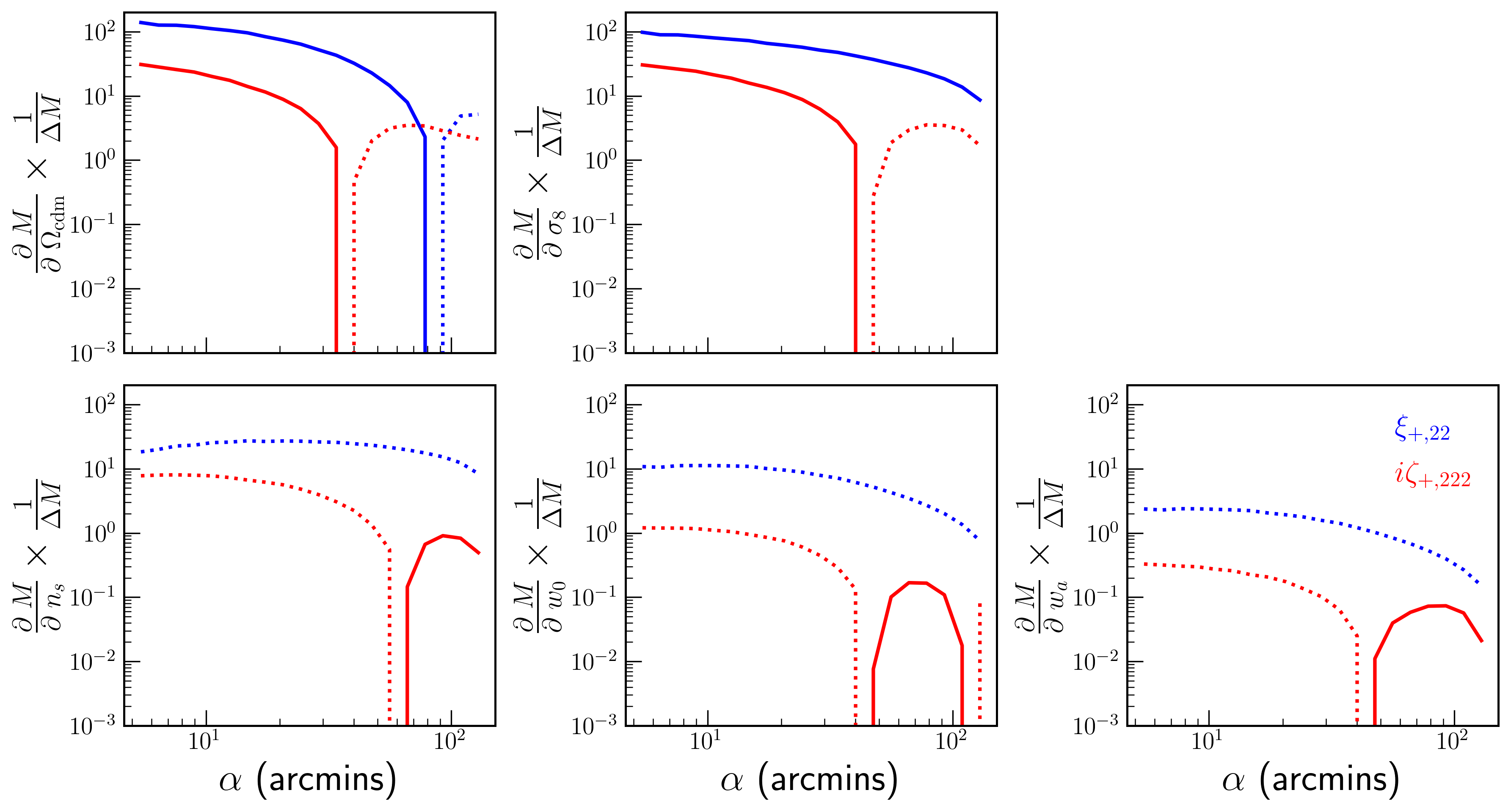}
    \caption{Same as Figure \ref{fig:derivatives_z1} but for $\xi_{+,22}$ (blue) and $i\zeta_{+,222}$ (red) shear correlation functions for source redshift bin $z_2 = 1.0334$.}
    \label{fig:derivatives_z2}
\end{figure*}
\begin{figure*}
  \centering
  \begin{minipage}[b]{\columnwidth}
    \includegraphics[width=\columnwidth]{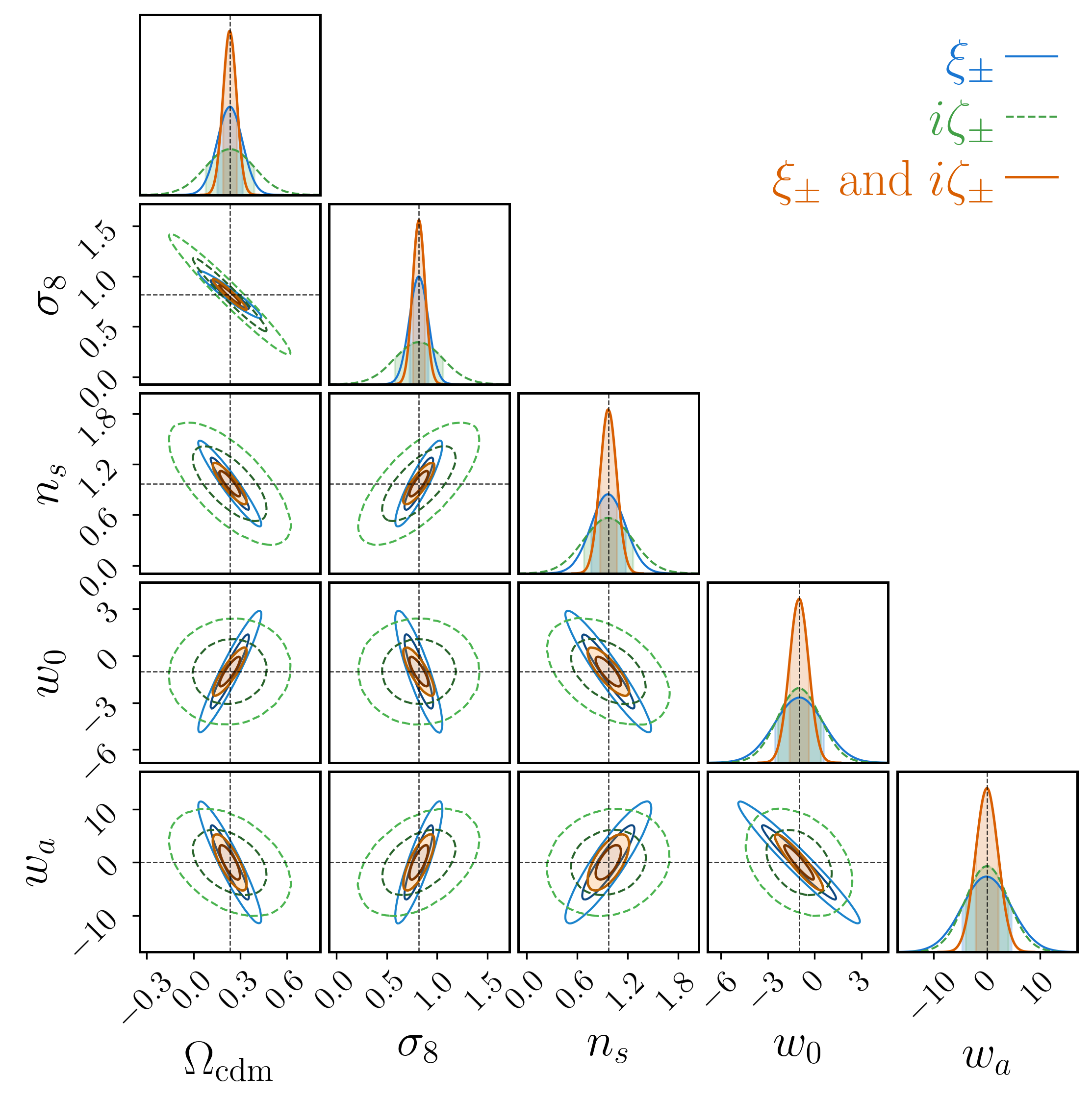}
  \end{minipage}
  \hfill
  \begin{minipage}[b]{\columnwidth}
    \includegraphics[width=\columnwidth]{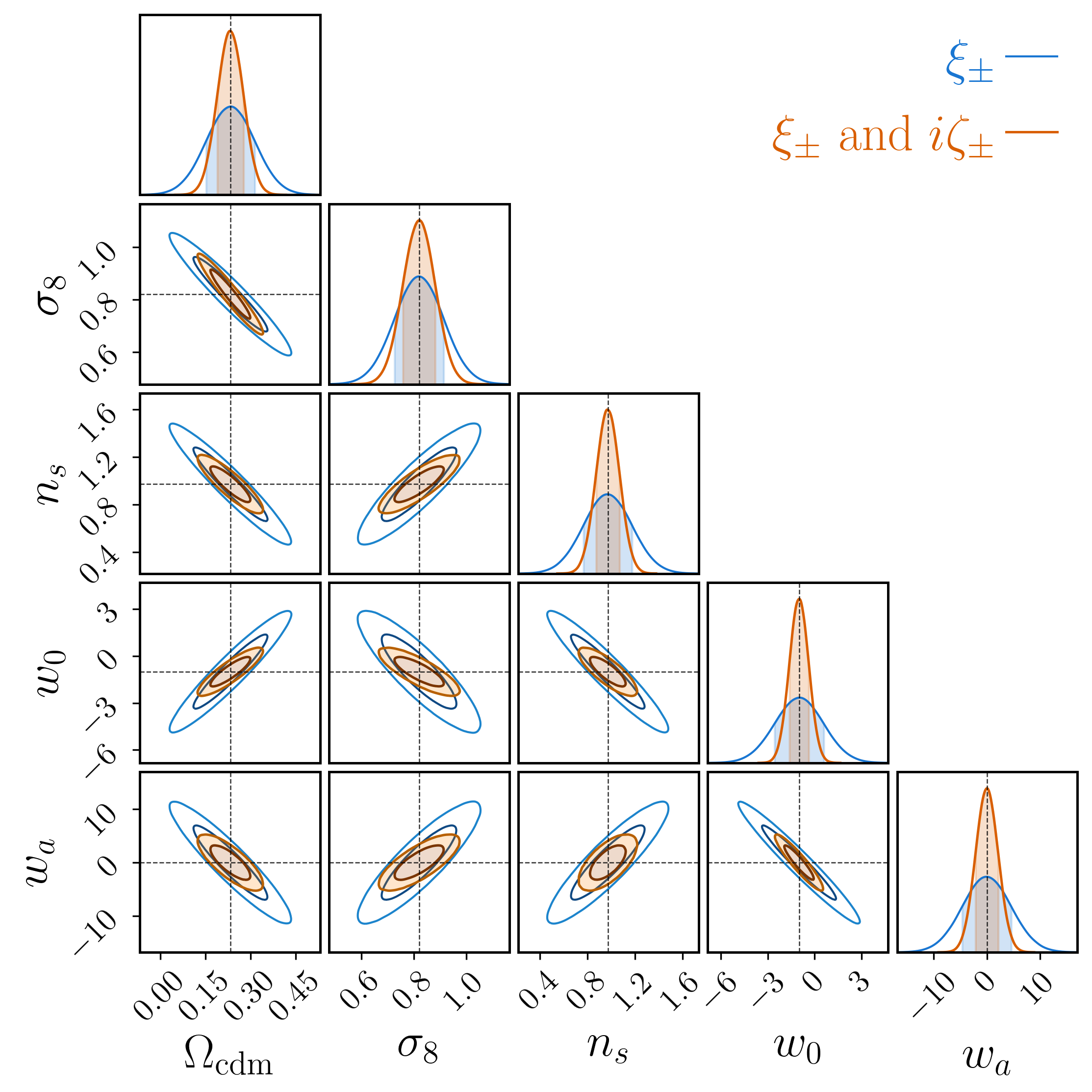}
  \end{minipage}
  \caption{\textbf{Left panel}: Fisher contours for the 5 cosmological parameters $\boldsymbol{\pi} = \{ \Omega_{\mathrm{cdm}}, \sigma_8, n_s, w_0, w_a \}$ for the model vectors --- $\xi_{\pm}$ (blue), $i\zeta_{\pm}$ (green dashed) and their joint model vector (orange) using the FLASK DES-like covariance matrix with realistic shape-noise in a two tomographic source redshift bin setting with $z_1 = 0.5739$ and $z_2 = 1.0334$. The contours are centred around the fiducial parameter values (black dotted lines) and are computed after imposing the angular scale-cuts on the model vectors (see Figure \ref{fig:xi_iZ}). \textbf{Right panel}: Same as left panel but zoomed in and only showing the $\xi_{\pm}$ (blue) and the joint contours (orange).}
  \label{fig:fisher_contours}
\end{figure*}
Having validated our theory model for the integrated 3-point shear correlations --- $i\zeta_{+}$ on all angular scales that we are interested in and $i\zeta_{-}$ on large angular scales --- we shall now address the Fisher information content of this statistic on cosmological parameters when analysed jointly with the 2-point shear correlation function. The Fisher information matrix $\mathbf{F}$ for a model vector $M$ which depends on a set of parameters $\boldsymbol{\pi}$ reads \citep{dodelson2020modern,Huterer2002}
\begin{equation} \label{eq:fisher_model_vector}
    F_{ij} = \left( \frac{\partial M(\boldsymbol{\pi})}{\partial \pi_{i}} \right)^{\mathrm{T}} \mathbf{C}^{-1} \left( \frac{\partial M(\boldsymbol{\pi})}{\partial \pi_{j}} \right). 
\end{equation}
where $F_{ij}$ corresponds to an element of $\mathbf{F}$ for the model parameters $\pi_{i}$ and $\pi_{j}$. The partial derivative of the model vector with respect to a model parameter $\pi_{i}$ can be computed using a 4-point central difference quotient\footnote{We prefer to use 4-point to 2-point central difference quotient for obtaining more accurate first derivatives (see also \citealp{yahiacherif2020}).} (also known as 5-point stencil derivative) \citep{abramowitz_stegun}:
\begin{equation}
    \frac{\partial M(\boldsymbol{\pi})}{\partial \pi_{i}} = \frac{-M(\pi_{i} + 2\delta_{i}) + 8M(\pi_{i} + \delta_{i}) - 8M(\pi_{i} - \delta_{i}) + M(\pi_{i} - 2\delta_{i})}{12 \delta_{i}}
\end{equation}
where $\delta_{i}$ is a small change of the parameter $\pi_{i}$ about its fiducial value, and $M(\pi_{i} \pm \delta_{i})$ means evaluating the model vector at the changed parameters $\pi_{i} \pm \delta_{i}$ while keeping all other parameters fixed. For our purpose we shall be interested in the cosmological parameters $\boldsymbol{\pi} = \{ \Omega_{\mathrm{cdm}}, \sigma_8, n_s, w_0, w_a \}$ where $w_0$ and $w_a$ indicate the dynamical dark energy equation of state parameters in the CPL parametrization \citep{chevallier_2001, Linder_2003} adopted by the Dark Energy Task Force \citep{Albrecht2006} to compare different dark energy probes. The fiducial values for our cosmological parameters are the same as that of the T17 simulations i.e. $\boldsymbol{\pi} = \{0.233, 0.82, 0.97, -1, 0 \}$. For the first four parameters we choose the step sizes $\delta_{i}$ to be 4, 2, 10 and 8 per cent of the corresponding fiducial values. For the $w_a$ parameter we adopt $\delta_{w_a} = 0.16$. We keep other parameters such as $\Omega_{\mathrm{b}}$, $h$ fixed to their fiducial (T17) values and keep the flatness of the Universe unchanged. This means when varying $\Omega_{\mathrm{cdm}}$, the amount of dark energy in the Universe is adjusted accordingly. These step sizes were motivated from \citealp{yahiacherif2020} who proposed optimal steps for the 5-point stencil derivative for  Fisher analysis with the galaxy power spectrum. For our analysis we use slightly larger steps than them but within the proposed range of steps for the parameters. Our steps were found as a trade off between neither being too big\footnote{To ensure that the steps were not too large, we verified that the $\delta_{i}$ were smaller than one-third of the 1-sigma marginalized Fisher constraints on the parameters (see Table \ref{tab:analysis_comparison}) for the joint model in our final analysis.} (in order to obtain accurate derivatives i.e. have low truncation errors) nor being too small (such that the derivatives are not dominated by numerical noise i.e. have low rounding-off errors). We do not impose any priors on these 5 cosmological parameters.

The inverse of the Fisher matrix gives the parameter covariance matrix $\mathbf{C}_{\boldsymbol{\pi}}$ under the assumptions that the measured data vector is drawn from a multi-variate Gaussian distribution\footnote{To go beyond the assumption that the data vector is drawn from a multi-variate Gaussian distribution, one can also perform cosmological parameter inference with the integrated 3-point shear correlation function in a likelihood-free inference setup \citep{Alsing_2018} as advocated recently for other weak lensing summary statistics by \citealp{Jeffrey_2020}.} and that the dependence of $M$ on the parameters $\boldsymbol{\pi}$ is close to linear \citep{trotta2017bayesian, Uhlemann_2020}:
\begin{equation} \label{eq:parameter_covariance_matrix}
    \mathbf{C}_{\boldsymbol{\pi}} = \mathbf{F}^{-1} \ .
\end{equation}
\begin{table*}
\caption{Comparison of our work in real space using shear 2-point and \textbf{integrated} 3-point correlations $\xi_{\pm},i\zeta_{\pm}$ against previous works in Fourier space by \citealp{Takada_2004, Kayo2013, Sato_2013} who used the convergence power spectrum $P_{\kappa}$ and the \textbf{full} convergence bispectrum $B_{\kappa}$. Some of the symbols used in the table that have not been defined in the text earlier are: the total number of source galaxies over all tomographic bins $n_g = \int \mathrm{d} z \; p(z)$ where $p(z)$ is the entire source galaxy distribution, $\alpha_s$ is the spectral running index parameter and $A_s = \delta_{\zeta}^2$ is the normalization parameter of the primordial power spectrum. We present the marginalized 1-sigma constraints on the cosmological parameters $\sigma(\pi_i)$ along with the dark energy figure-of-merit (FoM) for our work using $\xi_{\pm}$, $i\zeta_{\pm}$ and the combined (joint) data vector of the two, respectively (with scale-cuts on the data vector). The step sizes $\delta_i$ (for computing the derivatives of the model with respect to the parameters) when specified in per cent are relative to the fiducial parameter values. We also show corresponding values reported in the other works. The `---' indicates values which are not explicitly reported or inapplicable to the other works.}
\label{tab:analysis_comparison}
\centering
\begin{adjustbox}{width=1\textwidth}
\begin{tabular}{|c|c|c|c|c|}
\hline
 & Our work & \citealp{Takada_2004} & \citealp{Kayo2013} & \citealp{Sato_2013}\\
\hline
\hline
Total \# source galaxies $n_g$ (per arcmin$^2$) & 10 & 100; $p(z)$ following \citealp{Huterer2002} & 20 & 25\\
$\sigma_{\epsilon}$ & 0.3 & 0.4 & --- & 0.22\\
Area coverage (square degrees)& 5000 & 4000 & 1500 & 1100 \\
\# source redshift bins for tomography & 2 & 2 & 2 & 3\\
Source redshifts $z_i$ & $z_1 = 0.5739, z_2 = 1.0334$ & $ 0 \leq z_1 \leq 1.3 , z_2 > 1.3$ & --- & $z_1 = 0.6, z_2 = 1.0, z_3 = 1.5$ \\
Type of source redshift bin $p_i(z)$ & Dirac-$\delta$ function & Equal $n_{g,i}$ in each bin from $n(z)$ & Top-hat function & Dirac-$\delta$ function\\
\hline
\hline
Field & cosmic shear $\gamma$ & convergence $\kappa$ & convergence $\kappa$ & convergence $\kappa$\\
Analysis in real or Fourier space & real & Fourier & Fourier & Fourier \\
Data vectors (DVs) & $\xi_{\pm}(\alpha), i\zeta_{\pm}(\alpha), \mathrm{joint}$ & $P_{\kappa}(l), B_{\kappa}(l_1,l_2,l_3), \mathrm{joint}$ & $P_{\kappa}(l), B_{\kappa}(l_1,l_2,l_3), \mathrm{joint}$ & $P_{\kappa}(l), B_{\kappa}(l_1,l_2,l_3), \mathrm{joint}$
\\ Minimum and maximum scales & $5' < \alpha < 140'$ & $50 \leq l_i \leq 3000$ & $10 \leq l_i \leq 2000$ & $72 \leq l_i \leq 2000$ \\
Data-covariance & Lognormal simulations & Theoretical (only Gaussian covariance) &  Theoretical & Theoretical \\
Cross-covariance for joint DV & yes & no & yes & yes \\
Fisher analysis parameters & $\Omega_{\mathrm{cdm}}, \sigma_8, n_s, w_0, w_a$ & $\Omega_{\mathrm{de}}, \Omega_{\mathrm{b}}, h, n_s, \sigma_8, w_0, w_a$ & $\Omega_{\mathrm{de}}, \Omega_{\mathrm{m}}h^2, \Omega_{\mathrm{b}}h^2, n_s, \alpha_s, \delta_{\zeta}, w_0, w_a$ &  $\Omega_{\mathrm{de}}, \Omega_{\mathrm{cdm}}h^2, n_s, A_s, w_0, w_a$ \\
Derivative step sizes $\delta_i$ & 4\%, 2\%, 10\%, 8\%, 0.16 & $\delta_{w_a}=0.1$; 5\% for other parameters & --- & $\delta_{w_a}=0.5$; 10\% for other parameters \\
Analysis with flat or non-flat Universe & flat & flat & non-flat & flat \\
Priors in analysis & none & Planck priors on $\Omega_{\mathrm{b}}, h, n_s$ & Planck priors on all parameters & none \\
\hline
\hline
Marginalized $\sigma(\Omega_{\mathrm{cdm}})$ & 0.08 , 0.16 , 0.04 & --- & --- & ---\\
Marginalized $\sigma(\sigma_8)$ & 0.09 , 0.24, 0.06 & --- & --- & ---\\
Marginalized $\sigma(n_s)$ & 0.20 , 0.29 , 0.10 & --- & --- & ---\\
Marginalized $\sigma(w_0)$ & 1.55 , 1.36 , 0.62 & 0.34 , 0.32 , 0.11 & 0.51 , 0.62 , 0.38 & ---\\
Marginalized $\sigma(w_a)$ & 4.56 , 4.01 , 2.11 & 0.93 , 0.91 , 0.36 & 1.30 , 1.60 , 0.94 & --- \\
Dark energy FoM & 0.78 , 0.19 , 2.28 & --- & 11 , 7.2 , 20 & 5 , 15 , 25 \\
\hline
\end{tabular}
\end{adjustbox}
\end{table*}
Hence, using the derivatives and the expected data-covariance matrix (for a DES-sized survey) we can compute this parameter covariance matrix and forecast error contours on the cosmological parameters that we are interested in. In Figures \ref{fig:derivatives_z1} and \ref{fig:derivatives_z2} we show the derivatives of some components of our model vector with respect to the 5 cosmological parameters, normalised by the standard deviation for each component obtained from the \verb|FLASK| covariance matrix (in other words, dividing the derivative of a statistic by the corresponding grey shaded error in Figure \ref{fig:xi_iZ} for a given separation bin $\alpha$). This gives a visual estimate of the shape and amplitude of the ingredients of the Fisher matrix. It is clear that the way in which the amplitudes and shapes of $i\zeta_{+}$ derivatives (as a function of $\alpha$) differ from one parameter to another is different compared to $\xi_{+}$ which results in slightly altered orientations of the error contours of each statistic in the parameter planes. This can be seen in Figure \ref{fig:fisher_contours}. The error contours from $\xi_{\pm}$ are shown in blue, the contours from $i\zeta_{\pm}$ are shown in green dashed ellipses and the joint contours of the two together in orange. For clarity, we also remove the integrated 3-point function contours and show only the $\xi_{\pm}$ and the joint contours on the right hand panel of the Figure. Although the $i\zeta_{\pm}$ alone has larger contours compared to $\xi_{\pm}$  --- due to the lower amplitudes of the derivatives (see Figures \ref{fig:derivatives_z1}, \ref{fig:derivatives_z2}) which partly stems from the low $S/N$ of $i\zeta_{\pm}$ (see Table \ref{tab:S_N_ratio}) --- the degeneracy directions are slightly different. A joint analysis of $\xi_{\pm}$ along with $i\zeta_{\pm}$ thus helps to alleviate some of the parameter degeneracies  present in $\xi_{\pm}$ alone and result in a significant decrease in the contour sizes. The contribution from $i\zeta_{\pm}$ to the joint contours is significant with respect to the $w_0, w_a$ parameters. This can be reasoned by investigating the derivatives (see Figures \ref{fig:derivatives_z1} and \ref{fig:derivatives_z2}) of the statistics with respect to the dark energy equation of state parameters. The derivatives change more significantly for the different source redshifts with respect to $w_0$, $w_a$ for $i\zeta_{+}$ compared to $\xi_{+}$. This can be attributed to the fact that the 2-point shear correlation is a projection of the 3D power spectrum along the line-of-sight with a weighting of $\frac{q^2(\chi)}{\chi^2}$ (see equations \eqref{eq:2pt_shear_correlation_power_spectrum} and \eqref{eq:convergence_power_spectrum}), whereas the integrated 3-point shear correlation function has a factor of $\frac{q^3(\chi)}{\chi^4}$ (see equation \eqref{eq:integrated_3_point_function_plus_bispectrum_relation}) implying that the latter is weighted more heavily at lower redshifts (or smaller $\chi$), especially in the dark energy dominated era. This sensitivity of the projected integrated 3-point function to $w_0$ and $w_a$ shows potential in probing the dynamical dark energy equation of state from cosmic shear data\footnote{Interestingly, \citealp{Byun2017} found that the 3D integrated bispectrum is relatively insensitive in constraining the dynamical dark energy parameters compared to the 3D power spectrum. However, as we find, the sensitivity to $w_0$ and $w_a$ is different for the projected 2D integrated bispectrum compared to the projected 2D power spectrum --- arising due to the different geometric projection kernel weighting terms in their respective line-of-sight projections.}. Quantitatively, this can also be seen from the marginalized 1-sigma constraints $\sigma(\pi_i) = \sqrt{C_{\boldsymbol{\pi},ii}}$ of the $w_0$ and $w_a$ parameters for our analysis reported in the second column of Table \ref{tab:analysis_comparison}. The constraints obtained from the joint analysis of $\xi_{\pm}$ and $i\zeta_{\pm}$ i.e. $\sigma(w_0) = 0.62$, $\sigma(w_a) = 2.11$ are significantly smaller than those present in the individual analysis of $\xi_{\pm}$ i.e. $\sigma(w_0) = 1.55$, $\sigma(w_a) = 4.56$ or in $i\zeta_{\pm}$ i.e. $\sigma(w_0) = 1.36$, $\sigma(w_a) = 4.01$. The same is true for the other cosmological parameters. Alternatively, one often quotes the dark energy figure of merit (FoM) defined as \citep{Albrecht2006,Sato_2013}:
\begin{equation}
\label{eq:FOM}
    \mathrm{FoM} \equiv \frac{1}{\sqrt{\mathrm{det}\left( \mathbf{C}_{\boldsymbol{\pi}}[w_0,w_a]\right)}}
\end{equation}
to characterize the power of a survey to constrain these two parameters. The higher the FoM, the stronger are the constraints in the $w_0-w_a$ plane. For a DES-like survey, our joint analysis has a FoM = 2.28 which is almost 3 times larger than the FoM = 0.78 that we get from $\xi_{\pm}$ shear correlations alone; visually, this is reflected from the smaller size of the orange contours in the right-hand panel of Figure \ref{fig:fisher_contours} compared to the blue contours. The above quoted numbers are with the scale-cuts assumed in our analysis. We expect that including smaller angular scales will show more improvement on the marginalized constraints and also on the FoM\footnote{For example, assuming that our model is correct on all angular scales and without imposing any scale-cuts, we find that the FoM for the joint data vector improves by over a factor of 2.}. However this needs the development of more accurate models down to small angular scales. We also show for comparison, the marginalized constraints and FoM from previous works by \citealp{Takada_2004, Kayo2013, Sato_2013} who investigated the convergence power spectrum and the full convergence bispectrum. Their reported constraints are significantly better than ours which we associate to several differences in their analysis settings to ours e.g. higher $n_g$, no assumed scale-cuts, and for \citealp{Kayo2013} they assumed priors on the parameters of their Fisher analysis (see their Figure 1) whereas we do not impose any priors. Most importantly, these works investigate the constraining power of the full convergence bispectrum whereas we study only an integrated quantity of the bispectrum. The full bispectrum can be targeted to probe general bispectrum configurations thereby probing more information than integrated quantities of the bispectrum. Of course, this is also true in real space for the full 3-point shear correlation function $\gamma$-3PCF or the generalized third-order aperture mass statistics \citep{Schneider2005}. All these statistics should ideally be able to constrain the dark energy equation of state parameters better than the integrated 3-point shear correlation function. However, all of them rely on the accurate measurement of the full $\gamma$-3PCF (or the bispectrum) from data which is still unexplored in current wide-area weak lensing surveys. The integrated 3-point shear correlation function is much easier to measure and holds potential to improve upon the parameter constraints obtained from 2-point shear analyses alone. On the theory side, we expect that including other effects such as galaxy intrinsic alignments, baryonic feedback, impact of massive neutrinos etc. should be easier to tune into the $i\zeta_{\pm}$ model compared to including them for the full shear 3-point correlation function. From both observational and theoretical aspects, this makes the integrated 3-point shear correlation function a promising statistic to explore in current and future cosmic shear data.

\section{Conclusions}
In this paper we propose a higher-order statistic --- \textit{the integrated 3-point shear correlation function} --- which can be measured directly from the cosmic shear field observed in current wide-area weak-lensing surveys such as DES, KiDS, HSC and future surveys like Rubin Observatory Legacy Survey of Space and Time (LSST) and EUCLID\footnote{see \url{https://www.lsst.org} and \url{https://www.euclid-ec.org} .}.
The following are the key results of this work:
\begin{itemize}
    \item The integrated 3-point shear correlation function $i\zeta_{\pm}$ can be measured by dividing a large survey area into several top-hat patches (each having an area of a few square degrees) and correlating the \textit{position-dependent (local) 2-point shear correlation function} inside each patch with the \textit{aperture mass statistic} evaluated at the centre of the corresponding patch using a compensated filter. For fixed filter sizes, the $i\zeta_{\pm}(\alpha)$ is a function of a single variable --- the separation scale $\alpha$ at which the local 2-point shear correlation function is measured. This makes it analogous to the full shear 2-point correlation function $\xi_{\pm}(\alpha)$ which is widely measured in weak lensing surveys (see Figure \ref{fig:xi_iZ}).
    \item We develop a theoretical model for $i\zeta_{\pm}$ which is the real space counterpart of \textit{the integrated convergence bispectrum} as introduced by \citealp{munshi2020estimating} in Fourier space. The authors however, formulated the integrated bispectrum using equal-sized top-hat patches on the convergence field. Working in real space with cosmic shear, we instead propose the usage of a combination of compensated (for the aperture mass statistic) and top-hat filters (for the local 2-point shear correlation) of different sizes allowing for the evaluation of the statistic directly from cosmic shear data without any need for constructing a convergence map. We compute our theoretical models using the \citealp{Gil_Marin_2012} bispectrum fitting formula with the \verb|revised halofit| non-linear matter power spectrum \citep{Takahashi2012} implementation in the \verb|CLASS| software \citep{lesgourgues2011,blas2011}.
    \item We validate our model for the integrated 3-point function using the weak lensing shear simulations from \citealp{Takahashi2017}. We find that our theoretical predictions are in excellent agreement for the measured `$+$' integrated 3-point functions $i\zeta_{+}$ (analogous to the $\xi_{+}$ shear 2-point correlation function) within the scatter of the simulations for multiple source redshifts and the cross-correlations thereof (see Figure \ref{fig:xi_iZ}). However, our model for the `$-$' integrated 3-point shear correlation functions $i\zeta_{-}$ (analogous to the $\xi_{-}$ shear correlation function) agree with the simulations on large angular scales but over predict the simulation results on small scales. We associate this with the over estimation of the bispectrum by the \citealp{Gil_Marin_2012} fitting formula for the highly squeezed configurations of the bispectrum which the $i\zeta_{-}$ correlation function is mainly sensitive to. A more theoretically motivated formalism e.g. using the response function approach to modelling the squeezed lensing bispectrum as recently studied by \citealp{Barreira_2019} --- who also formulated the effect of baryons on the squeezed bispectrum --- may help to accurately model the $i\zeta_{-}$ correlation functions down to smaller angular scales. This also shows the potential in encoding effects of non-linear processes (e.g. baryonic feedback) in the integrated 3-point function which we expect to be easier compared to modelling them for the full 3-point shear correlation function. This is left as a direction for future work. In $w$CDM cosmologies, one can use a more accurate 3D matter bispectrum fitting function such as \verb|bihalofit| \citep{Takahashi_2020} to achieve improved modelling of the $i\zeta_{-}$ correlations (see Appendix \ref{app:tests_on_iB}).
    \item Making appropriate scale-cuts on the model vectors, we use the Fisher matrix formalism to forecast constraints on cosmological parameters for a DES Year 6 sized survey with realistic shape-noise in a 2-redshift bin tomographic setting (see Figure \ref{fig:fisher_contours}). For the data-covariance matrix we use a set of lognormal simulations using the \verb|FLASK| tool \citep{xavier2016}. We find that the joint analysis of the integrated 3-point function and the 2-point shear correlation functions can allow for a significant improvement in the parameter constraints compared to those obtained from 2-point shear correlation functions alone (see Table \ref{tab:analysis_comparison}). This is because the responses of the integrated 3-point shear correlations to the cosmological parameters are different from that of 2-point shear correlations thereby resulting in slightly different degeneracy directions in the parameter planes (see Figures \ref{fig:derivatives_z1} and \ref{fig:derivatives_z2}). In particular, we find that the integrated 3-point function has the potential to significantly improve the dark energy figure of merit on a combined analysis with 2-point shear correlation functions. This arises due to the derivatives of the integrated 3-point function (with respect to the dark energy equation of state parameters) varying considerably in shape and amplitude compared to the derivatives of the 2-point shear correlation. This can be attributed to the fact that the line-of-sight projection kernel in the expression for the convergence bispectrum is weighted considerably more heavily down to low redshifts (in the late-time dark-energy dominated era) compared to the convergence power spectrum (see equations \eqref{eq:convergence_bispectrum} and \eqref{eq:convergence_power_spectrum}). This can be very useful for probing the dark energy equation of state parameters from cosmic shear data alone and makes the integrated 3-point shear correlation function a promising method to probe higher-order information content of the shear field and thereby complement 2-point shear analysis.
\end{itemize}
Theoretically, the integrated 3-point function (or the integrated bispectrum) of the lensing convergence field should be easier to work with than the $i\zeta_{\pm}$ shear correlation function that we investigate in this paper. However, observationally, the former requires one to go through the convergence map making process from the cosmic shear field. This process becomes challenging when the observed shear field has complicated masks and survey geometry. Although our analysis involves a simulated setup with simplifying assumptions such as a circular survey footprint without masks and holes, accounting for the masking effects is straight forward as our statistic is designed to be measured directly from the cosmic shear data (where the masking effects are inherent) without the need for any map making. The integrated 3-point shear correlation function with its ease of measurement through the 2-point position-dependent shear correlation function and the 1-point aperture mass statistic is tailor-made for application to real data.

Although we have concentrated on the integrated 3-point function of the cosmic shear field, we provide a general framework of equations in chapter \ref{chap:theory_general_formalism} which can be used for computing the integrated 3-point function for any projected field e.g. galaxy counts field and its cross-correlations with the shear field. This will be explored in future works.

\section*{Acknowledgements}

We sincerely thank Alexandre Barreira, Daniel Gruen and Eiichiro Komatsu for helpful discussions and suggestions at various stages of the project. We remain grateful to Ryuichi Takahashi for making the T17 simulation suite publicly available and for clarifying our queries. OF gratefully acknowledges support by the Kavli Foundation and the International Newton Trust through a Newton-Kavli-Junior Fellowship and by Churchill College Cambridge through a postdoctoral By-Fellowship. This research was supported by the Excellence Cluster ORIGINS which is funded by the Deutsche Forschungsgemeinschaft (DFG, German Research Foundation) under Germany's Excellence Strategy - EXC-2094-390783311. Some of the numerical calculations have been carried out on the computing facilities of the Computational Center for Particle and Astrophysics (C2PAP). The results in this paper have been derived using the following publicly available libraries and software packages: \verb|gsl| \citep{gough2009gnu}, \verb|healpy| \citep{Zonca2019}, \verb|treecorr| \citep{Jarvis_2004}, \verb|CLASS| \citep{lesgourgues2011}, \verb|FLASK| \citep{xavier2016} and \verb|NumPy| \citep{harris2020numpy}. We also acknowledge the use of \verb|matplotlib| \citep{Hunter:2007} and \verb|ChainConsumer| \citep{Hinton2016} python packages in producing the Figures shown in this paper.

\section*{Data Availability}

The data for the N-body simulations used in this article were accessed from the public domain: \url{http://cosmo.phys.hirosaki-u.ac.jp/takahasi/allsky_raytracing/} . The lognormal simulations used in this work were generated using the publicly available \verb|FLASK| software: \url{http://www.astro.iag.usp.br/~flask/} .



\bibliographystyle{mnras}
\bibliography{references} 




\appendix

\section{Fourier and Hankel Transforms}
\label{app:hankel}
The forward and inverse Fourier transforms of a field $f$ in the 2D sky-plane can be written as
\begin{equation} \label{eq:ft_2D}
\begin{split}
    f(\boldsymbol{l}) & = \mathcal{F}_{\mathrm{2D}}[f(\boldsymbol{\theta})] \equiv \int \mathrm{d}^2 \boldsymbol{\theta} \;  f(\boldsymbol{\theta}) e^{-i\boldsymbol{l}\cdot \boldsymbol{\theta}} \qquad (\text{forward FT}) \\ f(\boldsymbol{\theta}) & = \mathcal{F}_{\mathrm{2D}}^{-1}[f(\boldsymbol{l})] \equiv \int \frac{\mathrm{d}^2 \boldsymbol{l}}{(2\pi)^2} \; f(\boldsymbol{l}) e^{i\boldsymbol{l}\cdot \boldsymbol{\theta}} \qquad (\text{inverse FT})
\end{split}
\end{equation}
where $\boldsymbol{l} = (l_x,l_y)$ is the 2D Fourier wave-vector. If the field $f$ is real i.e. $f^*(\boldsymbol{\theta}) = f(\boldsymbol{\theta})$, then it follows from the above equation that $f^*(\boldsymbol{l}) = f(-\boldsymbol{l})$.

If a function (e.g. correlation function) $\xi(\boldsymbol{\alpha})$ defined in the 2D sky plane is independent of the direction of the vector $\boldsymbol{\alpha}$ i.e. $\xi(\boldsymbol{\alpha}) = \xi(\alpha)$, then it follows from the Fourier transformation equation \eqref{eq:ft_2D} and from the properties of ordinary Bessel functions that \citep{Schneider_2006, dodelson2020modern}:
\begin{equation} \label{eq:ht_2D}
\begin{split}
    P(l) \equiv \mathcal{F}_{\mathrm{2D}}[\xi(\alpha)] & = \int \mathrm{d}^2 \boldsymbol{\alpha}\; \; \xi(\alpha) \; e^{-i\boldsymbol{l}\cdot \boldsymbol{\alpha}} \\ 
    & = 2\pi \int \mathrm{d} \alpha\; \alpha \; \xi(\alpha) \;J_0(l \alpha) \ , \\ 
    \xi(\alpha) \equiv \mathcal{F}_{\mathrm{2D}}^{-1}[P(l)] & = \int \frac{\mathrm{d}^2 \boldsymbol{l}}{(2 \pi)^2} \;P(l)  \; e^{i\boldsymbol{l}\cdot \boldsymbol{\alpha}} \\
    & = \int \frac{\mathrm{d} l\; l}{2\pi}  \;P(l) \;J_0(l \alpha)
\end{split}
\end{equation}
where $J_0(x)$ is the zeroth-order Bessel function of the first kind.

On the other hand, the 2D Fourier transform of $\xi(\alpha)$ with a complex phase factor $e^{4i\phi_{\boldsymbol{\alpha}}}$ and its inverse transform reads
\begin{equation} \label{eq:ht_phase_2D}
\begin{split}
    P(l) \equiv \mathcal{F}_{\mathrm{2D}}[\xi(\alpha)e^{4i\phi_{\boldsymbol{\alpha}}}] & = \int \mathrm{d}^2 \boldsymbol{\alpha}\; \; \xi(\alpha) \; e^{-i\boldsymbol{l}\cdot \boldsymbol{\alpha}} e^{4i\phi_{\boldsymbol{\alpha}}} \\ 
    & = 2\pi \int \mathrm{d} \alpha\; \alpha \; \xi(\alpha) \;J_4(l \alpha) \ , \\ 
    \xi(\alpha) \equiv \mathcal{F}_{\mathrm{2D}}^{-1}[P(l)e^{-4i\phi_{\boldsymbol{l}}}] & = \int \frac{\mathrm{d}^2 \boldsymbol{l}}{(2 \pi)^2} \;P(l)  \; e^{i\boldsymbol{l}\cdot \boldsymbol{\alpha}} e^{-4i\phi_{\boldsymbol{l}}}\\
    & = \int \frac{\mathrm{d} l\; l}{2\pi}  \;P(l) \;J_4(l \alpha)
\end{split}
\end{equation}
where $\phi_{\boldsymbol{\alpha}}$ is the polar angle of $\boldsymbol{\alpha}$, $\phi_{\boldsymbol{l}}$ is the polar angle of $\boldsymbol{l}$ and $J_4(x)$ is the fourth-order Bessel function of the first kind. These equations are Hankel transformations.

\section{T17 simulations power spectra correction formulae}
\label{app:takahashi}
\citealp{Takahashi2017} found that the convergence power spectra that was measured from the mean of the 108 simulated sky-maps in their simulation suite, slightly underestimated the theoretical power spectrum calculated with \verb|revised halofit| (which we are also using in this paper). They associated 3 effects that caused the underestimation and provided correction factors to the theory formulae to take them into account:
\begin{enumerate}
    \item \textbf{Finite simulation-box-size effect} : In Appendix B of \citealp{Takahashi2017}, the authors report that in order to consider the effect of density fluctuations larger than the simulation-box-size $L$ on the angular power spectrum, one needs to impose the condition that for $k < 2 \pi / L$, the matter power spectrum $P_{\delta}^{\mathrm{3D}}(k,\eta) \overset{!}{=} 0$, as the box does not include fluctuations larger than $L$.
    \item \textbf{Finite lens-shell effect} : The T17 lensing maps were produced by ray tracing through lens shells of finite thickness in the simulation boxes (see section \ref{sec:T17_sims}). The finite thickness affects the angular power spectrum of surface density fluctuations on a shell. To account for this, \citealp{Takahashi2017} suggest to convolve the matter power spectrum with the window function of the shell (see their Appendix B). They provide a fitting formula for the convolved power spectrum:
    \begin{equation}
        P_{\delta}^{\mathrm{3D}}(k,\eta) \longrightarrow \frac{(1+c_1 k ^{-\alpha_1})^{\alpha_1}}{(1+c_2 k ^{-\alpha_2})^{\alpha_3}} P_{\delta}^{\mathrm{3D}}(k,\eta)
    \end{equation}
    with $c_1 = 9.5171 \times 10^{-4}$, $c_2 = 5.1543 \times 10^{-3}$, $\alpha_1 = 1.3063$, $\alpha_2 = 1.1475$, and $\alpha_3 = 0.62793$ which they find to be in good agreement with the analytically computed convolved power spectrum up to redshift $z < 7.1$ which is well within the range considered in this paper.
    \item  \textbf{Finite angular resolution of sky-maps} : For a given \verb|NSIDE| of a \verb|Healpix| map, the angular power spectrum $C(l)$ measured from the sky-map is underestimated compared to the theoretical power spectrum at large $l$ due to lack of angular resolution. To account for this in the theory spectrum, \citealp{Takahashi2017} suggest a damping factor at small scales (high-$l$) given by
    \begin{equation}
        C(l) \longrightarrow \frac{C(l)}{1+\left( l/l_{\mathrm{res}}\right)^2} 
    \end{equation}
    where $l_{\mathrm{res}} = 1.6 \cdot \verb|NSIDE|$ .
\end{enumerate}

\section{3D matter bispectrum}
\label{app:3D_bispectrum_modelling}
The 3D matter bispectrum at leading order (tree-level) in density perturbations as computed with standard Eulerian perturbation theory (PT) for Gaussian initial conditions is written as \citep{Bernardeau_2002,dodelson2020modern}:
\begin{equation} \label{eq:tree_level_bispectrum}
\begin{split}
    B_{\delta,\mathrm{tree}}^{\mathrm{3D}}(\boldsymbol{k}_1,\boldsymbol{k}_2,\boldsymbol{k}_3,\eta) & = 2 \;  F_2(\boldsymbol{k}_1,\boldsymbol{k}_2,\eta) \; P_{\delta,L}^{\mathrm{3D}}(k_1,\eta) \; P_{\delta,L}^{\mathrm{3D}}(k_2,\eta) \\ & \qquad \qquad + \; \text{cyclic  permutations}  
\end{split}
\end{equation}
where 
\begin{equation}
    P_{\delta,L}^{\mathrm{3D}}(k,\eta) = D_{+}^2(\eta)P_{\delta,L}^{\mathrm{3D}}(k,\eta_0)
\end{equation}
is the 3D linear matter power spectrum today evolved to time $\eta$ using the linear growth factor $D_{+}(\eta)$ which is normalised to unity today i.e. $D_{+}(\eta_0) = 1$. $F_2(\boldsymbol{k}_i,\boldsymbol{k}_j,\eta)$ is a symmetrized two-point mode coupling kernel which in a general $\Lambda$CDM universe takes the form \citep{Friedrich_2018}:
\begin{equation}
    F_2(\boldsymbol{k}_i,\boldsymbol{k}_j,\eta) = \mu(\eta) + \frac{1}{2} \cos(\phi_{ij}) \left( \frac{k_i}{k_j} + \frac{k_j}{k_i} \right) + [1-\mu(\eta)]\cos^2(\phi_{ij})
\end{equation}
where $\phi_{ij}$ is the angle between the two wave-vectors $\boldsymbol{k}_i$ and $\boldsymbol{k}_j$. In an Einsten-de Sitter (EdS) universe, the function $\mu(\eta)$ is a constant and takes the value $\mu(\eta) = \frac{5}{7}$. However, this form of the bispectrum only works in the linear regime (large physical scales) and fails in the non-linear regime. To improve upon this, one can go on to include higher-order PT corrections but calculating the higher order terms are cumbersome. Another way of predicting the non-linear matter bispectrum is to propose a fitting formula for the bispectrum and calibrate the function's parameters using the bispectra measured from cold dark matter N-body simulations. This approach was first taken by \citealp{Scoccimarro_1999} and later improved by \citealp{Scoccimarro_2001} and \citealp{Gil_Marin_2012}. In this paper we use the bispectrum fitting formula of \citealp{Gil_Marin_2012}:
\begin{equation}
\begin{split}
    B_{\delta}^{\mathrm{3D}}(\boldsymbol{k}_1,\boldsymbol{k}_2,\boldsymbol{k}_3,\eta) & = 2 \;  F_2^{\mathrm{eff}}(\boldsymbol{k}_1,\boldsymbol{k}_2,\eta) \; P_{\delta}^{\mathrm{3D}}(k_1,\eta) \; P_{\delta}^{\mathrm{3D}}(k_2,\eta) \\ & \qquad \qquad + \; \text{cyclic  permutations}  
\end{split}
\end{equation}
where $P_{\delta}^{\mathrm{3D}}(k,\eta)$ is the 3D non-linear matter power spectrum (e.g. obtained using \verb|revised halofit| \citep{Takahashi2012}) and the effective mode coupling kernel $F_2^{\mathrm{eff}}(\boldsymbol{k}_1,\boldsymbol{k}_2,\eta)$ is a modified version of the EdS $F_2$ kernel and reads
\begin{equation}
\begin{split}
    F_2^{\mathrm{eff}}(\boldsymbol{k}_i,\boldsymbol{k}_j,\eta) & = \quad \frac{5}{7}a(k_i,\eta)a(k_j,\eta)  \\ & \quad + \; \frac{1}{2} \cos(\phi_{ij}) \left( \frac{k_i}{k_j} + \frac{k_j}{k_i} \right)b(k_i,\eta)b(k_j,\eta) \\ & \quad + \;  \frac{2}{7}\cos^2(\phi_{ij})c(k_i,\eta)c(k_j,\eta) \ .
\end{split}
\end{equation}
The functions $a(k,\eta)$, $b(k,\eta)$ and $c(k,\eta)$ are fitting formulae calibrated with N-body simulations to interpolate the results between the linear (tree-level bispectrum) and the non-linear regime bispectrum measured from the simulations:
\begin{equation}
\begin{split}
    a(k,\eta) & = \frac{1+\sigma_8^{a_6}(\eta)[0.7Q_3(n_{\mathrm{eff}})]^{1/2}(qa_1)^{n_{\mathrm{eff}}+a_2}}{1+(qa_1)^{n_{\mathrm{eff}}+a_2}} \\
    b(k,\eta) & = \frac{1+0.2a_3(n_{\mathrm{eff}}+3)(qa_7)^{n_{\mathrm{eff}}+3+a_8}}{1+(qa_7)^{n_{\mathrm{eff}}+3.5+a_8}} \\
    c(k,\eta) & =  \frac{1+4.5a_4/[1.5+(n_{\mathrm{eff}}+3)^4](qa_5)^{n_{\mathrm{eff}}+3+a_9}}{1+(qa_5)^{n_{\mathrm{eff}}+3.5+a_9}} \ .
\end{split}
\end{equation}
Although these functions have been expressed in terms of conformal time, it is completely equivalent to replace $\eta$ with the corresponding redshift $z$ as the time argument in the above expressions. $\sigma_8(\eta)$ is the standard deviation of matter density fluctuations today linearly evolved to time $\eta$ i.e. $\sigma_8(\eta) = D_{+}(\eta)\sigma_8(\eta_0)$. The effective logarithmic slope of the linear matter power spectrum today $n_{\mathrm{eff}}(k)$ reads
\begin{equation}
    n_{\mathrm{eff}}(k) = \frac{\mathrm{d} \log P_{\delta,L}^{\mathrm{3D}}(k,\eta_0)}{\mathrm{d} \log k} \ .
\end{equation}
$q \equiv k/k_{\mathrm{nl}}$ is defined with the scale $k_{\mathrm{nl}}(\eta)$ at which non-linearities start to become important and is defined as
\begin{equation}
   \frac{k_{\mathrm{nl}}^3 \;  P_{\delta,L}^{\mathrm{3D}}(k_{\mathrm{nl}},\eta)}{2 \pi^2} \equiv 1 
\end{equation}
and the function $Q_3(n_{\mathrm{eff}})$ is defined as
\begin{equation}
    Q_3(n_{\mathrm{eff}}) \equiv \frac{4-2^{n_{\mathrm{eff}}}}{1+2^{n_{\mathrm{eff}}+1}} \ .
\end{equation}
The values for the parameters calibrated using simulations as found by \citealp{Gil_Marin_2012} are:
\begin{equation*}
\begin{split}
  & a_1 = 0.484, \; a_2 = 3.740, \; a_3 = -0.849, \; a_4 = 0.392, \;  a_5 = 1.013,\\ & a_6 = -0.575, \; a_7 = 0.128, \; a_8 = -0.722,  \; a_9 = -0.926 \ . 
\end{split}
\end{equation*}
As reported by \citealp{Gil_Marin_2012}, the fitting formula with these parameter values works reasonably well for $z < 1.5$ and for $k < 0.4 \; \mathrm{ Mpc^{-1} h}$ in $\Lambda$CDM cosmologies. However, in this paper we use this fitting function for non-$\Lambda$CDM cosmologies, in particular to predict the bispectrum for cosmologies with varying dark-energy equation of state parameters by encoding the information of the latter into the fitting formula through the linear and non-linear (\verb|revised halofit|) power spectra and $\sigma_8(\eta)$ obtained using \verb|CLASS|. Our approach is similar to what has previously been done by \citealp{Sato_2013} who verified that the GM formula reasonably described the lensing bispectrum measured in N-body simulations with dynamical dark energy. Another approach, for $w$CDM cosmologies (i.e. $w_0$ = constant and $w_a$ = 0) can be taken by using the recently introduced \verb|bihalofit| fitting function for the matter bispectrum by \citealp{Takahashi_2020} which is more accurate than the GM fitting function especially in predicting the highly squeezed configurations of the matter bispectrum which the GM formula overestimates. We show results of modelling $i\zeta_{\pm}$ using \verb|bihalofit| in Appendix \ref{app:tests_on_iB}.

\section{Integrated shear bispectrum using different approximations}
\label{app:tests_on_iB}
\begin{figure}
    \centering
	\includegraphics[width=0.8\columnwidth]{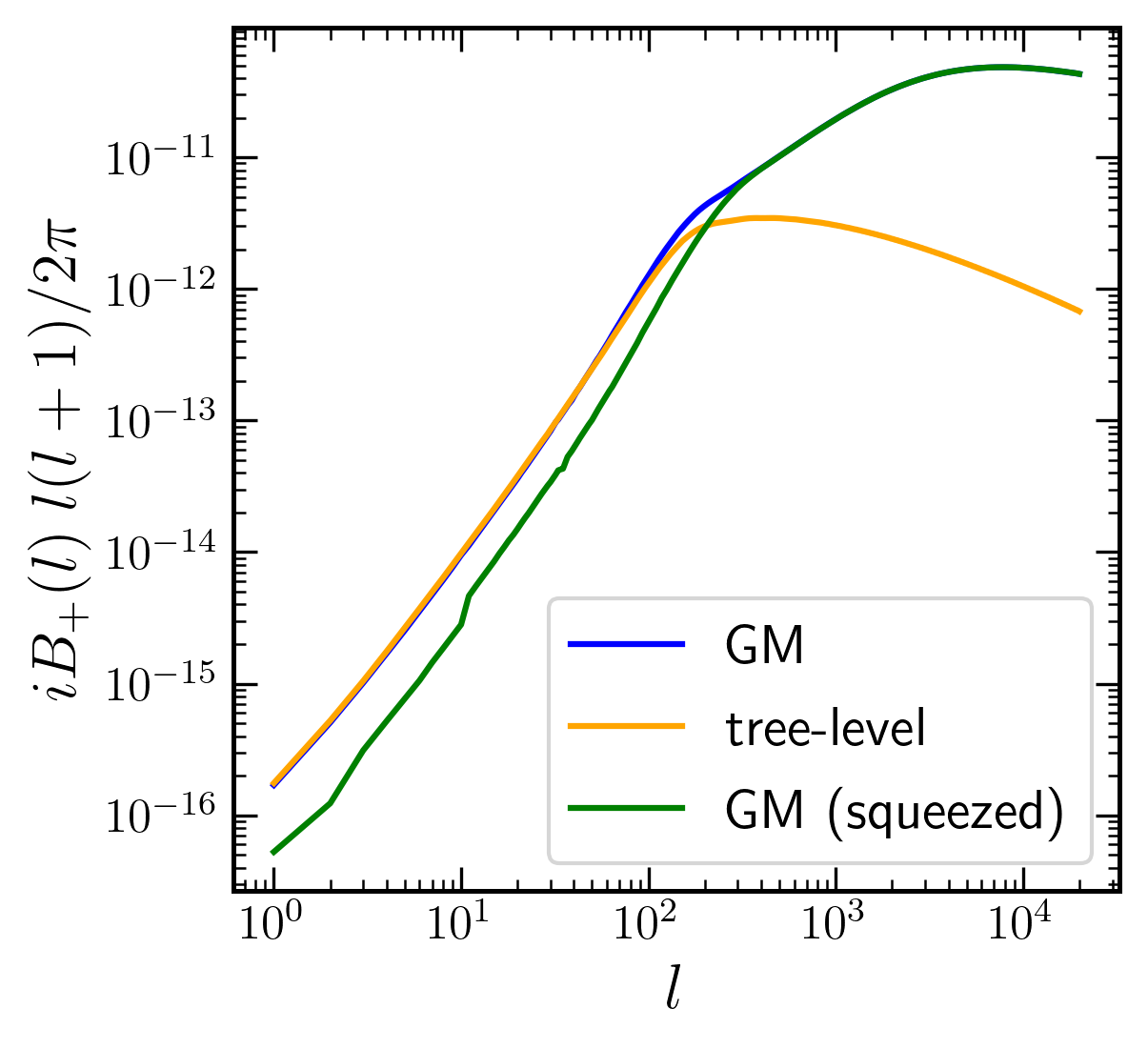}
    \caption{The scaled integrated shear bispectrum $iB_{+,222}(l)$ for source redshift bin $z_2 = 1.0334$ computed with equation \eqref{eq:integrated_bispectra} using the Gil-Marin (GM) fitting formula (blue), the tree-level bispectrum (orange) and the GM formula but only when considering squeezed configurations (green). The non-smoothness in the green curve on low-$l$ are numerical artefacts arising from the integration routine being forced to exclude sampled points in the integration volume for non-squeezed configurations. The computations for $iB_{+}$ were performed using a compensated filter of size $\theta_{\mathrm{ap}} = 70'$ and top-hat window of radius $\theta_{\mathrm{T}} = 75'$.}
    \label{fig:iB_plus_tests}
\end{figure}
\begin{figure}
	\includegraphics[width=0.9\columnwidth]{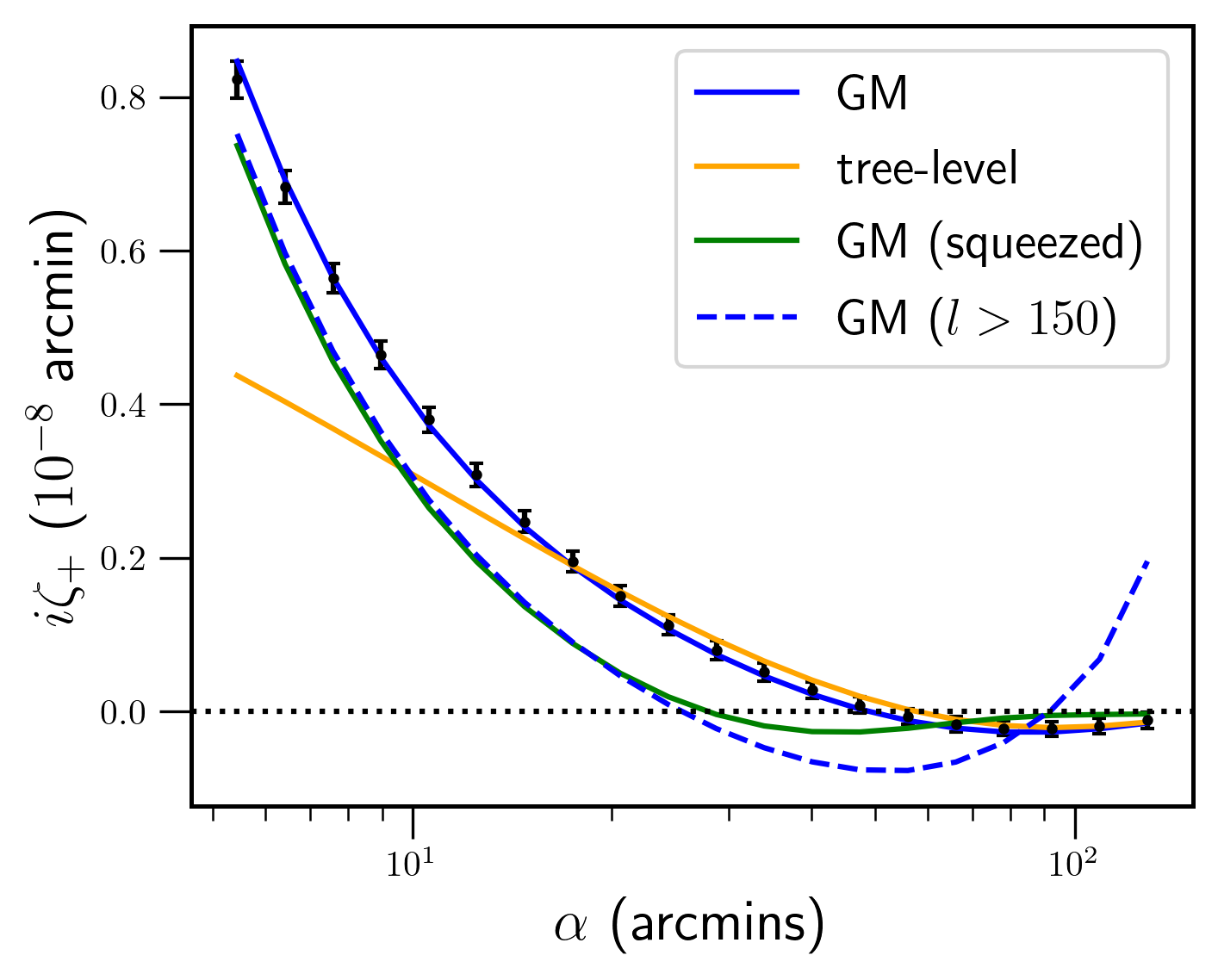}
    \caption{The integrated 3-point function $i\zeta_{+,222}(\alpha)$ for source redshift bin $z_2 = 1.0334$. The black dots with the error bars show the mean and the 1-sigma standard deviation of the measurements from the 108 T17 maps, respectively. The blue-solid curve shows the  model prediction using the GM bispectrum and in blue-dashed the prediction using only $l > 150$. The orange curve shows the predicted signal using the tree-level bispectrum and in green the signal using only the squeezed configurations of the GM formula. The theory curves include the corrections needed to account for finite angular resolution, simulation box size and shell thickness effects in the T17 simulations. The computations use a compensated filter of size $\theta_{\mathrm{ap}} = 70'$ and top-hat window of radius $\theta_{\mathrm{T}} = 75'$.}
    \label{fig:iZ_plus_tests}
\end{figure}
\begin{figure*}
	\includegraphics[width=\textwidth]{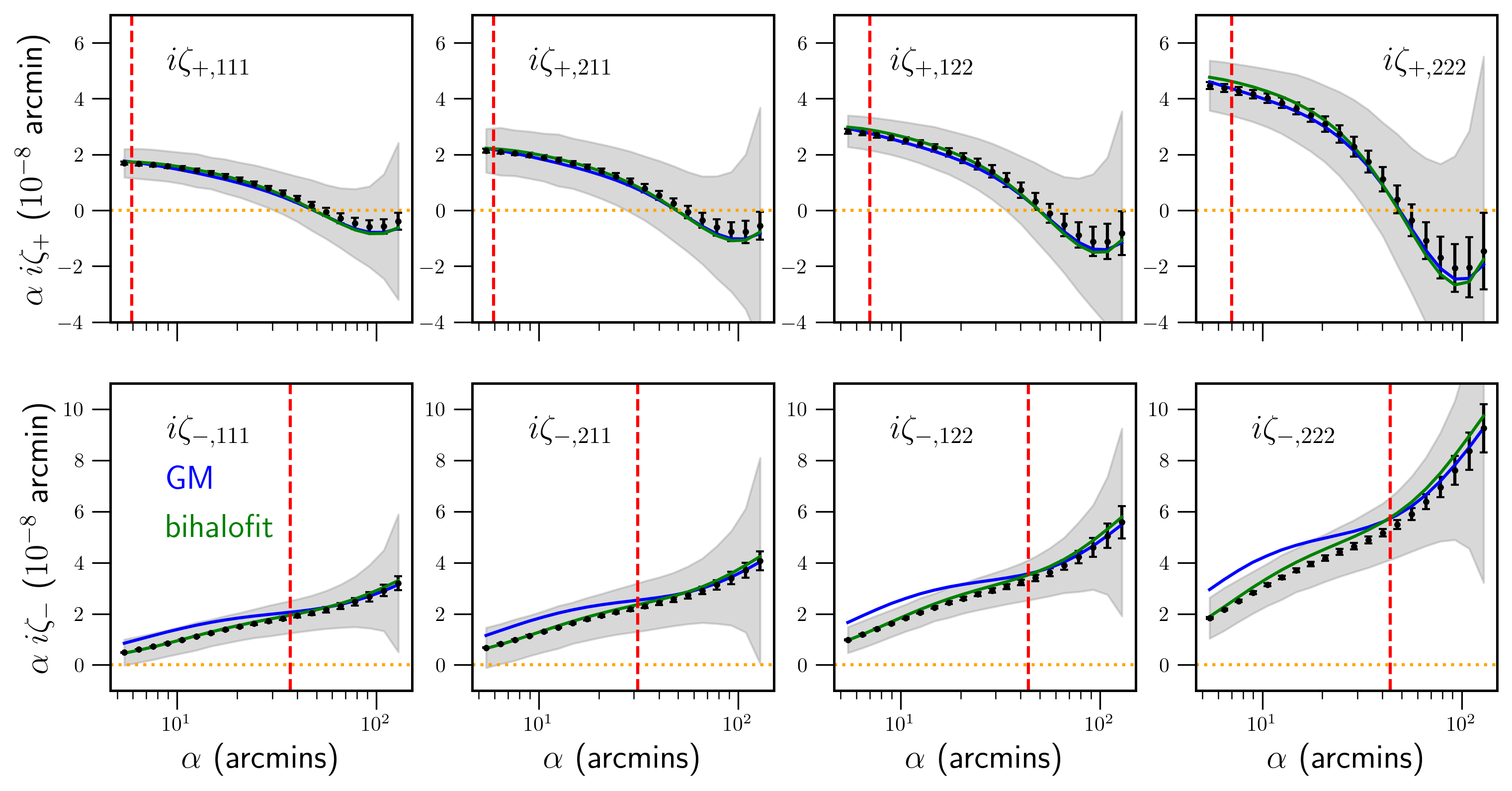}
    \caption{Same as the lower two panels of Figure \ref{fig:xi_iZ} depicting the integrated 3-point shear correlation functions $i\zeta_{\pm}(\alpha)$ for two tomographic source redshift bins $z_1 = 0.5739$ and $z_2 = 1.0334$. The blue curves show the theoretical model predictions for the statistics using the 3D matter bispectrum fitting formula by \citealp{Gil_Marin_2012} (GM --- our fiducial model for Fisher analysis). The green curves show the model predictions using a more recent 3D bispectrum fitting formula by \citealp{Takahashi_2020} (bihalofit). As described in the text, although the bihalofit fitting formula enables more accurate modelling of the $i\zeta_{-}$ correlations than the GM formula on small angular scales, we do not use it for our Fisher analysis as it is currently not applicable to cosmologies with dynamical dark energy.}
    \label{fig:iZ_GM_bihalofit}
\end{figure*}
Here we compare the results of computing the integrated bispectrum $iB_{+,222}(l)$ and correspondingly the integrated 3-point shear correlation function $i\zeta_{+,222}(\alpha)$ for the source redshift $z_2 = 1.0334$ when using different 3D matter bispectrum approximations in equations \eqref{eq:integrated_bispectra} and \eqref{eq:integrated_3_point_function_plus_bispectrum_relation}. In Figure \ref{fig:iB_plus_tests} we plot the integrated bispectrum  when computed with the GM bispectrum fitting formula (as already shown in Figure \ref{fig:iB_plus}) along with the prediction when using the tree-level bispectrum (see equation \eqref{eq:tree_level_bispectrum}). We also plot the integrated bispectrum with the GM bispectrum but only when considering elongated/squeezed configurations i.e. when two modes of the bispectrum are at least larger than two times the smallest mode. 

From the Figure, it is clear that the tree-level bispectrum and the GM formula match on low-$l$ ($l \ll 100$) corresponding to large angular scales, but differ significantly on small scales which correspond to the non-linear regime (high-$l$). On the other hand, when computing the $iB_{+}$ with only the squeezed configurations of the GM bispectrum, we find that the result matches with the full GM result only in the high-$l$ end, indicating that most of the $iB_{+}$ signal is dominated by squeezed configurations for $l$ much larger than the characteristic mode corresponding to the diameter of the patch within which the position-dependent shear correlation is measured i.e. $l \gg 2\pi / (2 \theta_{\mathrm{T}}) \approx 145$. However, for low-$l$ modes corresponding to scales approximately the size of the patch or larger, the squeezed configuration result underestimates the full GM result. This shows that our statistic probes not only the squeezed but partly also other bispectrum configurations. This also explains the non-smoothed behaviour of the squeezed $iB_{+}$ result on low-$l$ as the integration routine is forced to exclude sampled points in the integration volume for the non-squeezed configurations which contribute mostly at low-$l$. The non-smoothness is insignificant and does not affect the computation of the $i\zeta_{+}$ signal which we discuss next.

In Figure \ref{fig:iZ_plus_tests} we show the corresponding real space $i\zeta_{+,222}(\alpha)$ predictions by Hankel transforming (actually using equation \eqref{eq:Stebbins_conversion}) the integrated shear bispectra computed above and compare them with the result of the T17 simulations. The GM bispectrum computed prediction matches well with the simulations as already seen in Figure \ref{fig:xi_iZ}. The tree-level bispectrum computed $i\zeta_{+}(\alpha)$ signal only captures the result on the largest angular scales but heavily deviates in the non-linear regime. The squeezed configuration calculation of the GM bispectrum follows the trend of the simulation on small scales while slightly underestimating the measured signal. This can be attributed to the fact that at a given small angular separation $\alpha$, $i\zeta_{+}(\alpha)$ receives contributions not only from the high-$l$ end of $iB_{+}(l)$ but also from the low-$l$ end that correspond to scales larger than the separation scale (see equation \eqref{eq:Stebbins_conversion}). As seen in Figure \ref{fig:iB_plus_tests}, the squeezed bispectrum $iB_{+}(l)$ underestimates the full GM bispectrum result in the low-$l$ end thereby explaining the slight deficit. On larger scales, the squeezed bispectrum fails to describe the simulation results showing that the squeezed-limit approximation does not hold as $\alpha$ approaches the size of the patch. We also show the result of the $i\zeta_{+}(\alpha)$ signal computation using the GM bispectrum but restricting the Hankel summation of $iB_{+}(l)$ to include only $l > 150$ i.e. modes corresponding to scales much smaller than the size of the patch. Although the result does not describe the simulations, the signal matches the squeezed bispectrum results on the small scales confirming that the $i\zeta_{+}$ signal is indeed described by the squeezed limit bispectrum on these scales. However, it is worth noting that on very small scales (smaller than $5'$) the $i\zeta_{+}$ prediction with the full GM bispectrum will eventually fail to describe the T17 simulations as the integrated bispectrum result at extremely small scales (very high-$l$) receives most contribution from highly squeezed bispectrum configurations which are known to be overestimated by the GM formula \citep{Namikawa2018, Takahashi_2020}. This was apparent for the $i\zeta_{-}$ signals (see Figure \ref{fig:xi_iZ}) which are already sensitive to the very high-$l$ values of the integrated bispectrum for angular separations around $30'$ (due to the $J_4$ Bessel function weighting).

To show that one can indeed improve the modelling at smaller angular scales for the $i\zeta_{-}$ correlations, we use the recently introduced \verb|bihalofit| fiting function by \citealp{Takahashi_2020} for the 3D matter bispectrum and compare it against the results obtained using the GM fitting function which we have adopted as our fiducial modelling choice. The results comparing them both to the T17 simulations are shown in Figure \ref{fig:iZ_GM_bihalofit}. As already depicted in Figure \ref{fig:xi_iZ}, the GM model predictions for $i\zeta_{\pm}$ are shown in blue and the corresponding angular scale-cuts in the red-dashed vertical lines. The green-solid curves show the theoretical predictions using the \verb|bihalofit| fitting formula and demonstrates the significant improvement achieved in modelling the $i\zeta_{-}$ simulation results at the smaller angular scales. This is exactly due to the fact that the squeezed bispectrum configurations are more accurately predicted by \verb|bihalofit| than the GM fiiting function \citep{Takahashi_2020}. Using \verb|bihalofit| would thus allow to push the currently imposed angular scale-cuts down to even smaller scales and retain larger parts of the data vector. Nevertheless, we still use the $i\zeta_{\pm}$ model predictions with the GM formula for the Fisher analysis because currently \verb|bihalofit| is only applicable to $w$CDM cosmologies (i.e. $w_0$ = constant and $w_a$ = 0) whereas a major goal of our analysis (see section \ref{sec:results_Fisher}) is to investigate the constraining power of $i\zeta_{\pm}$ for cosmologies with $w_a \neq$ 0. The GM fitting function in combination with the \verb|revised halofit| non-linear power spectrum has no such restriction and has previously been validated for cosmologies with dynamical dark energy \citep{Sato_2013}. Furthermore, as we only use those parts of the GM computed $i\zeta_{\pm}$ model vectors which have been validated on the T17 simulations (ensured with the imposed scale-cuts shown in Figure \ref{fig:xi_iZ}) our parameter constraints are more on the conservative side (see Figure \ref{fig:fisher_contours}). Including smaller angular scales of the $i\zeta_{-}$ data vector with improved modelling is expected to only improve the overall constraining power. However, further improved modelling in cosmologies with $w_a \neq$ 0 is beyond the scope of this paper and is left as a direction for future work.

\section{Impact of systematic offset between model and data vectors on parameter constraints}
\label{app:systematic_check}
\begin{figure}
	\includegraphics[width=\columnwidth]{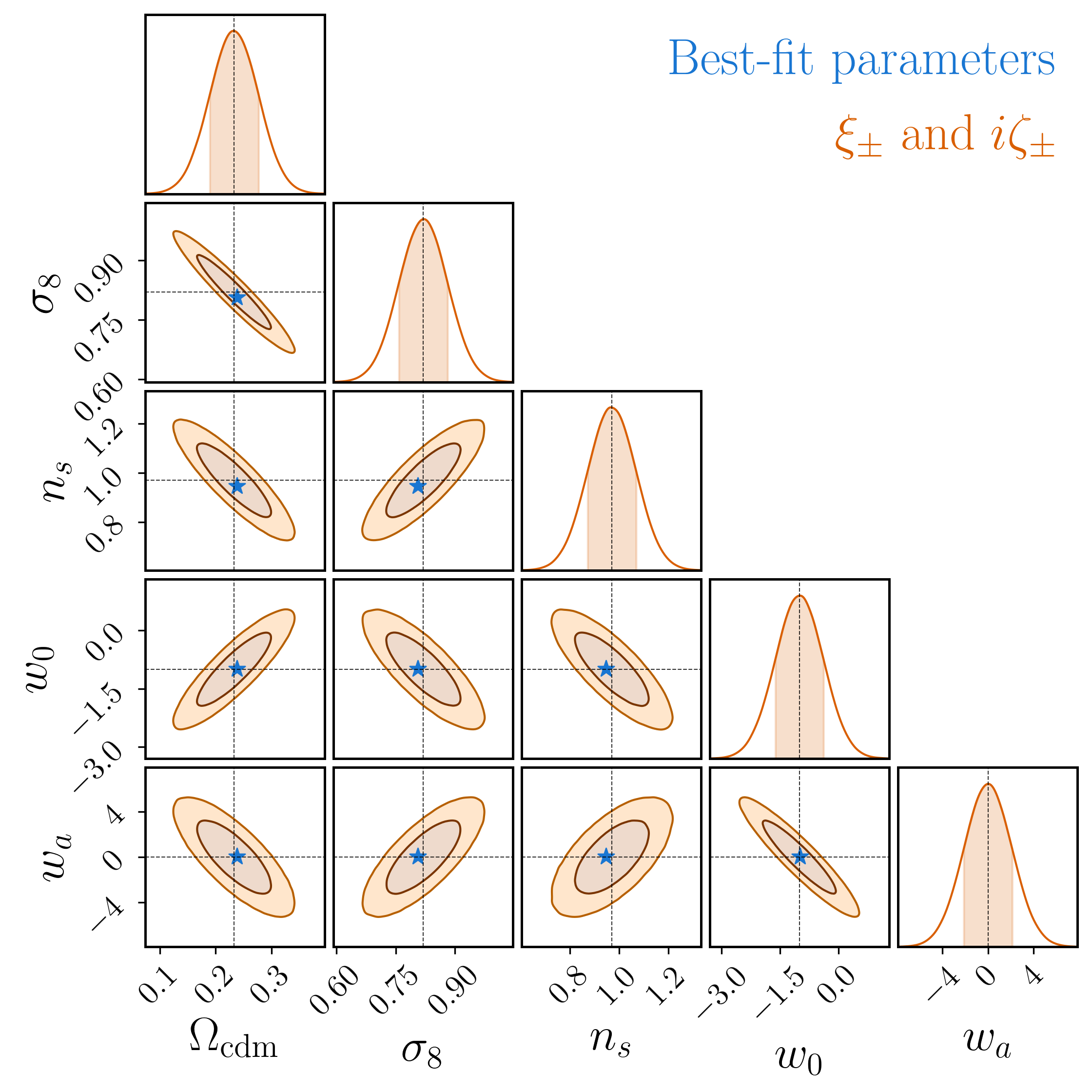}
    \caption{Offsets between fiducial parameters (black dotted lines) and best-fitting parameters (blue stars) in the parameter planes. The Fisher contours expected from the analysis of the entire model vector (after imposing scale-cuts) are shown in orange centred around the fiducial parameters.}
    \label{fig:fisher_joint_best_fit_params}
\end{figure}

Here we discuss the impact of the remaining systematic offset between the model $M$ and data vector $\overline{D}$ (see Figure \ref{fig:xi_iZ}) after imposing the angular scale-cuts in our analysis. A systematic offset would amount to a bias in our parameter constraints which would cause the Fisher contours in Figure \ref{fig:fisher_contours} to be centred around the wrong cosmological values $\boldsymbol{\pi}^{0}$ --- in our case the fiducial parameters. In other words, we want to explore how much the best-fitting\footnote{MP stands for maximum posterior in the notation of \citealp{friedrich2020}.} parameters $\boldsymbol{\pi}^{\mathrm{MP}}$ of the model describing the data vector is off from $\boldsymbol{\pi}^{0}$. In order to do so, we need to minimize the $\chi^2(\boldsymbol{\pi})$ as a function of the parameters (see equation \eqref{eq:chi_squared}) between the data and model. We already saw in section \ref{sec:results_validation} that the $\chi^2(\boldsymbol{\pi} = \boldsymbol{\pi}^{0})$ between $\overline{D}$ and $M(\boldsymbol{\pi}^{0})$ has a value of 1.08. We now want to find the parameters $\boldsymbol{\pi}^{\mathrm{MP}}$ which describe the data vector with the lowest $\chi^2_{\mathrm{MP}}$. We adopt the approach of \citealp{friedrich2020} (see their section 5.1) and study a linearized approximation of the model vector as a function of the parameters $M(\boldsymbol{\pi})$ around the fiducial parameters $\boldsymbol{\pi}^{0}$. This allows us to write the best-fitting parameters as (see equation 32 of \citealp{friedrich2020}):
\begin{equation}
    \boldsymbol{\pi}^{\mathrm{MP}} = \boldsymbol{\pi}^{0} + \mathbf{F}^{-1} \mathbf{x} 
\end{equation}
where we have assumed no priors on the parameters. $\mathbf{F}$ is the Fisher matrix (see equation \eqref{eq:fisher_model_vector}) of the model vector and $\mathbf{x}$ is another vector with components:
\begin{equation}
    x_i = \left(\overline{D}-M(\boldsymbol{\pi^{0}})\right)^{\mathrm{T}}\mathbf{C}^{-1} \left( \frac{\partial M(\boldsymbol{\pi})}{\partial \pi_{i}} \right) 
\end{equation}
where $\mathbf{C}$ is the data-covariance matrix (see equation \eqref{eq:data_covoriance_independent_realisations}) and $\frac{\partial M(\boldsymbol{\pi})}{\partial \pi_{i}}$ are the derivatives of the model with respect to the parameters,  evaluated at the fiducial values $\boldsymbol{\pi} = \boldsymbol{\pi}^{0}$. We show our best-fitting parameters for the model describing the entire T17 data vector (after imposing the scale-cuts) in Figure \ref{fig:fisher_joint_best_fit_params} which can be seen to scatter very closely around the fiducial parameters. We also plot the orange contours (see Figure \ref{fig:fisher_contours}) of the parameters from the Fisher analysis for the entire data vector (with $\xi_{\pm}$ and $i\zeta_{\pm}$ including the assumed scale-cuts). The absolute offsets of the best-fitting parameters from the fiducial values in units of the marginalized 1-sigma Fisher constraints for the 5 parameters $\Omega_{\mathrm{cdm}}, \sigma_8, n_s, w_0, w_a$ are 0.12, 0.22, 0.25, 0.02, and 0.02 respectively. As these offsets are smaller than one-third the marginalized 1-sigma constraints in the parameter planes, we conclude that our fiducial model after imposing the angular scale-cuts describes the T17 data-vector very well and there is no significant bias in our results. Ideally, one should include these offsets as a systematic error but as they are not significant we deem it safe to ignore for our analysis. 

\section{Validating the use of lognormal data-covariance on parameter constraints using the precision matrix expansion (PME)}
\label{app:precision_matrix_expansion}
In this Appendix we test whether the use of lognormal \verb|FLASK| simulations (see section \ref{sec:flask_sims}) as a model for computing the data-covariance matrix $\mathbf{\hat{C}}$ and its inverse, the precision matrix, causes any significant over/under-estimation of the Fisher constraints presented in Section \ref{sec:results_Fisher}. For our purpose we use the precision matrix expansion (PME) formalism developed in Section 3 of \citealp{Friedrich2017} which we explain below.

Let us assume that on one hand we know the true data-covariance $\mathbf{C}_{\mathrm{true}}$ (e.g. from N-body simulations) and on the other hand we have a model $\mathbf{C}$ for the covariance (e.g. lognormal model). We can then write:
\begin{equation}
\begin{split}
    \mathbf{C}_{\mathrm{true}} & =  \mathbf{C} + \mathbf{C}_{\mathrm{true}} - \mathbf{C} \\
    & = \left( \mathbf{1} + (\mathbf{C}_{\mathrm{true}} - \mathbf{C}) \mathbf{C}^{-1} \right) \mathbf{C} \\
    & = \left( \mathbf{1} + \mathbf{X} \right) \mathbf{C}
\end{split}
\end{equation}
where $\mathbf{1}$ is the identity matrix and $\mathbf{X} \equiv (\mathbf{C}_{\mathrm{true}} - \mathbf{C}) \mathbf{C}^{-1}$. The true precision matrix i.e. $\mathbf{C}_{\mathrm{true}}^{-1}$ can then be expressed as
\begin{equation}
\begin{split}
    \mathbf{C}_{\mathrm{true}}^{-1} & = \Big( \left( \mathbf{1} + \mathbf{X} \right) \mathbf{C} \Big)^{-1} \\
    & = \mathbf{C}^{-1} \left( \mathbf{1} + \mathbf{X} \right)^{-1} \\
    & = \mathbf{C}^{-1} \left( \mathbf{1} - \mathbf{X} + \mathbf{X}^2 + \mathcal{O}[\mathbf{X}^3]\right)
\end{split}
\end{equation}
where in the last line we have used the geometric series expansion of $\left( \mathbf{1} + \mathbf{X} \right)^{-1}$. We can now write estimates for the true precision matrix for different orders in $\mathbf{X}$. Up to the zeroth order we have:
\begin{equation}
\begin{split}
    \mathbf{C}_{\mathrm{true},0\mathrm{th}}^{-1} & \equiv \mathbf{C}^{-1} 
\end{split}
\end{equation}
which is the inverse of the model covariance matrix. This is exactly what we have used as our fiducial precision matrix throughout the main text of the paper for computing the $\chi^2$ values and for our Fisher analysis. To remind ourselves, we first estimate the model covariance matrix $\mathbf{\hat{C}}$ from \verb|FLASK| lognormal simulations and we then write (see equation \eqref{eq:Hartlap_correction}) an unbiased estimate for the inverse model covariance $\mathbf{C}^{-1}$ using the \citealp{Hartlap2007} correction factor which assumes that the estimated covariance matrix $\mathbf{\hat{C}}$ is distributed according to a Wishart distribution. Now, going up to 1st order in PME we have:
\begin{equation}
\begin{split}
\label{eq:1st_PME}
    \mathbf{C}_{\mathrm{true},1\mathrm{st}}^{-1} & \equiv \mathbf{C}^{-1} - \mathbf{C}^{-1} \mathbf{X} \\
    & = \mathbf{C}^{-1} - \mathbf{C}^{-1} (\mathbf{\hat{C}}_{\mathrm{true}} - \mathbf{C}) \mathbf{C}^{-1} \\
    & = 2\mathbf{C}^{-1} - \mathbf{C}^{-1} \mathbf{\hat{C}}_{\mathrm{true}} \mathbf{C}^{-1} \ .
\end{split}
\end{equation}
The second term in the last line is the leading order correction to our model precision matrix $\mathbf{C}^{-1}$ and as depicted, is written using a direct estimate $\mathbf{\hat{C}}_{\mathrm{true}}$ of the true covariance matrix. This is a crucial point because even if one has only a few N-body mocks to estimate $\mathbf{C}_{\mathrm{true}}$, the correction term does not involve the inversion of this estimate. This is applicable for us as we only have a finite number of T17 mocks to estimate $\mathbf{\hat{C}}_{\mathrm{true}}$. In order to do this, we first add shape-noise to all the 108 T17 simulations (similar to what we did for the \verb|FLASK| lognormal maps as described in section \ref{sec:flask_sims}) and then cut out two big circular footprints of 5000 square degrees (approximately the size of the DES footprint) in each hemisphere of a given map. This gives us effectively 216 DES sized true mocks from which we estimate $\mathbf{\hat{C}}_{\mathrm{true}}$. To ensure that equation \eqref{eq:1st_PME} is an unbiased estimator, we evaluate the two $\mathbf{C}^{-1}$ terms appearing in the $\mathbf{C}^{-1} \mathbf{\hat{C}}_{\mathrm{true}} \mathbf{C}^{-1}$ from 2 independent sets of \verb|FLASK| lognormal simulations. We also symmetrize the final term after its evaluation.

\begin{figure}
	\includegraphics[width=\columnwidth]{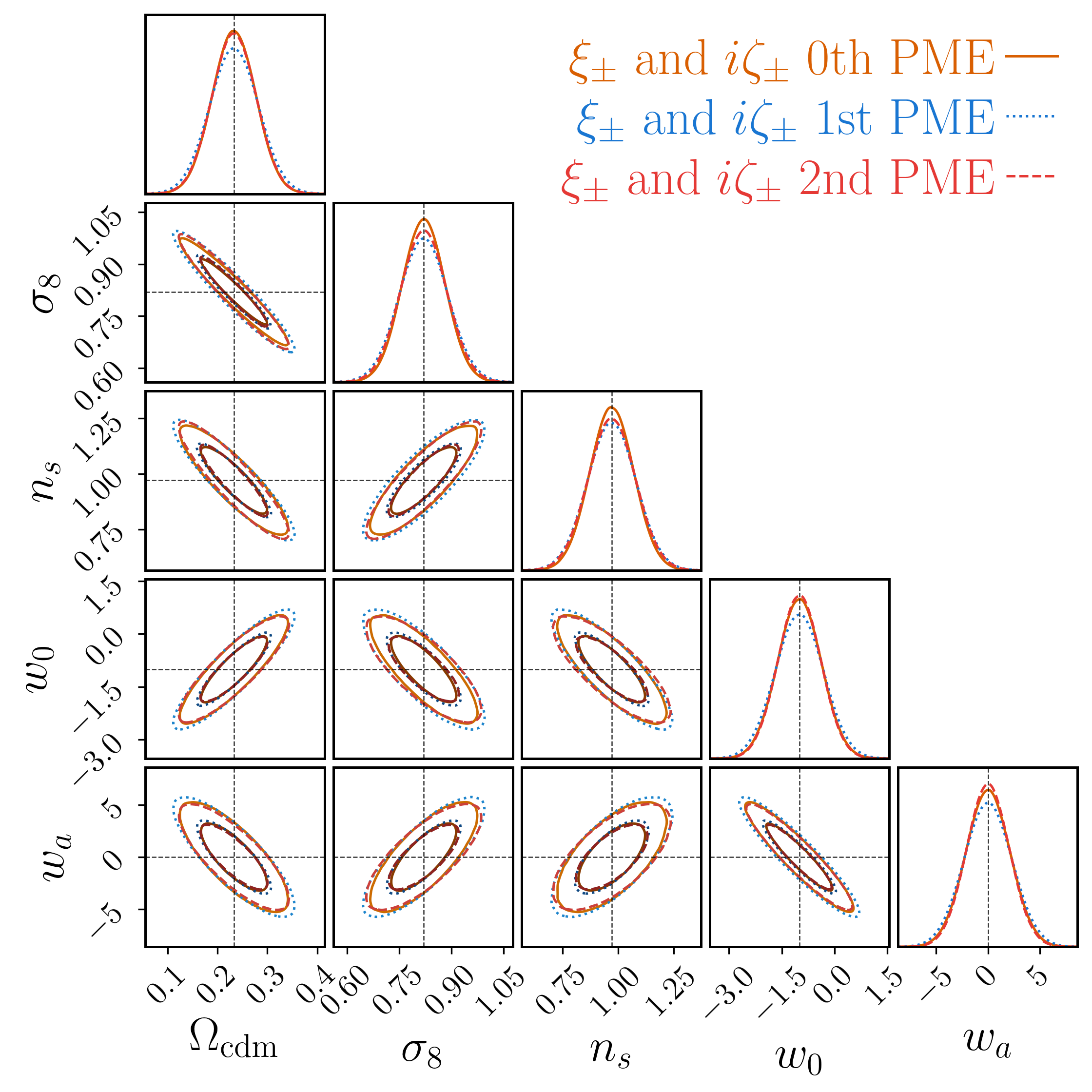}
    \caption{The Fisher contours expected from the analysis of the entire model vector (after imposing scale-cuts) for 3 different use cases of the inverse covariance matrix --- zeroth order precision matrix expansion PME (orange solid --- same as the constraints shown in right panel of Figure \ref{fig:fisher_contours}), 1st order PME (blue dotted) and 2nd order PME (red dashed). The higher order PME terms hardly change the contours obtained when using the zeroth order precision matrix which is estimated from lognormal simulations.}
    \label{fig:fisher_joint_PME}
\end{figure}

Practically, we can use the $N_r = 216$ T17 footprints estimated true data-covariance $\mathbf{\hat{C}}_{\mathrm{true}}$ and invert it to obtain the precision matrix with the Hartlap correction formula (see equation \eqref{eq:Hartlap_correction}). This is possible as $N_r$ is larger than the size of our entire data-vector after imposing scale-cuts $N_d = 182$ (see Table \ref{tab:S_N_ratio}). However, as discussed in \citealp{Taylor2013}, the uncertainty that one encounters in the estimation of the precision matrix goes as $\sqrt{2/(N_r-N_d-4)}$ which upon using the T17 covariance would result in a very large error of roughly 25 per cent on the Fisher matrix. Furthermore, even when the covariance estimate can be safely inverted (i.e.\ when $N_r \gg N_d$) the noise in the precision matrix estimate still leads to a significant additional scatter in maximum-likelihood parameters in actual likelihood analyses, unless $N_r - N_d \gg N_d$ (cf.\ \citealp{Dodelson2013} as well as Figure 1 of \citealp{Friedrich2017}). Keeping this in mind as well as the fact that in future applications we may add further redshift bins and hence consider even bigger data vectors, we opt for the strategy of \citealp{Friedrich2017} instead of standard inversion.

Following equation (12) of \citealp{Friedrich2017}, we can also write an unbiased estimator of the true precision matrix up to second order:
\begin{equation}
\begin{split}
\label{eq:2nd_PME}
    \mathbf{C}_{\mathrm{true},2\mathrm{nd}}^{-1} & \equiv \mathbf{C}^{-1} - \mathbf{C}^{-1} \mathbf{X} + \mathbf{C}^{-1} \mathbf{X}^2\\
    & = 3\mathbf{C}^{-1} - 3\mathbf{C}^{-1} \mathbf{\hat{C}}_{\mathrm{true}} \mathbf{C}^{-1} & \\
    & \quad + \mathbf{C}^{-1} \frac{\nu^2\mathbf{\hat{C}}_{\mathrm{true}}\mathbf{C}^{-1}\mathbf{\hat{C}}_{\mathrm{true}} - \nu\mathbf{\hat{C}}_{\mathrm{true}}\mathrm{tr}\left( \mathbf{C}^{-1}\mathbf{\hat{C}}_{\mathrm{true}} \right)}{\nu^2+\nu-2}\mathbf{C}^{-1}
\end{split}
\end{equation}
where $\nu=N_r-1$ with $N_r$ being the number of mock realizations used for estimating $\mathbf{\hat{C}}_{\mathrm{true}}$ and $\mathrm{tr}\left( . \right)$ stands for evaluating the trace of a matrix. Just as the correction factor that \citealp{Hartlap2007} advocated for the unbiased estimation of the inverse of a matrix that is Wishart distributed, the final term in the above equation also stems from the same assumption i.e. $\mathbf{\hat{C}}_{\mathrm{true}}$ is Wishart distributed. For more details, the reader is referred to section 3 and Appendix B of \citealp{Friedrich2017}. For evaluating the final term, as mentioned earlier, we again ensure that each of the involved $\mathbf{C}^{-1}$ are estimated from independent sets of \verb|FLASK| lognormal simulations.

In Figure \ref{fig:fisher_joint_PME} we show the effect on the Fisher parameter contours of the entire data-vector (with $\xi_{\pm}$ and $i\zeta_{\pm}$) when using the precision matrix as evaluated with the zeroth order PME (orange solid), 1st order PME (blue dotted) and 2nd order PME (red dashed) expressions. The correction induced to the parameter constraints at 1st order in PME hardly changes the results obtained using the 0th order PME (our fiducial lognormal model precision matrix). The result of adding even higher order terms up to second order in PME shows 
remarkable agreement with the fiducial contours (compare the red dashed and orange solid ellipses). 

We also evaluated the FoM in the $w_0-w_a$ plane for $\xi_{\pm}$, $i\zeta_{\pm}$ and their combined data vector. The values are reported in Table \ref{tab:FOM_PME}.
\begin{table}
\caption{The dark energy figure of merit (FoM) as measured from the marginalized Fisher constraints in the $w_0-w_a$ plane (see equation \eqref{eq:FOM}) where the parameter covariance $\mathbf{C}_{\boldsymbol{\pi}}$ is measured in the Fisher analysis with 3 different versions of the precision matrix: the 0th order PME (fiducial lognormal precision matrix), 1st order PME and 2nd order PME. The results are shown for $\xi_{\pm}$, $i\zeta_{\pm}$ and their combined data vector. The factor of improvement achieved in the FoM of the joint $\xi_{\pm}$ and $i\zeta_{\pm}$ compared to $\xi_{\pm}$ alone are also shown.}
\label{tab:FOM_PME}
\centering
\begin{tabular}{|c|c|c|c|}
\hline
Data vector & 0th PME & 1st PME & 2nd PME \\
\hline
\hline
FoM $\xi_{\pm}$ & 0.78 & 0.77 & 0.76 \\
FoM $i\zeta_{\pm}$ & 0.19 & 0.19 & 0.18 \\
FoM $\xi_{\pm}$ and $i\zeta_{\pm}$ & 2.28 & 1.86 & 2.40 \\
\hline
\hline
Factor improvement & 2.9 & 2.4 & 3.2\\
\hline
\end{tabular}
\end{table}
One sees that the relative factor of improvement of the FoM on performing a Fisher analysis of the joint data vector as compared to only $\xi_{\pm}$ is larger than 2 in all cases --- irrespective of whether one uses the fiducial lognormal model precision matrix or after correcting the model precision matrix with leading order PME terms estimated from the T17 simulations. All these results lead us to conclude that our analysis with the \verb|FLASK| lognormal covariance matrix is well justified and gives us robust qualitative and quantitative estimates for the significant improvement achieved in the parameter constraints upon complementing 2-point shear analysis with the measurement of integrated 3-point shear correlation functions.

\bsp	
\label{lastpage}
\end{document}